\documentclass[]{emulateapj}
\usepackage{graphicx}
\usepackage{rotating}
\usepackage{mathtools}
\usepackage[dvips]{color}
\usepackage[title]{appendix}
\usepackage{subfigure}

\begin{document}
\newcommand{\kms}{\mbox{km~s$^{-1}$}}
\newcommand{\s}{\mbox{$''$}}
\newcommand{\mloss}{\mbox{$\dot{M}$}}
\newcommand{\mdot}{\mbox{$\dot{M}$}}
\newcommand{\my}{\mbox{$M_{\odot}$~yr$^{-1}$}}
\newcommand{\ls}{\mbox{$L_{\odot}$}}
\newcommand{\um}{\mbox{$\mu$m}}
\newcommand{\ujy}{\mbox{$\mu$Jy}}
\newcommand{\ms}{\mbox{$M_{\odot}$}}

\newcommand{\vexp}{\mbox{$V_{\rm exp}$}}
\newcommand{\vsys}{\mbox{$V_{\rm sys}$}}
\newcommand{\vlsr}{\mbox{$V_{\rm LSR}$}}
\newcommand{\tex}{\mbox{$T_{\rm ex}$}}
\newcommand{\teff}{\mbox{$T_{\rm eff}$}}
\newcommand{\tmb}{\mbox{$T_{\rm mb}$}}
\newcommand{\trot}{\mbox{$T_{\rm rot}$}}
\newcommand{\tkin}{\mbox{$T_{\rm kin}$}}
\newcommand{\dens}{\mbox{$n_{\rm H_2}$}}
\newcommand{\bri}{\mbox{erg\,s$^{-1}$\,cm$^{-2}$\,\AA$^{-1}$\,\newcommand{\kms}{\mbox{km~s$^{-1}$}}$^{-2}$}}
\newcommand{\brib}{\mbox{erg\,s$^{-1}$\,cm$^{-2}$\,arcsec$^{-2}$}}
\newcommand{\flux}{\mbox{erg\,s$^{-1}$\,cm$^{-2}$\,\AA$^{-1}$}}
\newcommand{\ha}{\mbox{H$\alpha$}}

\title{High-Velocity Bullets from V Hydrae, an AGB Star in Transition: Ejection History and Spatio-Kinematic Modeling}
\author{S. Scibelli\altaffilmark{1,2}, R. Sahai\altaffilmark{1}, \& M. R. Morris\altaffilmark{3}}


\altaffiltext{1}{Jet Propulsion Laboratory, MS\,183-900, California Institute of Technology, Pasadena, CA 91109, USA}
\altaffiltext{2}{Steward Observatory, University of Arizona, Tucson, AZ 85721}
\altaffiltext{3}{Department of Physics and Astronomy, UCLA, Los Angeles, CA 90095-1547, USA}

\email{raghvendra.sahai@jpl.nasa.gov}
\begin{abstract}

The carbon star V Hydrae (V Hya) provides new insight into the nature of the launching mechanism of jet-like outflows that are believed to be the cause of the poorly understood transition phase of AGB stars into aspherical planetary nebulae. V Hya has been shown to periodically eject collimated gas blobs at high velocities (``bullets"). By analyzing data from HST/STIS 2-D spectra, obtained at six epochs spaced over a decade that show 4 successively ejected bullets with a spacing of ~8.5 years, we have created kinematic models of the dynamical evolution of a specific bullet (\#1) for the first three observed epochs (2002, 2003, 2004) using a 3D spatio-kinematic code, SHAPE. Using these models, we fit the observed morphology, line-of-sight velocity, proper motion and intensity for the extended, gaseous bullet as a function of time over a period of 2 years, in order to constrain its 3D movement and the evolution of its physical properties over this period. Our results suggest that although bullet \#1's motion is predominantly ballistic, there are small but significant changes in the position angle and inclination angle of the long (symmetry) axis of the bullet that tilt it progressively towards the symmetry axis of the bipolar molecular nebula around V\,Hya. In contrast, bullet\#3 shows strong acceleration soon after ejection. We discuss the possibilities that bullet acceleration is caused by either a non-radial magnetic field and/or by hydrodynamic interaction with the ambient gas through which the bullet is traveling.

\end{abstract}


\section{Introduction}

As asymptotic giant branch (AGB) stars age, over 10$^4$-10$^5$ years, they eject over half or more of their mass in slow winds, and then, in a short 100-1000 year period, are transformed into planetary nebulae (PNe) with a dazzling variety 
of morphologies and widespread presence of point-symmetry \citep{2011AJ....141..134S}. 
Recent morphological studies with HST support the idea that directed, high-speed outflows initiated during the very late AGB phase play a crucial role in the transformation to PNe (\citealt{1998AJ....116.1357S}, \citealt{2002ARA&A..40..439B}, \citealt{2007ApJ...658..410S}). Circumstellar envelopes around AGBs typically have a spherically symmetric shape and undergo uniform expansion. Due to extensive mass loss and the presence of these collimated jet-like outflows, the once spherically structured envelopes are re-sculpted in the very late stage of AGB development, producing the variety of aspherical shapes that we observe. In fact, high-resolution 3D hydrodynamical simulations have shown that time-dependent ejections and/or precessing jets successfully reproduce the bipolar morphology of protoplanetary nebulae (pPNe) and PNe, such as Hen 3-1475 (\citealt{2004A&A...419..991V}, \citealt{2004ASPC..313..487R}) and IC4634 \citep{2008ApJ...683..272G}.

Jet-like outflows are probably not launched by the AGB stars themselves, but by compact main-sequence or white dwarf companions (e.g., \citealt{1987PASP...99.1115M}, \citealt{1990AJ.....99.1869S}, \citealt{2008MNRAS.391.1063A}, \citealt{2012Sci...338..773B}, \citealt{2014MNRAS.439.2014T}, \citealt{2016ApJ...820..134H}, etc.). In \citealt{2014A&A...561A.145R} they model a precessing jet with a time-dependent ejection velocity that is launched from the secondary star of a binary system, finding the [SII] intensity maps predicted for the sinusoidal ejection velocity models have morphologies that agree with the [SII] HST images of multipolar pPN CRL 618. Observational evidence of jets in AGB stars is rare, however, as this transition phase is so short. Observational evidence of binarity has also been lacking, but recent progress has been made using UV and X-ray observations to detect accretion activity associated with the gravitational capture of material from the primary's wind by a 
main-sequence companion (e.g., \citealt{2015ApJ...810...77S}, \citealt{2016JPhCS.728d2003S}, \citealt{2016MNRAS.461.3036O}, \citealt{2018ApJ...860..105S}).

The carbon star, V\,Hya, has been observed to possess high-speed, collimated outflows, a slow outflow, and equatorially-dense structures seen via millimeter-wave CO observations (e.g., \citealt{1997A&A...326..318K}, \citealt{2004ApJ...616L..43H}). Observations of CO 4.6\micron~vibration-rotation absorption lines in V\,Hya, using the KPNO 4-m FTS \citep{1988A&A...201L...9S}, followed by a detailed study covering 6 epochs \citep{2009ApJ...699.1015S}, showed the presence of several high-velocity shells ($V_{exp} \approx 70-120$ \kms) of warm gas ($\sim115-570$ K); 
the slow outflow ($V_{exp} \approx 10$ \kms) is also detected. V\,Hya also shows the presence of FUV emission that may be related to a hot, active accretion disk \citep{2008ApJ...689.1274S}.

Using STIS/HST, \citealt{2003Natur.426..261S}, hereafter Setal03, found in V\,Hya, (i) a newly launched,
high-speed bullet offset by $\sim0\farcs25$ from the central star with a projected line-of-sight (LOS) velocity of 240\,\kms, and measured its proper motion, and (ii) a hot, slowly expanding (10--15\,\kms) central disk-like structure of diameter $\sim0\farcs6$. Recently, \citealt{2016ApJ...827...92S}, hereafter Paper I, 
reported the detection of a sequence of emission-line blobs at different distances from the star, and used these to 
construct a detailed history of bullet-like mass ejections from V\,Hya (observations directly linked with the models presented in this paper) that suggests that the most likely mechanism to explain the observed bullet ejections is a binary model which includes a main-sequence star orbiting V\,Hya every $\sim$ 8.5 years.  



In this paper we present quantitative models of our HST STIS observations. These models were produced by the 3D spatio-kinematic code, SHAPE\,(\citealt{2006RMxAA..42...99S}, \citealt{2011ITVCG..17..454S}).
In \S\,\ref{obs} we briefly summarize the observations and data reduction process; in \S\,\ref{shapeModeling} we discuss the specifics of the modeling process and how we arrived at our best-fit model; in \S\,\ref{results} we discuss our model's implications, and finally concluding remarks are made in \S\,\ref{conclusion}. We include further discussions in the Appendices, i.e., instrumental corrections in Appendix A, and specifics on the preceding models that led to our best-fit model in Appendix B. Additionally, for those curious about the components of SHAPE, we include a detailed description of the creation of specific models in Appendix C. 

We have adopted, as in our previous studies of this object, a distance of $D=0.4$\,kpc.
Until recently, V\,Hya lacked a significant measurement of its trigonometric parallax (its Hipparcos parallax/error is $1.440$\,mas/$1.41$\,mas: \citealt{2007A&A...474..653V}). In the recent 
GAIA Data Release 2 (L. Lindegren et al. 2018, {\it in prep}, arXiv), its measured parallax is $2.09\pm0.13$\,mas, giving a distance of $0.48\pm0.03$\,kpc. Since this is roughly within 2.7$\sigma$ of our 
adopted value, we have conservatively kept our original distance estimate for this paper, especially considering the DR2 results have yet to be scrutinized by the astrophysical community. The slightly larger distance, if correct, would imply a $\sim$20\,\% increase in velocities perpendicular to the LOS, derived from proper motions, 
and a $\sim$44\,\% increase in estimated masses, and does not significantly affect our conclusions.

\section{Summary of Observations} \label{obs}

\begin{figure}
\vspace{2in}
\centering
\includegraphics[width=0.75\textwidth]{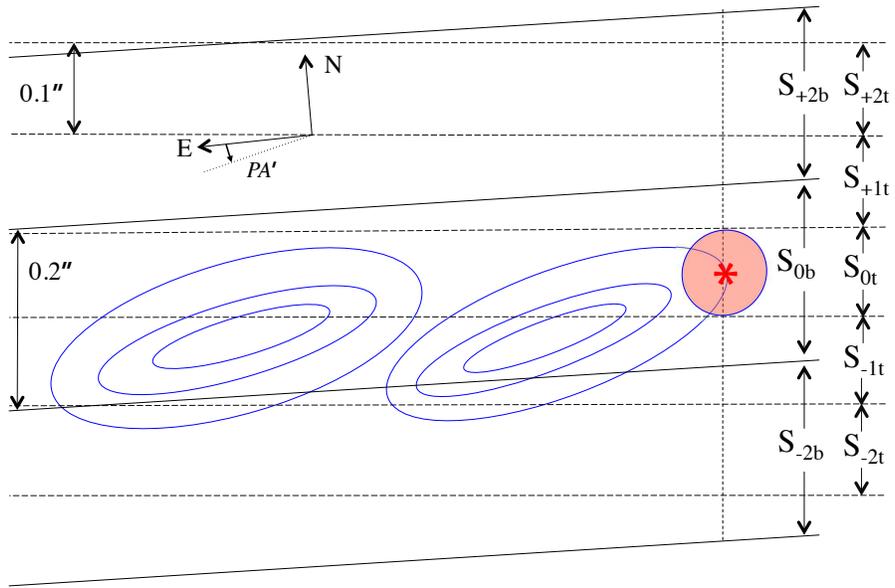} 
\caption{\label{slitorien} Schematic representation of the slit mosaics for different epochs and the orientation of bullet\#1 as it is ejected away from VHya (adapted from Paper I). Broad slits (i.e. with width = $0\farcs2$) used for for epoch 1 are shown as solid black lines and denoted with a `b' in the slit name. Thin slits (i.e. with width = $0\farcs1$) used for epochs 2 and 3 (in epoch 3, slit $S_{-2t}$ wasn't used)  are shown as dashed black lines and denoted with a `t' in the slit name.}
\end{figure}

An detailed description of observations and data reduction techniques is presented in Paper I. Spectra of V Hya from the STIS instrument on-board the \textit{Hubble Space Telescope (HST)} were taken during 6 epochs in two 3 year periods over the course of more than 10 years. In the present paper, we present models of the first 3 year period (2002, 2003, 2004) which covered epochs 1--3, collectively referred to as Period 1. The second set, Period 2, covered epochs 4--6 over another 3 year period (2011, 2012, 2013). The slit orientation changed slightly in declination from epoch to epoch. In epochs 1, 4, 5, and 6, the slits are $0\farcs2$ in width and spaced $0\farcs2$ apart, starting at $0\farcs0$, whereas in epochs 2 and 3 the slits are $0\farcs1$ in width and spaced $0\farcs1$ apart, starting at a declination offset of $0\farcs0$ (Fig.\,\ref{slitorien}). The on-center slit of width $0\farcs2$ is labeled $S_{0b}$ and the on-center slit of width $0\farcs1$ is labeled $S_{0t}$; the symbols ``b" or ``t" signify whether a broad ($0\farcs2$) or thin ($0\farcs1$) slit was used. The off-center slits are similarly named, but with the zero replaced by a number representing the declination offset from the star in units of tenths of an arcsecond, i.e., the most off-center slit to the south/negative declination direction for the first epoch is labeled $S_{-2b}$. In Table \,\ref{modeltable} we present central intensities for our observed blobs in the [SII]$\lambda$ $4069.7$ line (hereafter [SII]). Data were reduced and analyzed using IRAF.
Our source does not uniformly fill the slit, which produces an artificial shift along the dispersion direction, and we have applied a correction for this effect (Appendix A).

\section{Modeling} \label{shapeModeling}

The 3D spatio-kinematic code SHAPE was used to model the high-velocity bullet of Period 1 (2002, 2003, 2004) as a 3D axially-symmetric object. All velocities presented are in the heliocentric standard-of-rest frame. The emergent intensity is assumed to be proportional to the emission measure, and therefore the medium is assumed to be optically thin in the observed lines. We have not accounted for dust absorption due to the line-of-sight ambient circumstellar material to the blob and, given its complexity, any assumptions would be very uncertain. In reality the actual blob intensities would be larger, thus our estimates of the mass are likely lower limits.
The description of our modeling effort refers to ``levels" of models, each created and upgraded as morphology, central peak location, and relative intensities were better constrained. There is an increase in complexity as we move from ``level 0" to ``level 1" to ``level 2".


The overall size, opening angle, inclination angle $i$\footnote{Measured relative to the LOS.} and position angle $PA'$\footnote{Due to the HST STIS slit configuration, we measure $PA'$ counter-clockwise relative to a vector pointing east from the star -- different than the standard astronomy convention of position angle. Thus $PA'$=0$\arcdeg$ is due east and $PA'$=90$\arcdeg$ is due south.} are constrained by the presence (or absence) of the emission blob in the 3 slits that were used to observe it.  When fitting our models to the data, we compare the morphology, the peak emission location in position velocity (PV) diagrams, as well as the relative intensities in on-center and off-center slits.
As the $PA'$ of the (projected) axis of the modeled bullet decreases, the peak emission of the blob in both the on-center and off-center slits moves towards more negative spatial offsets in declination. Inclination is defined as the angle of the motion with respect to the LOS. When the inclination is increased by a few degrees, the peak emission in just the off-center slit moves towards negative spatial offsets in declination. First we adjust the $PA'$ of the bullet so the peak line emission in the central slit is fit correctly with respect to the observed emission, with a rough fit obtained for the off-center slit. Next we vary the inclination until both the on-center and off-center spatial peak emission positions are well fit. 

As the bullet moves during epoch 1 to epoch 3, there is a change in the LOS offset, which is calculated from LOS velocities and the elapsed time. Since we do not know the LOS velocity at the initial ejection of bullet\#1, we are unable to constrain the LOS offset at epoch 1 of our model. Observational constraints can, however, be used to determine this value since we know epoch 1 was observed 7.6 years after ejection (see \S\,\ref{radialacc}). We make the assumption for our model that the start of bullet\#1 in epoch 1 is at the origin (0,0). This assumption does not affect our results as the model PV plots of the emission blobs are insensitive to the LOS offset of the bullet from the central star.

The best-fit modeled peak intensity measurements, reported in Table\,\ref{modelparamtable}, were compared to observed intensities (Table \,\ref{modeltable}) by calculating relative intensity ratios (Table \ref{intratios}). These are the ratios of the intensity of a single line between different slit configurations. Epochs 2 and 3 were modeled only after parameters chosen for epoch 1 of the ``level 1" model created a best-fit to the observations (see \S\,\ref{evolution}).



\subsection{Best-Fit Model}\label{level2}
Our best-fit model, based on both qualitative and quantitative criteria, is described below and was constructed as follows. 
We began with a ``level 0" model (see Appendix B: \S\,\ref{lev0}) using a conical-shaped bullet, but found that it could not produce a good fit, irrespective of variations in the velocity and density laws. We therefore switched to a semi-ellipsoidal shape (our ``level 1" model: \S\,\ref{lev1model}), which improved the qualitative fit to the observed PV-morphology and the location of the centroid of the observed emission peak in PV-space, as well as the changes in these as a function of epoch. These improvements were made by adjusting the bullet length $\ell$, position angle $PA'$ and inclination $i$ (\S\,\ref{evolution}), our choice of velocity law (\S\,\ref{vellawsec}) and density law (\S\,\ref{density}). This model shows that from epoch 1 to 3 $PA'$ (inclination) steadily decreases (increases). However, this model still has some inadequacies when compared to measured quantities in our observations, leading to additional changes in only the density law and geometry of the bullet, resulting in our ``level 2" (best-fit) model. 

\begin{figure}
\vspace{5in}
\centering
\includegraphics[width=0.8\textwidth]{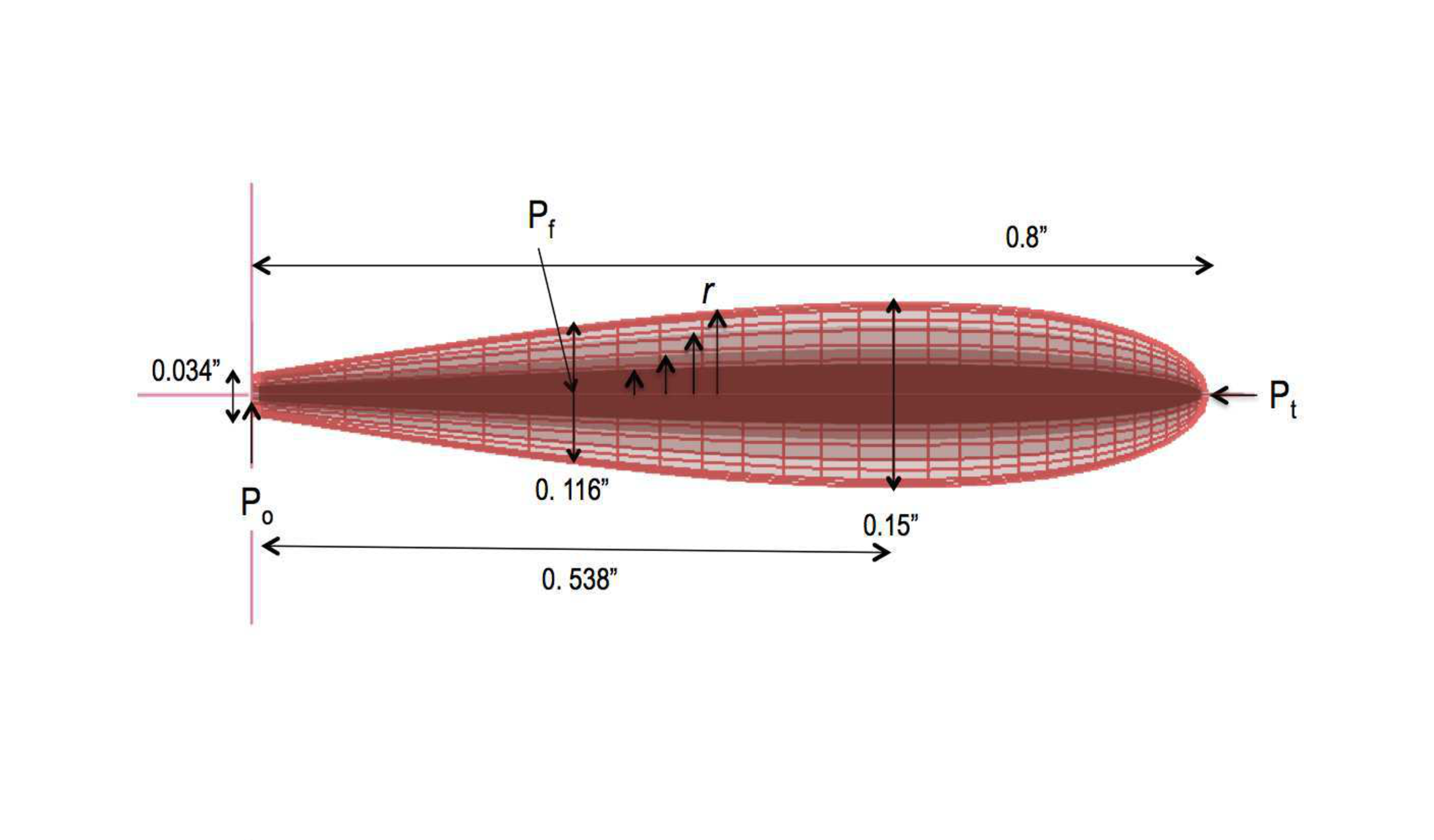}
\caption{\label{sketchnew}Schematic geometry of the axisymmetric bullet in the first epoch (January 2002), for 
our best-fit model (``level 2"). The bullet length is $\ell$ = $0\farcs8$, and the cross-sectional diameter at the base of the bullet is $0\farcs034$, increasing to $0\farcs15$ at its widest at a distance of $z=0\farcs538$ from the base. The density decreases away from the bullet axis, i.e., as a function of $r$ in our cylindrical coordinate system.}
\end{figure}

\begin{figure}
\centering
\includegraphics[width=0.55\textwidth]{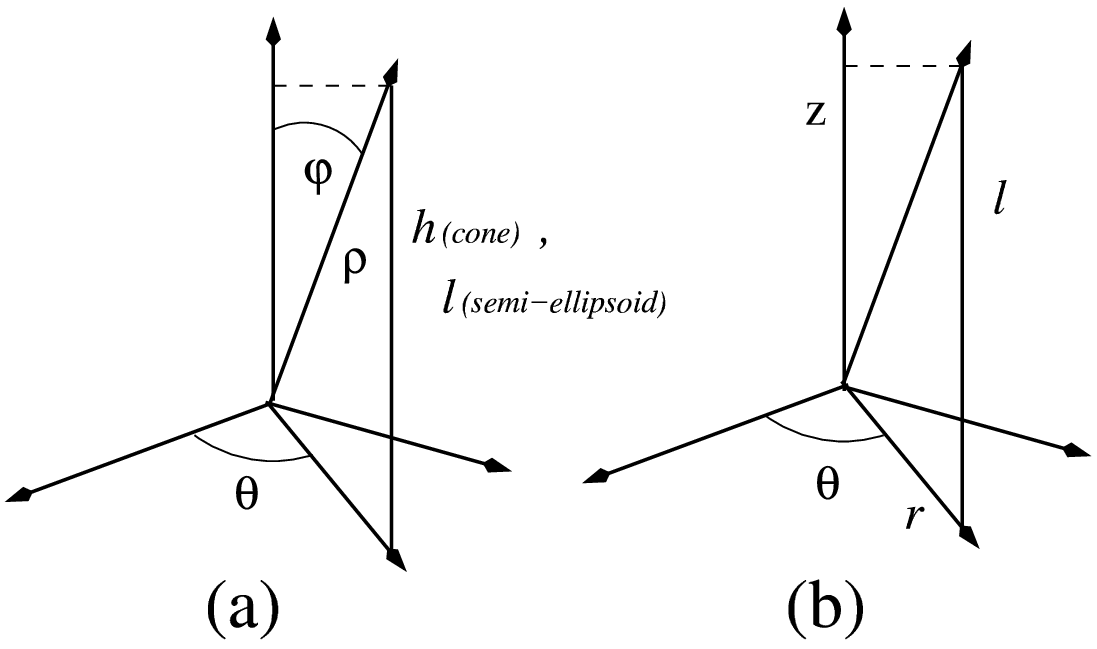}
\caption{\label{coordgrid} (a) Spherical coordinate system that shows the radial coordinate $\rho$ used to define the linearly-varying velocity laws for all our models, and 
for the density law in our ``level 0" model. The three different density laws we consider in ``level 1" are only functions of the spherical coordinate $\varphi$ (not $\rho$). 
(b) Cylindrical coordinate system that defines the variable $r$ for the density law in ``level 2". In each case the origin is at the star, V Hya. }
\end{figure}

The physical parameters characterizing our best-fit model are given in Table\,\ref{varymodelparams} and Table\,\ref{constantmodelparams}. A set of cardinal points $P_t$, $P_f$ and $P_o$ are defined as the tip, flattening and origin points of the bullet, respectively (labeled in Fig.\,\ref{sketchnew}). The bullet length is changed from epoch to epoch by using the amount of time that has elapsed between epochs and assuming ballistic motion for each of the cardinal points (justified later: \S\,\ref{radialacc}). Three separate velocity laws were tested - constant, linear and flattened - and it was found that the flattened law created the most consistent results. At the flattening point, $P_f$, the velocity becomes constant at a value of 245 km s$^{-1}$ (see Fig.\,\ref{vellaw}). The velocity  vectors are always directed radially away from the center in a spherical coordinate system (see Fig.\,\ref{coordgrid} a). Thus, all velocity vectors at different locations in the bullet are radially directed away from the bullet's origin, $P_o$. 

\begin{figure}
\centering
\includegraphics[width=0.5\textwidth]{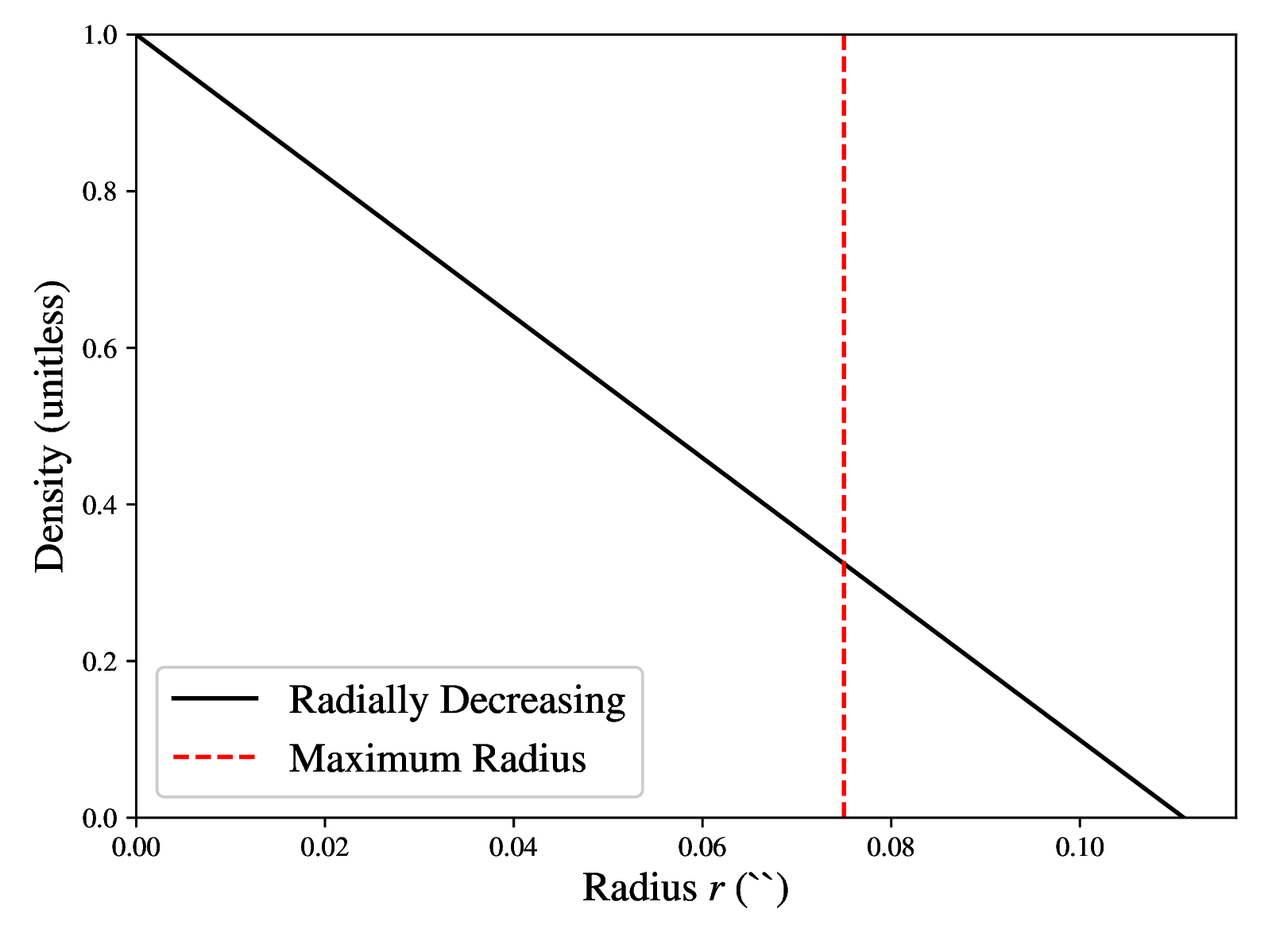}
\caption{\label{densitylaw2}Density law used for ``Level 2" model is linear in the radial direction, defined by the function $n(r)$ = $1 - 9.0r$ where $r$ is the radius of the ellipsoid in cylindrical coordinates. The maximum radius that the bullet has in epoch 1 is  $0\farcs075$.}
\end{figure}

The comparison of the ratio of peak emission intensities in the central and offset slits, defined as $R_I$, provides an important quantitative constraint on our models. Thus, although our ``level 1" model provided a reasonable fit to the morphology and peak location of the emission, the $R_I$'s had large discrepancies compared to the observed ones. Values of the geometrical parameters characterizing the shape were adjusted and a new linear density law was adopted (Fig.\,\ref{densitylaw2}), defined using  cylindrical coordinates (Fig.\,\ref{coordgrid} b). 
The bullet is squeezed at the origin, still bulging at the center and tapering off at its tip. The best-fit model density is defined as $n(r) = 1 - 9.0r$ ($r$ in cylindrical coordinates) between 0$<r<r_{max}$, where r$_{max}$ is the maximum radius of the bullet. The best-fit model produced better intensity ratios (Table \ref{intratios}, ``level 2") while keeping the overall PV-morphology and peak emission positions consistent with the observed emission. 



We compare our ``level 2" model to observations by providing two separate intensity ratio values, $R_I$, for epochs 2 and 3 (Table \ref{intratios}) because the model and observed intensity peaks in the PV plots lie at modestly different locations. The first value is at the location of the peak \textit{observed} intensity (this is true for all of the ``level 1" values as well). The second value is extracted at the location of the peak \textit{model} intensity. These remarks also apply to the equivalent rows in Table\,\ref{modelparamtable}.

\begin{turnpage}
\begin{figure}
\subfigure[]{\label{9100offsrc_lev2} \includegraphics[width=80mm]{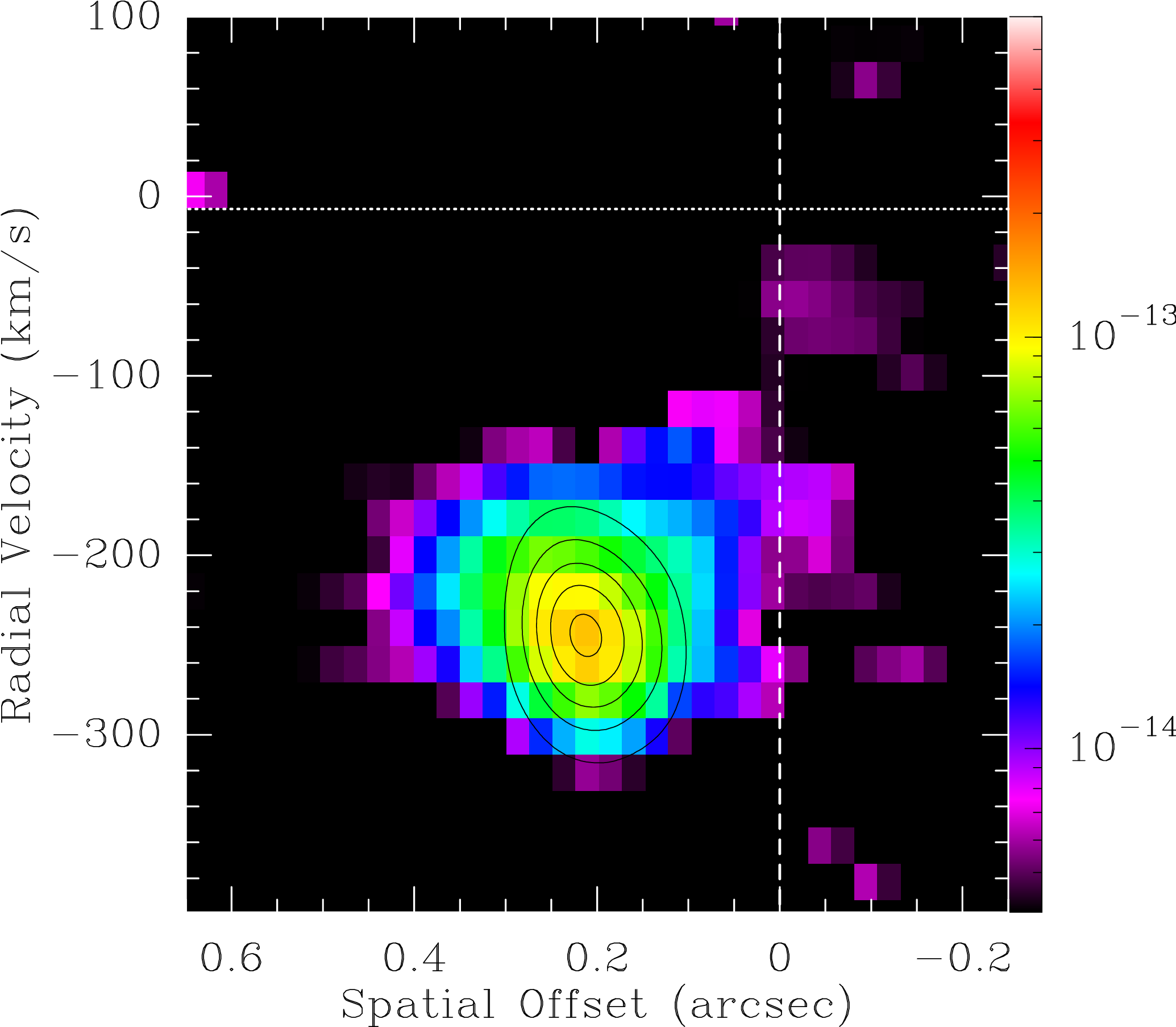}}
\subfigure[]{\label{9100onsrc_lev2} \includegraphics[width=80mm]{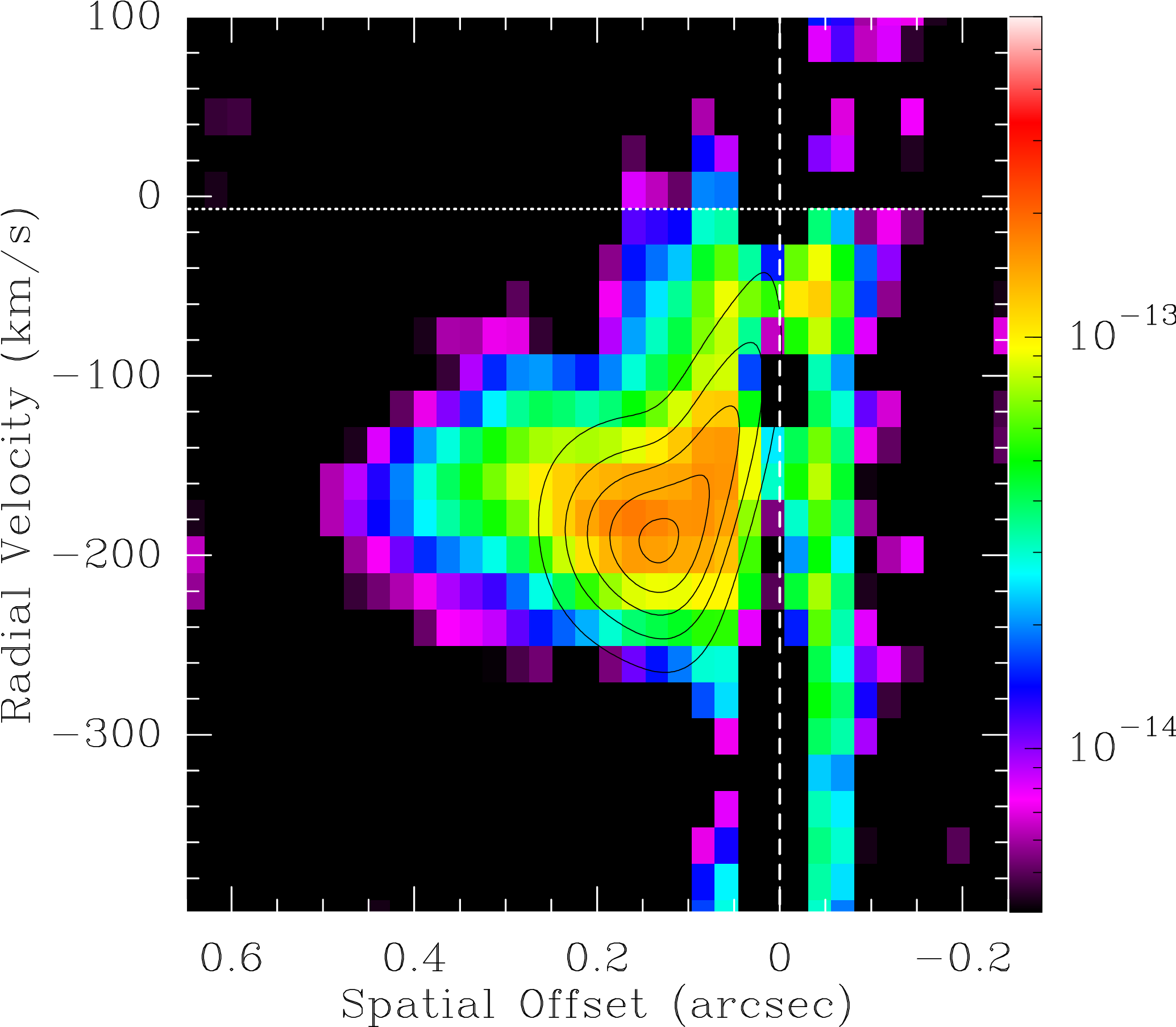}}
\caption{\label{9100modellev2} PV plot of ``level 2" model for the detached emission blobs in the first epoch, 01-28-2002; colorscale shows the observations and contours show the model. The 
observation and model results for slit $S_{-2b}$ are shown in (a), and results for slit $S_{0b}$ are shown in (b).}
\end{figure}
\end{turnpage}

\begin{turnpage}
\begin{figure}
\subfigure[]{\label{9632offsrc0.2_lev2}\includegraphics[width=80mm]{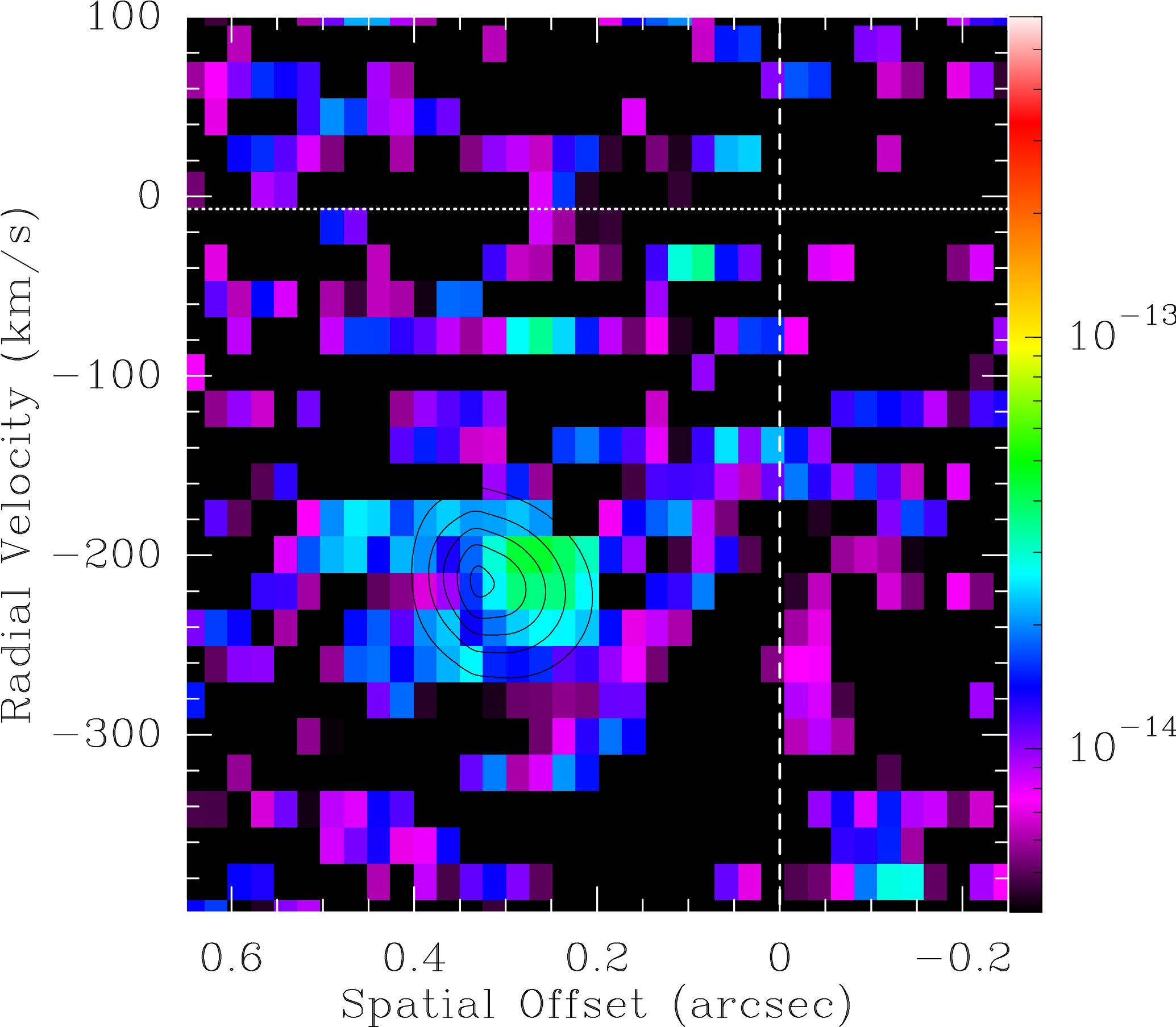}}
\subfigure[]{\label{9632offsrc0.1_lev2} \includegraphics[width=80mm]{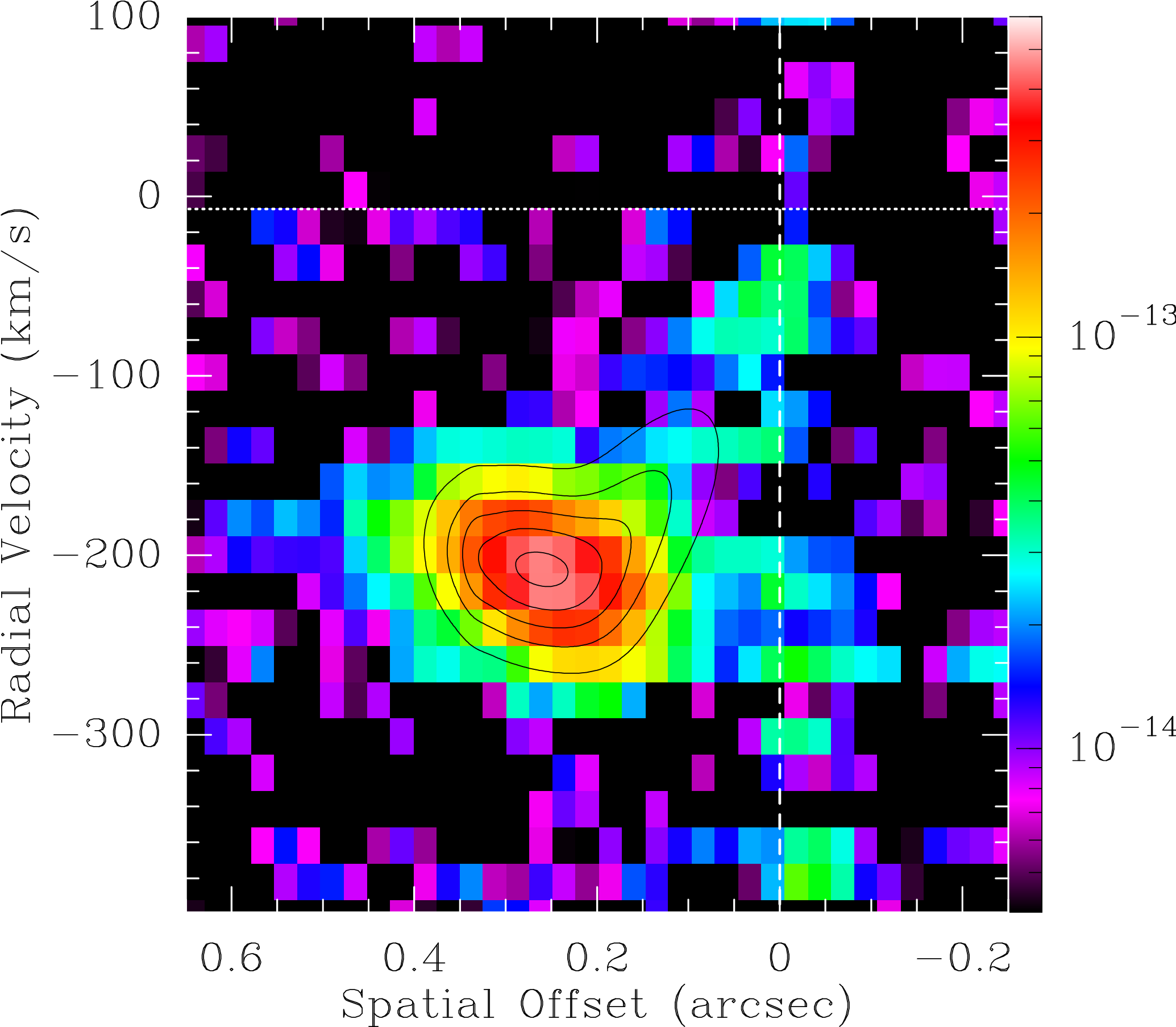}}
\subfigure[]{\label{9632onsrc_lev2} \includegraphics[width=80mm]{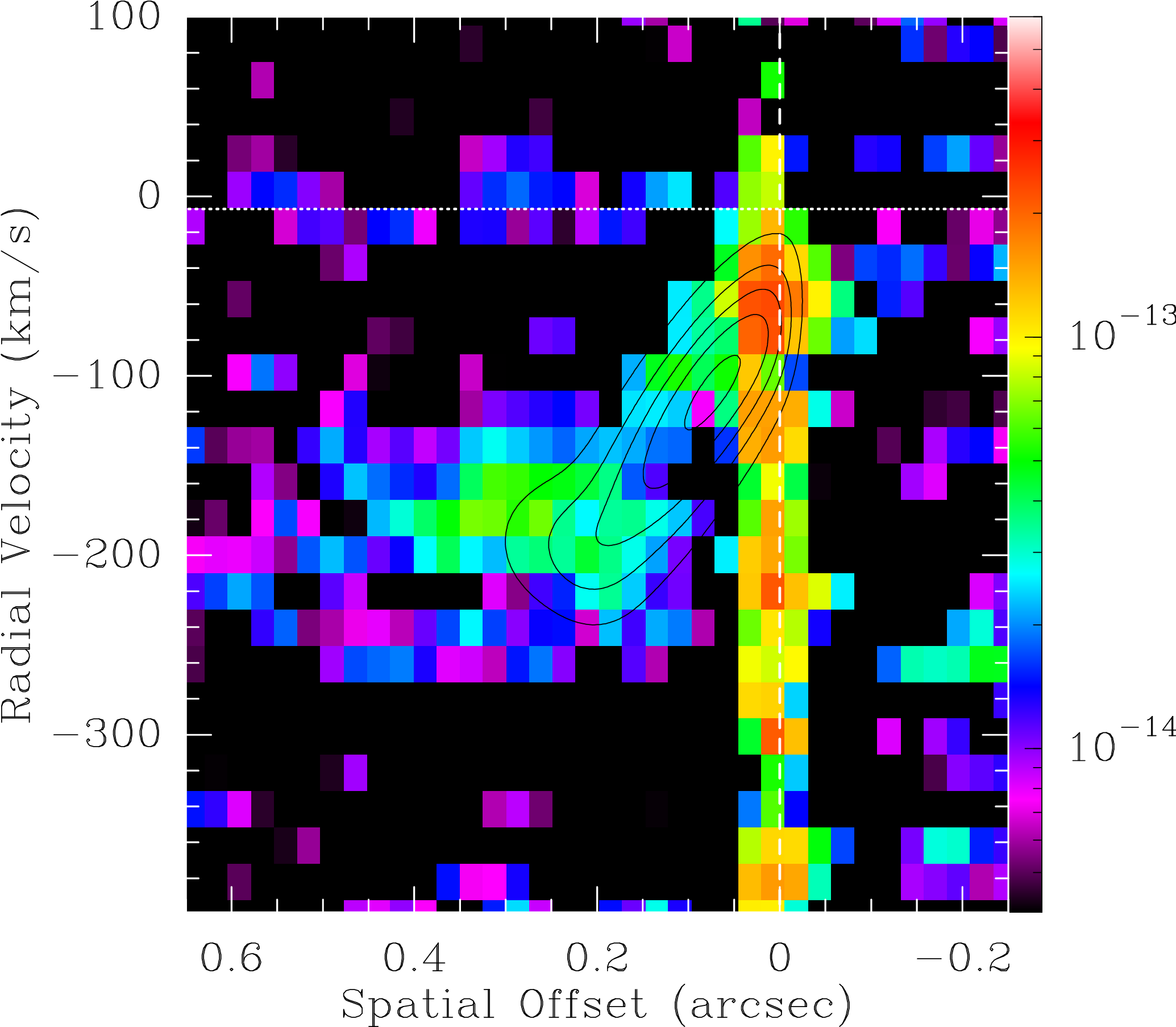}}
\caption{\label{9632modellev2} PV plot of ``level 2" model for the detached emission blobs in the second epoch, 12-29-2002; colorscale shows the observations and contours show the model. The 
observation and model results for slit $S_{-2t}$ are shown in (a), the results for slit $S_{-1t}$ are shown in (b), and the results for slit $S_{0t}$ are shown in (c).}
\end{figure}
\end{turnpage}

\begin{turnpage}
\begin{figure}
\subfigure[]{\label{9800offsrc_lev2} \includegraphics[width=80mm]{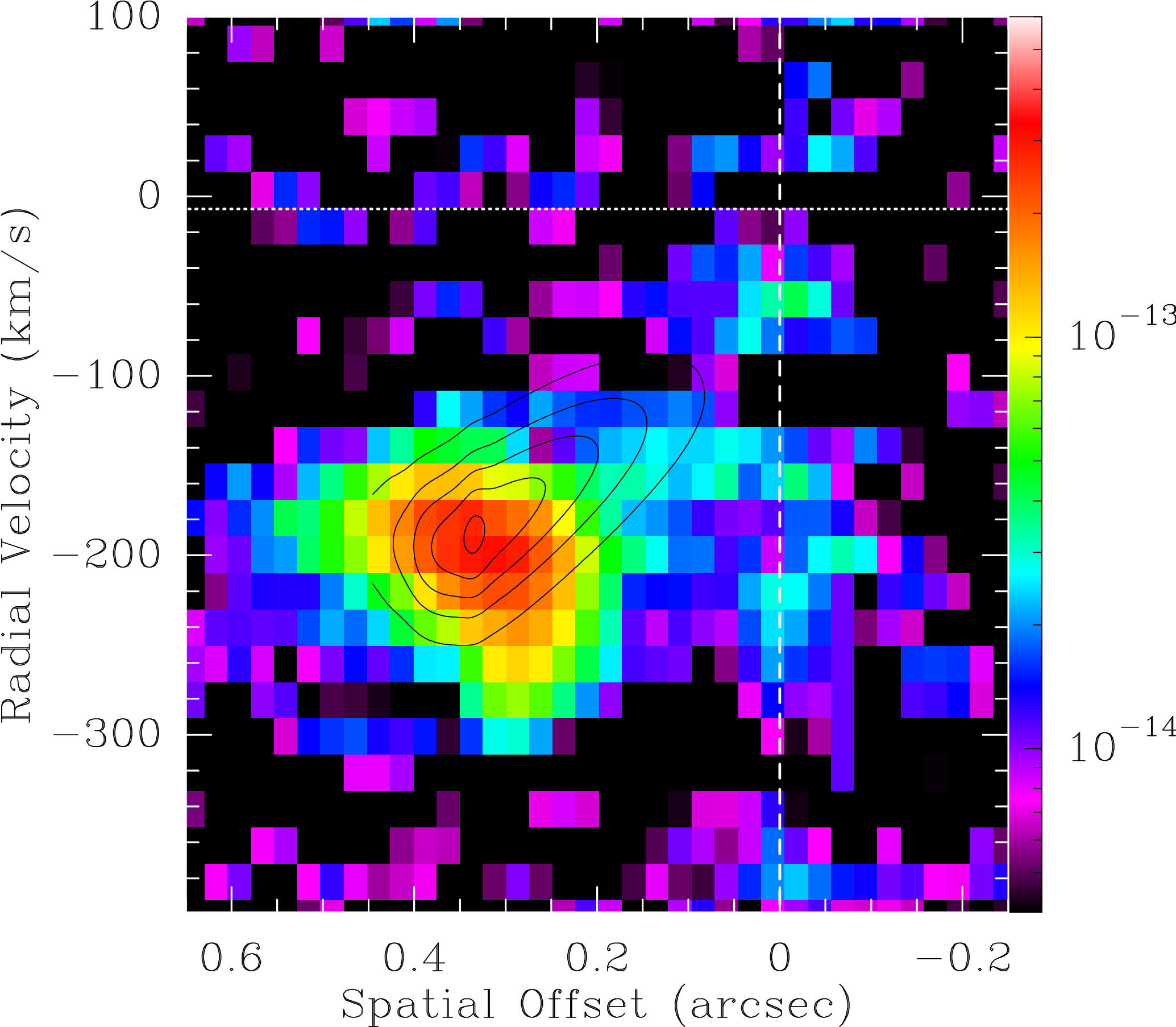}}
\subfigure[]{\label{9800onsrc_lev2} \includegraphics[width=80mm]{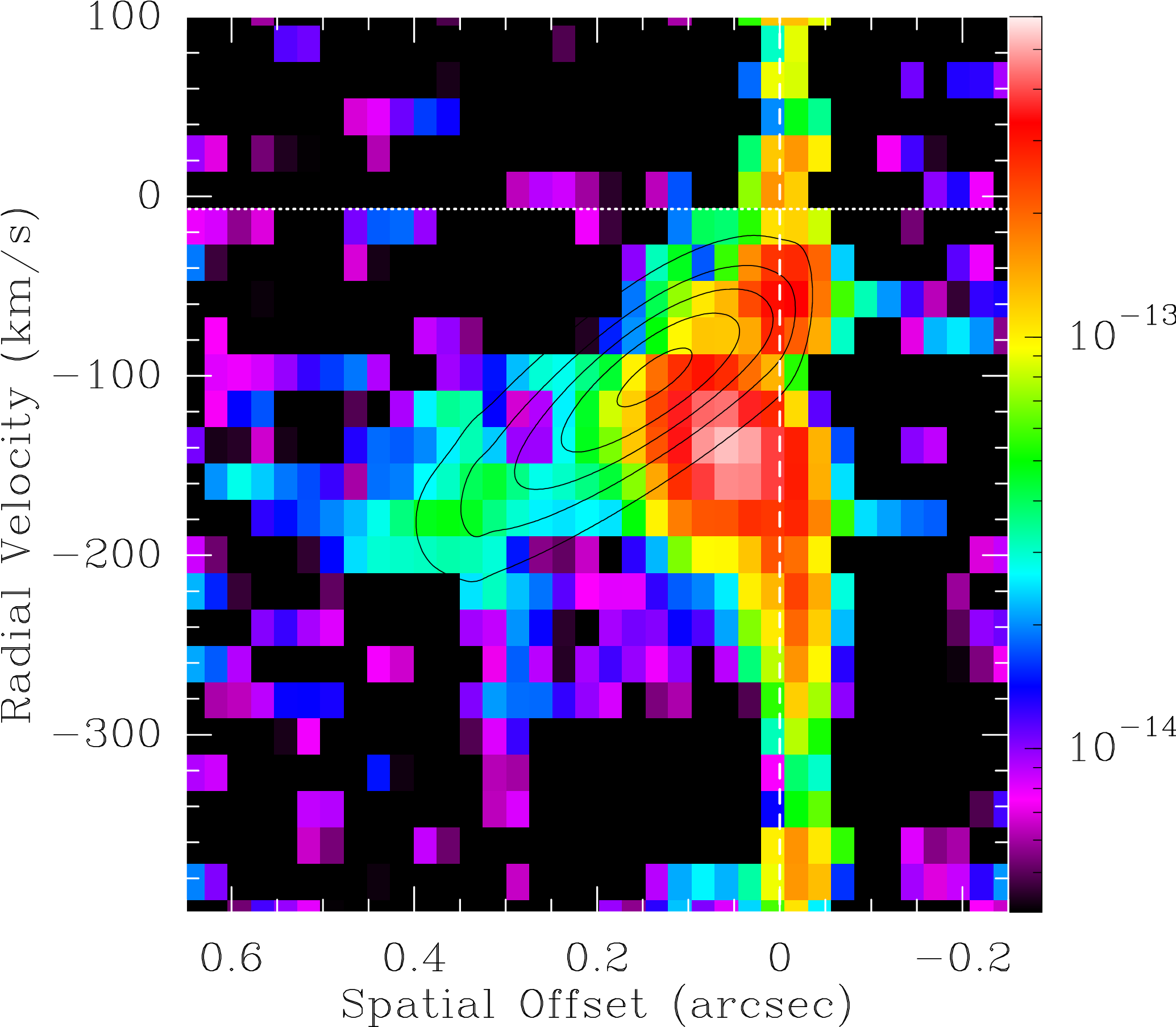}}
\caption{\label{9800modellev2} PV plot of ``level 2" model for the detached emission blobs in the last epoch, 01-12-2004; colorscale shows the observations and contours show the model. 
The observation and model results for slit $S_{-1t}$ are shown in (a), and results for slit $S_{0t}$ are shown in (b). }
\end{figure}
\end{turnpage}

\begin{turnpage}
\begin{figure}
\subfigure[]{\label{9100blob_lev2} \includegraphics[width=80mm]{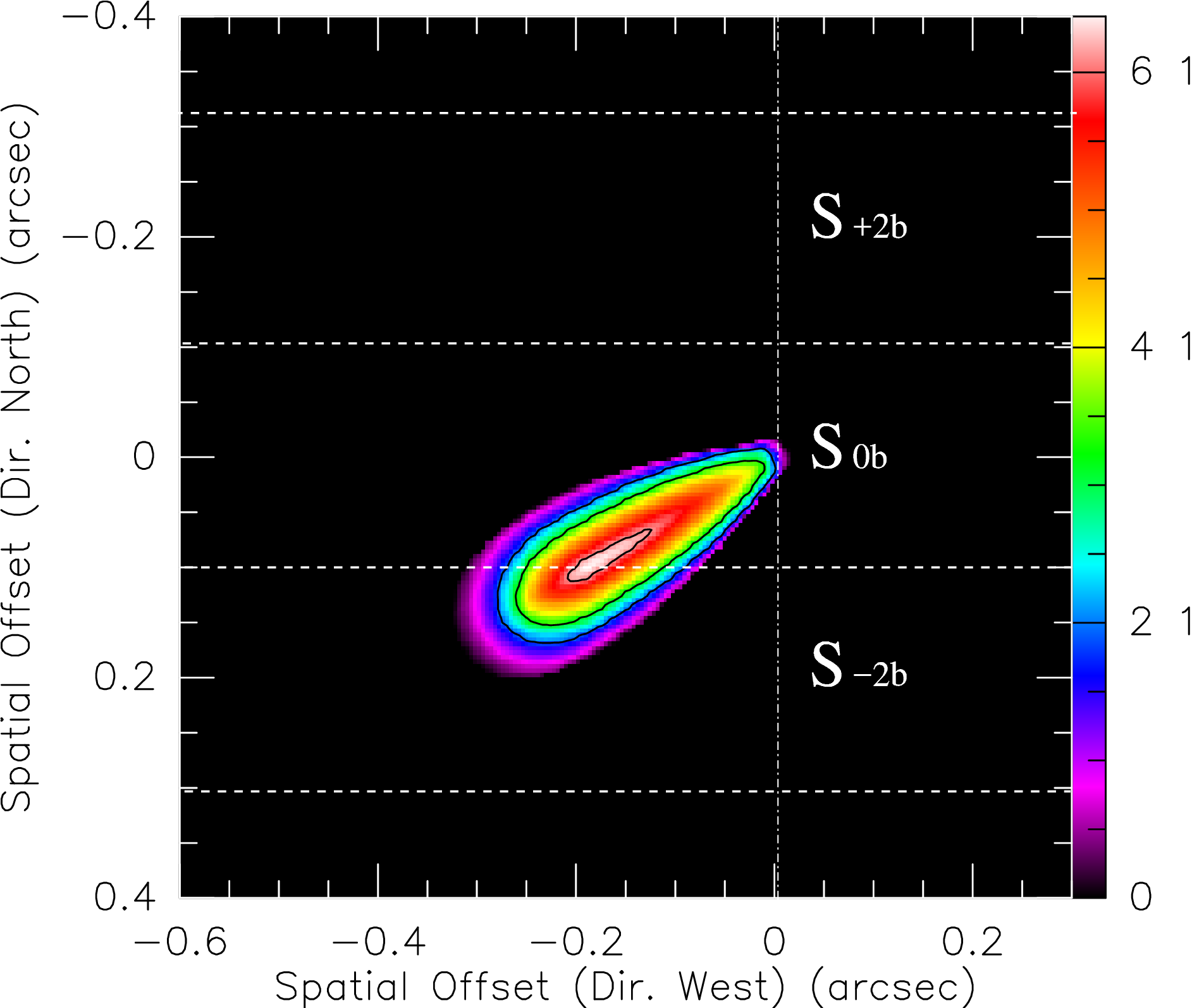}}
\subfigure[]{\label{9632blob_lev2}\includegraphics[width=80mm]{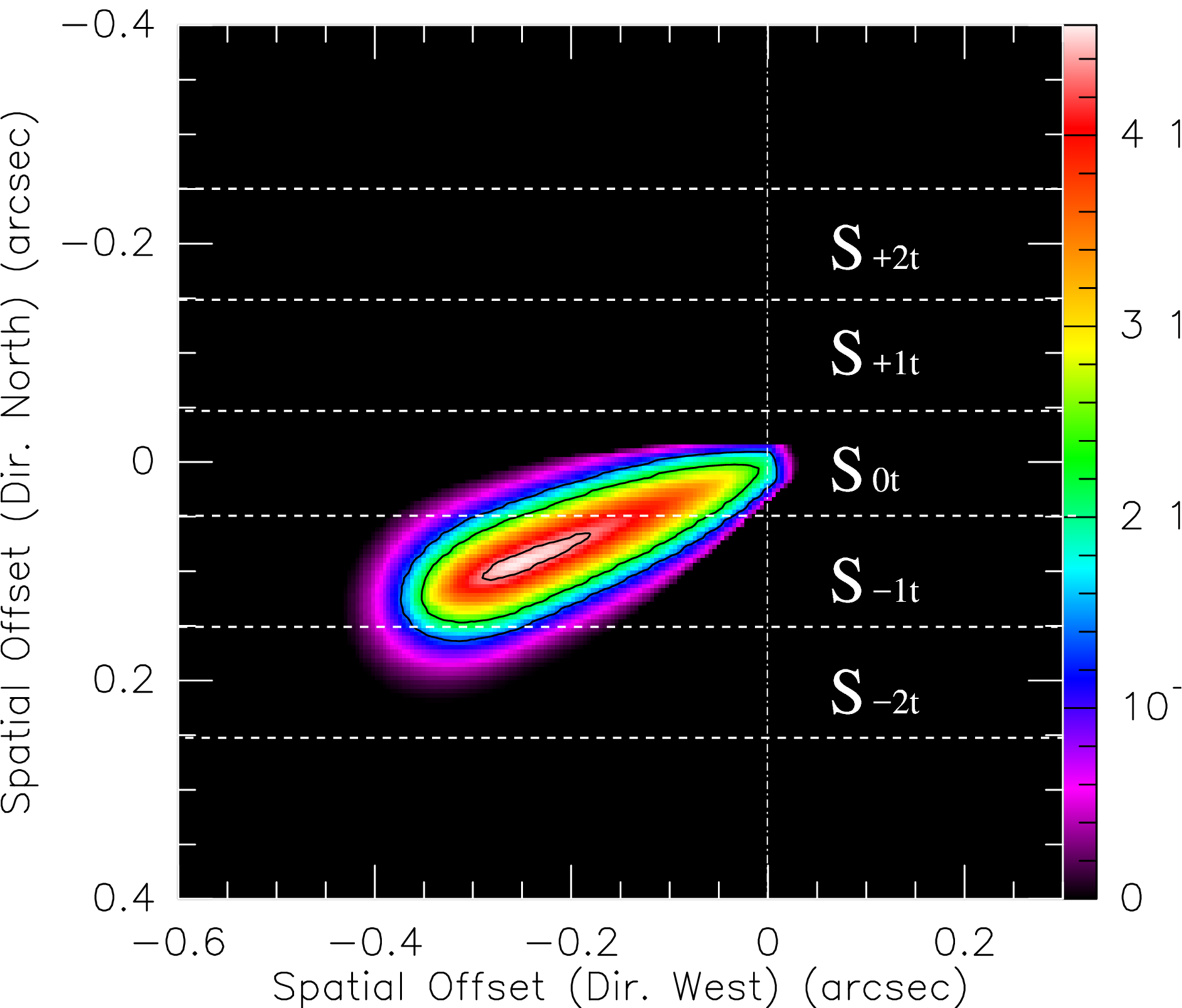}}
\subfigure[]{\label{9800blob_lev2} \includegraphics[width=80mm]{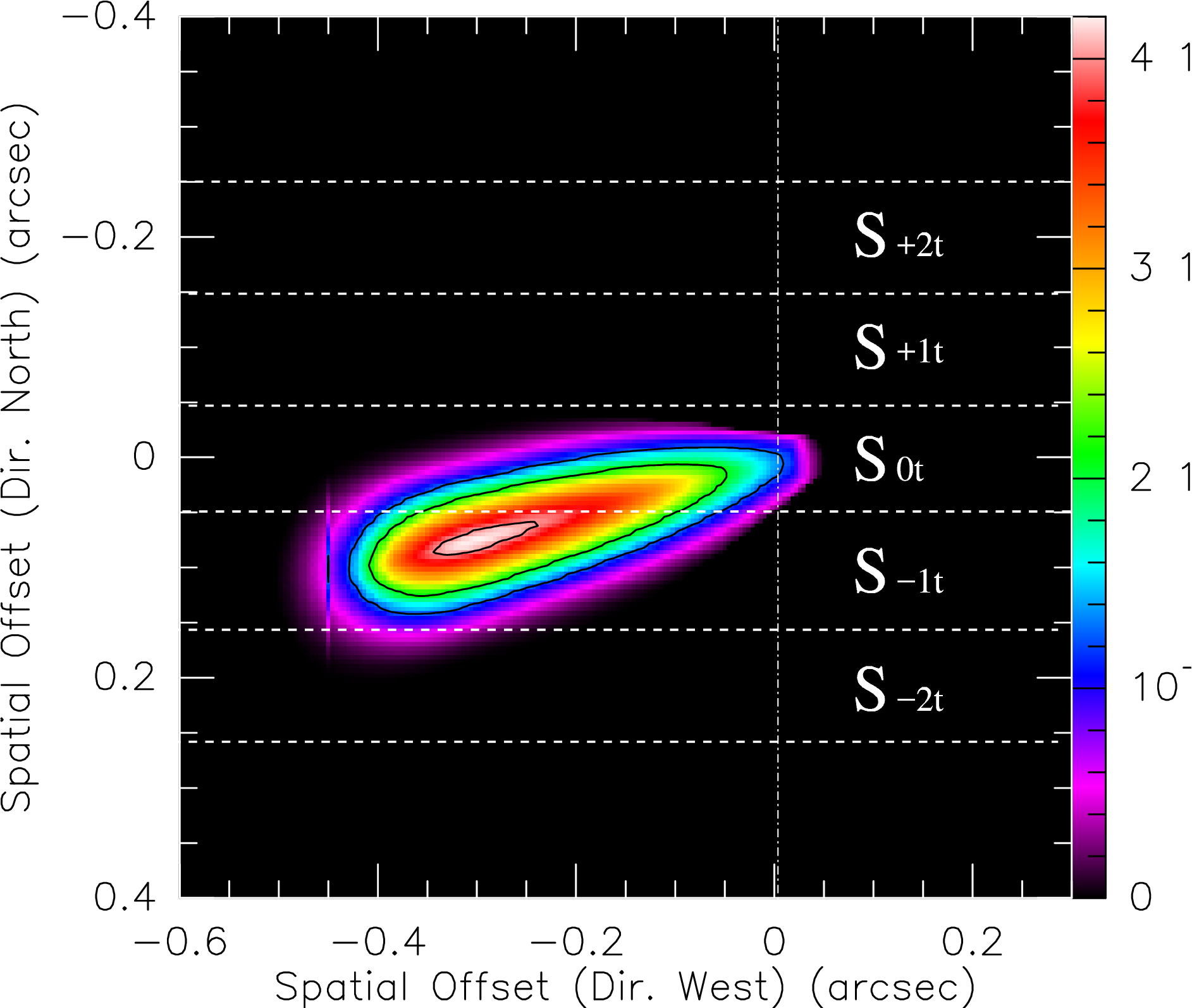}}
\caption{\label{blobs2}``Level 2" projected surface brightness distribution of the high-velocity jet in the first epoch 01-28-2002 (a), the second epoch 12-29-2002 (b), and the third
epoch 01-12-2004 (c).}
\end{figure}
\end{turnpage}

Overall, the ``level 2" PV plots of the [SII] emission for all three epochs (Fig.\,\ref{9100modellev2},\,\ref{9632modellev2}, and\,\ref{9800modellev2}) show peak-emission locations, morphological shapes, and intensity ratios that are in fairly good agreement with the observations. The surface brightness distribution of the bullet at each of the three epochs is shown in Fig.\,\ref{blobs2}. 

  
Uncertainties in observed peak velocities are a few \kms\, and the uncertainties in the spatial offsets are a few hundredths of an arcsecond (a few AU). Velocity laws modeled within SHAPE are also good to $\sim10$ \kms,\ since we can measure and fit the centroid of the peak emission to within about 5$\%$. Our mesh grid is broken into cells of roughly a few hundredths of an arcsecond across, leading to a precision error of a few $\%$ in length and therefore $\sim15\%$ error in volume. There are likely systematic modeling uncertainties as well that affect the absolute values of the various physical quantities that we infer (e.g., velocity-law, volume, mass). However, since systematic effects are likely to affect the different epochs in a similar way, we believe that the relative changes in these quantities from epoch to epoch are not significantly larger than $\sim15\%$.

\subsubsection{Verification of our 3D Velocity Law}

One could argue that we might be able to fit the decrease observed in the LOS velocity of the peak emission as seen in the PV-morphology plots simply by 1) changing the inclination angle, 2) changing the $PA'$ and 3) adjusting the velocity law so our observations fit a new set of parameters. This would introduce a degeneracy in our models.
We found this is not the case, and in fact our  ``level 1" and ``level 2" models are strongly dependent on the precise inclination, $PA'$ and velocity law chosen (see Fig.\,\ref{vel_inc_rel}). In the case of slit $S_{-2t}$ of epoch 2, by moving the inclination angle back towards the LOS and lowering the velocity at the flattening point ($P_f$), we can roughly match the LOS velocity in the peak emission. However, the spatial offset is completely misaligned. We can try to fix this problem by lowering the $PA'$, but it can not be lowered enough to fix the offset before there is no longer any emission seen in this slit. When we adjust the $PA'$ and inclination back to the starting model values of slit $S_{-2t}$ of epoch 2, the peak emission's LOS velocity is still misaligned. We conclude that we cannot fit the observations accurately without the changes in inclination and $PA'$ from epoch to epoch, so the velocity law is reasonably robust.
 
\section{Discussion}\label{results}

\subsection{Changes in Bullet Mass and Volume} \label{masschange}

Our modeling of bullet\#1 shows that its projected shape evolves with time, becoming progressively fatter (Fig.\,\ref{blobs2}).
Although the change in inclination accounts for some of this effect, we find  
quantitative increases in the mass and volume of the ionized gas in the bullet. In our ``level 2" best-fit model, the mass increases by 50$\%$ from epoch 1 to epoch 2 and by 7$\%$ from epoch 2 to epoch 3. As a result, we find that there is a significant increase in both momentum ($p$) and kinetic energy ($KE$) from epoch 1 to epoch 2 ($\sim150\%$ and $\sim200\%$, respectively) and from epoch 2 to epoch 3 ($\sim50\%$ and $\sim40\%$, respectively). The increase in $p$ and $KE$, from epoch 2 to epoch 3 is mostly due to an increase in mass, whereas from epoch 1 to epoch 2, both mass and velocity increases contribute. We provide a physical explanation for these increases in \S\,\ref{magfield}.

We integrated over the mesh for each epoch's bullet provided by SHAPE (i.e., for epoch 1 see Fig.\,\ref{sketchnew}), in order to estimate the actual mass and volume, and thus the average density of our bullet for each epoch in SHAPE's arbitrary units. By then scaling this density to the average density estimated in Paper I ($3.0 \times 10^{5}$ cm$^{-3}$) we derived the mass, volume and peak density of the bullet in  physical units. 

We find that, from epoch 1 to 3, the bullet mass increases from 10$^{27}$\,g to $2 \times 10^{27}$\,g, and then to $3 \times 10^{27}$\,g, and the volume  increases from $2 \times 10^{45}$\,cm$^{-3}$ to $3 \times 10^{45}$\,cm$^{-3}$, and then to $5 \times 10^{45}$\,cm$^{-3}$. Although the bullet's mass and volume increase\footnote{The 
[SII] emission from the bullets traces only hot material; it is likely there is a substantial amount of cool mass ($<$ 10,000 K), not visible in our STIS data that 
has also accumulated.}, its density remains unchanged, with the peak at $3.5 \times 10^{5}$ cm$^{-3}$, suggesting that the bullet is entraining and heating the much slower moving ambient outflow material as it moves away from V Hya. The smaller accumulation of mass between epochs 2 and 3, compared to that between epochs 1 and 2, could be the result of a radially-decreasing ambient density, as would be expected for a red giant wind with constant mass loss rate, outside the acceleration zone.
Additionally, as the bullet plows through the ambient gas, its leading edge experiences the strongest compression and deceleration, providing a physical explanation for the flattening of the 3D velocity profile that we derive for the leading half of our model bullet.



\subsection{Bullet Trajectories and Velocities}\label{nonballmotion}

The geometric bullet model presented in \S\,\ref{level2} of this paper applies only to bullet\#1, but in this section we place it in the context of the other bullets. We can quantitatively characterize the movement of bullets 1--4, i.e., by estimating their LOS and sky-plane offsets (x and y, respectively) as well as velocities ($V_x$, $V_t$) obtained from observations. We describe how each of these four quantities is calculated, detailing specific cases for different bullets.

In Paper I, we reported that the ejection axis of the bullets flip-flops around an average direction, both in and perpendicular to the sky-plane. For example, in Period 1 the `detached'\footnote{Bullet located at sky-plane offsets of $\sim0\farcs15$ to $0\farcs3$, see Paper I.} emission blob (bullet\#1) is seen predominantly in the slit immediately south of center (i.e. $S_{-2b}$ or $S_{-1t}$), whereas in Period 2 the detached emission blob (bullet\#2) is seen predominantly in the slit north of center (i.e. $S_{+2b}$). 

The LOS offsets of the bullets from V Hya at some specific time $t_1$ (x coordinate, i.e., along the vertical direction  in Fig.\,\ref{distantblobsch}) require knowledge of the LOS offset x$_0$ at an initial time $t_0$, and the mean LOS velocity $V_x$ between $t_0$ and $t_1$ (Table\,\ref{centroid}).
For bullet\#3 we observe the earliest stage of the ejection, where the bullet is coincident with the star, which makes the calculation much more straightforward, i.e. $t_0$=0 at epoch 4 with $V_x$ = 153\,\kms\,\ and the LOS displacement of the bullet between epochs 4 and 5 is 36 AU. In the case of bullet\#2, we estimate that the time between ejection and epoch 3 is 1 year and that the LOS velocities at these two epochs are the same, i.e., 132\,\kms, therefore the LOS offset from ejection to epoch 3 is 28 AU. For bullet\#1, we assume an initial LOS velocity of 93\,\kms~at ejection (i.e., x$_0=0$) based on our modeled velocity law, which when averaged with the LOS velocity of 199\,\kms~observed in epoch 1, gives a mean velocity of 146\,\kms~over a time span of $t_1-t_0=7.6$\,yr (1994.5 to 2002.1). The LOS offsets for epochs 2 and 3 then follow since we know the time intervals and mean velocities between epoch 1 and 2, and epochs 2 and 3. Bullet\#0's LOS offset during its 
`distant'\footnote{Bullet located at sky-plane offsets of $\sim0\farcs8$ to $1\farcs0$, see Paper I.} period was calculated utilizing LOS velocity measurements from ground-based observations when it was first ejected, i.e., 16 years prior to when it was observed with HST in epoch 1 (Table 6, Paper I).

The sky-plane offsets, i.e. orthogonal to the LOS in the sky-plane (towards the east, i.e., positive y-axis in Fig.\,\ref{distantblobsch}) were computed directly from the bullet's measured angular offsets from the star in the 2D spectra, using V Hya's adopted distance, D=400pc. In Table\,\ref{centroid}, column four lists the sky-plane offsets for each bullet at each epoch, so for example, for bullet\#1 in epoch 1 the offset  was measured at 0.16 arcseconds or 64 AU. 

The LOS velocity itself, $V_x$, is calculated from weighted $V_p$'s (Table 2, Paper I) based on the intensity of the observed emission in the slit. For example, in the case for bullet 1, epoch 1, the center slit's intensity-weighted $V_p$ is estimated as,

\begin{equation}
V_p =  \frac{V_{s1} I_{s1} + V_{s2} I_{s2}  }{ I_{s1} +  I_{s2} } 
 \end{equation}

where $V_{s1}$ and $V_{s2}$ are the velocities in \kms\,\ in slit's 1 and 2, respectively (i.e, $S_{-2b}$ and $S_{0b}$). Variables $I_{s1}$ and $I_{s2}$ are the corresponding intensities in cgs units. We can plug in these values from Table 2 in Paper I,

\begin{equation}
V_p =  \frac{(-241) 1.6\times 10^{-13} + (-180) 2.2\times 10^{-13} }{1.6\times 10^{-13}+2.2\times 10^{-13}} \,\textnormal{[\kms]}
 \end{equation}
 
 which gives $V_p = -205.7$ \kms. Therefore, we find the LOS velocity to be 198.7\,\kms\,\ in this case.

The sky-plane velocity, $V_t$, is calculated directly using the sky-plane offsets for each bullet and 
the time elapsed from epoch to epoch (1 year).

In Fig.\,\ref{distantblobsch} we present a schematic history of the locations and movement of bullets \#0, \#1, \#2 and \#3. 


\subsubsection{Acceleration and Deceleration }\label{radialacc}

\begin{figure}
\centering
\includegraphics[width=0.60\textwidth]{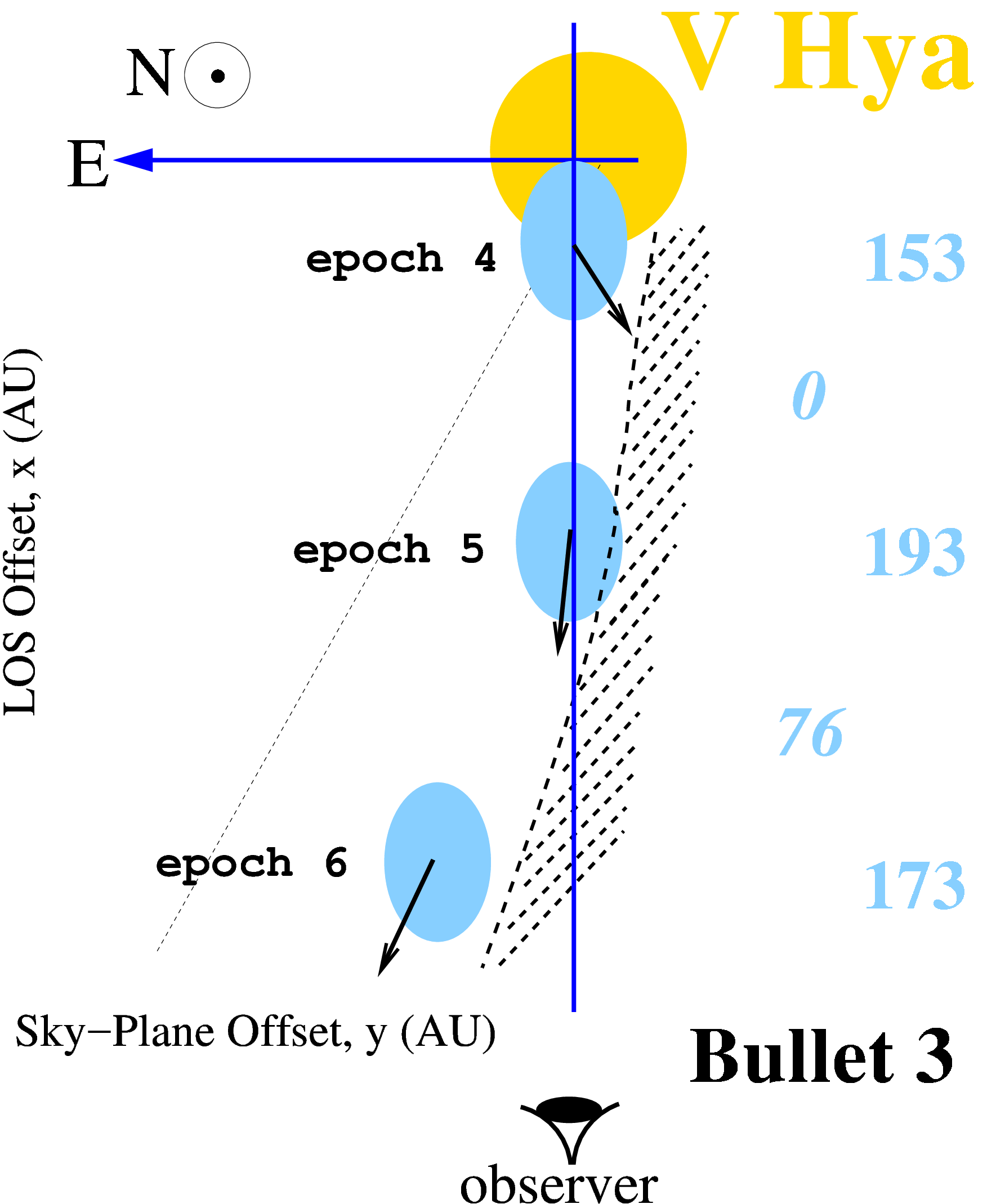}  
\caption{\label{bullet3_radacc} Schematic representation of bullet acceleration in the radial direction from epoch 4 to 5, resulting from the presence of dense ambient material with which the bullet interacts, changing the latter's 3D velocity vector such that it is oriented more towards our line-of-sight in epoch 5, compared to epoch 4. Numbers in bold and non-italicized are the associated LOS velocities in km s$^{-1}$, relative to V Hya's systemic radial velocity, $V_{hel} = 7$ km s$^{-1}$. Tangential, ``sky-plane", velocities are displayed in italics. }
\end{figure}

We now investigate whether there is evidence for non-ballistic motion of the individual bullets that have been observed at multiple epochs. The non-ballistic motion can be of two types, involving: 1) radial acceleration and 2) tangential acceleration. Bullet\#3 appears to accelerate along the LOS after it is first ejected, i.e. observations show the LOS velocity increasing\footnote{Increasing LOS velocity 
implies motion towards us, i.e. in the negative direction.} from 153\,\kms~to 193\,\kms~in 1 year (i.e., from epoch 4 to 5). 
During epochs 4 and 5 bullet\#3 could have been deflected due to a hydrodynamic interaction with ambient gas, such that its 3D velocity vector is aligned closer to the LOS in epoch 5 compared to epoch 4 (Fig.\,\ref{bullet3_radacc}). Here we characterize bullet\#3's movement by quantifying its path from epoch to epoch. 
We can use the LOS offsets (i.e., between epochs 4--5 and 5--6) and sky-plane offset during epoch 5--6 (Table\,\ref{centroid}) to estimate the average geometric radius of curvature of the bullet trajectory $R$, defined as,
\begin{equation}
R = \frac{(1 + (\Delta\,y/\Delta\,x)^2)^{1.5}}{(\Delta^2y/\Delta\,x^2)}\,\text{cm}
\end{equation}
Then $\Delta\,y/\Delta\,x=0.183$, taken to be the average of the values of $\Delta\,y/\Delta\,x$ during epochs 4--5 and 5--6
($\Delta\,y/\Delta\,x_{45}$=0 and $\Delta\,y/\Delta\,x_{56}$=0.37). We set,
\begin{equation}
\Delta^2y/\Delta\,x^2=\frac{(\Delta\,y/\Delta\,x_{56}-\Delta\,y/\Delta\,x_{45})}{\Delta\,x_{45,56}}\,\text{cm$^{-1}$}
\label{diff}
\end{equation}
where $\Delta\,x_{45,56}$ is the x-distance between the midpoints of the epoch 4--5 and epoch 5--6 legs of the trajectory. Thus,
\begin{equation}
\Delta^2y/\Delta\,x^2=\frac{0.37}{(5.4\times10^{14})}\,\text{cm$^{-1}$}
\end{equation}
and $R=1.5\times10^{15}$\,cm.
The observed acceleration of the bullet is,
\begin{equation}
a=(a_x^{2}+a_y^{2})^{0.5}
\end{equation}
where $a_x$ ($a_y$) is the acceleration in the x- (y-) direction from epochs 4 to 6. Since $a_x=10$\,\kms yr$^{-1}$ and $a_y=38$\,\kms yr$^{-1}$, we find $a=39.2$\,\kms yr$^{-1}$=0.12\,cm\,s$^{-2}$.

Bullet\#2 was observed only during a single-epoch in its ``on-source"\footnote{Bullet located at a sky-plane offset of $<0\farcs1$ from the central star, see Paper I.} period, so unlike bullet\#3, its early 3D motion cannot be directly inferred. Comparing its radial velocity of 132\,\kms~during its on-source period and 173\,\kms~7.6 yrs later, it is plausible that it has experienced a change in its 3D velocity-vector similar to that of bullet\#3, and due to the same physical mechanism.

Observationally, the trajectory of bullet\#1 does not seem to curve. From Table\,\ref{centroid} we compute $\Delta\,y/\Delta\,x$ for bullet\#1 from epoch 1 to 2 ($\Delta\,y/\Delta\,x_{1,2}$) and from epoch 2 to 3 ($\Delta\,y/\Delta\,x_{2,3}$), and find that $\Delta\,y/\Delta\,x_{1,2}$ = $\Delta\,y/\Delta\,x_{2,3}$ = 1.4. However, our model of bullet\#1 suggests that, 7.6--9.6 years from ejection,  there are small but significant changes in the $PA'$ of the long axis of the 
bullet and inclination angles which affects the shape of the bullet, curving it towards the nebular axis. It is the shape of the bullet that changes from epoch to 
epoch in our models, not its trajectory, which is inferred to be ballistic. Our model suggests a decrease in $PA'$, from 28$\arcdeg$, 23$\arcdeg$ to 
17$\arcdeg$, and an increase in inclination, from 26$\arcdeg$, 32$\arcdeg$ to 35$\arcdeg$, over a period of 3 years. The changes in bullet\#1's inclination and $PA'$ from epoch 1 to 3, that we infer from our modeling, imply that its symmetry axis is becoming more closely aligned with the overall nebular axis as it evolves.


\subsection{Acceleration Mechanisms}

The non-ballistic movement and evolution of the high-speed bullets being ejected from V Hya may occur due to the presence of a magnetic field, and/or from a
hydrodynamic interaction.

\subsubsection{Magnetic Field} \label{magfield}
Magnetic fields are important because they cannot only shape the jets produced by evolved stars, but also the circumstellar envelope around them, creating asymmetries during the transition from a spherically symmetric star into a non-spherical PN. The high mass-loss rate from these bullet outflows we observe in V\,Hya and other late-stage AGB stars, coupled with the short evolutionary transition to PN, could be directly linked with the generation of a strong magnetic field from the primary star \citep{2001Natur.409..485B}. \cite{2001ApJ...553L.173S} argue that the ultra-fast, collimated jet from the pPN Hen 3-1475 with its linearly increasing velocity, requires the presence of a magnetic field. It has also been suggested that the presence of a binary companion as an additional source of angular momentum can maintain a magnetic field to launch these jet-like outflows continuously \citep{2004ApJ...614..737F}. 
If we assume that the force required for the lateral acceleration of bullet\#3 comes from the magnetic tension of curved field lines, we can set a lower limit on the field strength\footnote{If the field is not strong enough, the moving bullet will, if it is flux-frozen to the field, simply 
force the field lines into a radial configuration.}.
The acceleration due to the magnetic tension force is given by 

\begin{equation}
a_B = \frac{B^{2}}{4 \pi \sigma R}
\end{equation}

where $B$ is the strength of the magnetic field in Gauss, R is radius of field curvature, and $\sigma$ is the plasma density. 

The bullet plasma density is derived to be $\sigma=8\times10^{-18}$\,g\,cm$^{-3}$, assuming it to be fully ionized, from its number density, $\sim5\times10^6$\,cm$^{-3}$ (Paper I). Setting $a=a_B$, we get $B=130$\,mG for bullet\#3, at a distance of about $5.7\times10^{14}$\,cm (38\,AU) from the center. This value compares favorably with those derived by 
\citealt{2013A&A...554A.134L}, from observations of H$_2$O masers in three AGB stars -- these authors find magnetic field strengths to be $B_{H_2O}$ = 
65--271\,mG, 131--194\,mG, and 44--413\,mG at radial 
distances of 10--29\,AU, 6--15\,AU and 12--30\,AU in IK Tau, RT Vir, and IRC+60370, respectively (see Table 5 of their paper). Hence, the magnetic field strength we infer around V\,Hya is consistent with the results of \citealt{2013A&A...554A.134L} if it is generated by the AGB primary. Other studies find similar strengths of H$_2$O masers in evolved stars (\citealt{2010A&A...509A..26A}, \citealt{2006ApJ...648L..59V}). 


We consider (but reject) the possibility that the magnetic field in V\,Hya may be generated by, and thus anchored in, the accretion disk around the companion. We can estimate the 
magnetic field strength at its footpoint in the 
disk ($B_c$) for different field geometries. Magnetocentrifugal wind models typically yield jet ejection speeds similar to the orbital speed at the field line's footpoint. Hence, 
setting the ejection speed to be comparable to that measured for bullet\#3's speed in epoch 4 (153\,\kms), we find that the footpoint lies at a radius of 
$\lesssim7\times10^{10}$\,cm from the companion. 
Hence, for a poloidal field geometry (i.e., $B(r) \propto r^{-2}$), $B_c>7.51\times10^{6}$\,G, a value that is implausibly large. 

We therefore conclude that if bullet\#3's non-ballistic trajectory and motion is due to a magnetic field, the latter must be due to the primary AGB star, and not the companion or the disk around it.

Additionally, we find evidence for a magnetic field embedded within the bullet, based on the inferred increase in $p$ from epoch 1 to epoch 3 of  bullet\#1. We propose that this increase is due to the relaxation of a toroidal magnetic field embedded within the hot bullet, as has been demonstrated via magneto-hydrodynamical simulations by  \cite{1999ApJ...517..767G}. These authors find a linear increase in the expansion velocity of their simulated high-velocity jets resulting from the presence of such a magnetic field. The presence of an embedded toroidal magnetic field within the bullet is supported by a numerical simulation of V Hya's bullet\#1 by Huang \& Sahai, in prep, which shows that such a field is required in order to restrict its lateral expansion with time as observed.

\subsubsection{Hydrodynamic Interactions}

\begin{figure}
\centering
\includegraphics[width=0.75\textwidth]{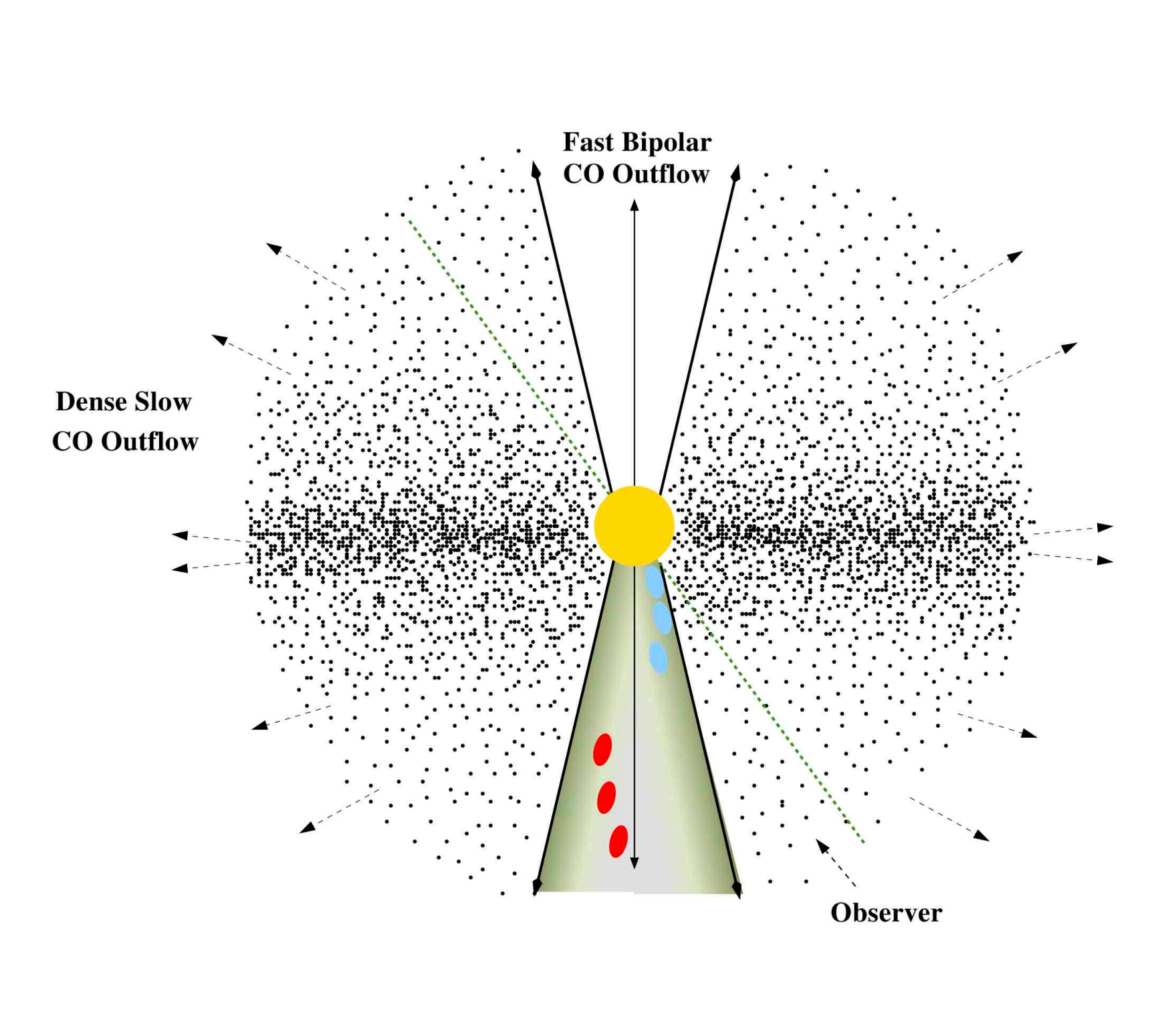}
\caption{\label{hydro} Schematic representation of the trajectories of various bullets as they move through a structured ambient circumstellar medium (``dense slow CO outflow"). A
several hundred year long history of such bullet ejections in V Hya's past has created a conical cavity that naturally shepherds the bullets towards the nebular axis, as defined 
by a ``fast bipolar CO outflow". Image not to scale. The conical shape of the cavity provides an inexact representation of the actual shape of the cavity walls, which are most likely curved inwards (e.g., the cavity more closely resembles a paraboloid than a cone). }
\end{figure}

The changes in the mass, volume, $PA'$ and inclination angle of bullet\#1 
(\S\,\ref{masschange}) during epochs 1--3, inferred from our SHAPE modeling, provide support for a strong hydrodynamic interaction of the bullet with its environment. Our modeling shows that the changes in $PA'$ and inclination angle are due to changes in the shape and orientation of the bullet. A latitudinal density gradient within the ambient material, such that the density decreases away from the nebular axis, can explain these 
changes. This density gradient could have been created by a continual sweeping up of material as more and more bullets are ejected, pushing material away from the 
vicinity of the nebular axis.  Bullet\#1 and, by inference, the other three bullets, must accumulate mass as they move away from V Hya, through interaction with ambient material. 
V Hya has likely been in this bullet-ejection stage for several hundred years (Paper I); during this period, the bullets have carved out a dense-walled conical cavity within the ambient circumstellar envelope 
created by the primary's mass-loss. The cavity walls naturally deflect the bullets towards the axis of the fast bipolar CO outflow in V Hya observed by \citealt{2004ApJ...616L..43H} (Fig.\,\ref{hydro}).

Further evidence for hydrodynamic interaction comes from the significant brightening of bullet\#3 from epoch 4--5. In this case, bullet 3's trajectory during epochs 4--6 (Fig.\,\ref{bullet3_radacc}) may be explained 
as a result of the bullet interacting with the walls of the cavity 
such that its velocity vector becomes aligned closer to the LOS, resulting in a larger LOS velocity in epoch 5 compared to that in epoch 4. Bullet\#3's path thus supports the idea that bullets can be pushed towards the symmetry-axis of the bipolar outflow (Fig.\,\ref{hydro}) by interaction with a cavity wall. The fast bipolar CO outflow likely consists of material from the originally ionized bullets which have cooled (and entrained ambient material) as they move away from the star.

\paragraph{Tangential Acceleration}

We discuss below the bullets' tangential offsets relative to the nebular axis, and find that these are not linear, suggesting that the bullets are not experiencing linear motion but instead are being deflected by the cavity walls. 

Bullet\#1's average tangential motion in the first 7.6 years following ejection, $\sim$8.4 AU yr$^{-1}$, is much slower than its average tangential motion during epochs 1--3, $\sim$30 AU yr$^{-1}$ (from Table\,\ref{centroid} column 4). This change of motion is perhaps due to bullet\#1 being ejected at a slightly smaller angle relative to the bipolar cavity axis compared to bullet\#3, and therefore encountering the dense cavity wall later along its trajectory, right before epoch 1. Due to this encounter, bullet\#1 is deflected by the wall, acquiring a larger tangential motion, qualitatively similar to bullet\#3. We expect bullet\#3 to cross (in projection) the nebular axis earlier than bullet\#1 (which has not yet crossed to the east side of the nebular axis), assuming a ballistic trajectory for bullet\#3 and a continued steady offset to the east of 14 AU yr$^{-1}$. The assumption that bullet\#3 will undergo a constant tangential motion in the near future is based on bullet\#1's consistent not linear movement, after the first 7.6 years, every year at the same rate as stated above -- i.e., $\sim$30 AU yr$^{-1}$ from epoch 1 to 2, and again from epoch 2 to 3. The most recent ballistic motion of bullet\#1 probably results from it being located within the central region of the cavity, away from the dense walls.

Bullet\#2 is observed in its `detached' period (epochs 4--6) to lie where bullet\#1 was in its `detached' period (epochs 1--3), i.e., on the side of the nebular axis opposite to the one where it was ejected (west side), suggesting that bullet\#2 was also deflected by a cavity wall and crossed the nebular axis (Fig.\,\ref{distantblobsch}). During its `distant' period, bullet\#0 (the oldest bullet observed) is on the same side of the nebular axis (east side) as the one where it was first ejected, perhaps because it has undergone two deflections, off opposite sides of the cavity walls during its early history.

We conclude that cavity walls can act as a ``deflecting surface" which forces the impinging bullets to rebound from the wall towards the nebular axis (i.e. $PA'$ will decrease until it crosses the nebular axis, and then it will increase). We note that when the bullets are moving within the tenuous interior of a biconical cavity, there is little mass to entrain or sweep up. The increase in the bullet mass is likely due to material entrained from the dense cavity wall, each time the bullet deflects off the latter. Hydrodynamic simulations are needed to determine the densities within the interior of the biconical cavities and its wall. Numerical simulations have shown that interactions of an outflow with an external dense molecular cloud cause the outflow to get increasingly deflected as the impact parameter\footnote{Defined as a fraction of the interacting cloud radius, and measured from the cloud center in the sky-plane.} decreases, since the collision with the dense external cloud becomes increasingly head-on  \citep{2009ApJ...690..944B}. Although these simulations are on timescales much longer than what we observe in V Hya (i.e., up to 4200\,yr), they demonstrate the effect of a deflecting cloud on a jet-like outflow. The projected curvature of a bullet trajectory could therefore be due to a change in its 3D velocity vector as the bullet is deflected by its interaction with the cavity wall (Fig.\,\ref{hydro}). 



\paragraph{Radial Deceleration} 


We observe in bullet\#1, a decrease in radial velocity from 186\,\kms~in epoch 2 to 179\,\kms~in epoch 3. From our model's 3D velocity law there is complementary evidence of an internal decrease in 3D velocity (see \S\,\ref{level2}). In our velocity law the flattening point, $P_f$, is changed so the bullet reaches it's maximum modeled speed of 245\,\kms at a later distance towards the tip of the bullet, i.e. for epoch 1 at $0.38\ell$, for epoch 2 at $0.46\ell$ and for epoch 3 at $0.64\ell$, where $\ell$ is the length of the bullet. 

\paragraph{Brightening}

\begin{turnpage}
\begin{figure}
\subfigure[]{\label{9632offsrc0.2_lev1} \includegraphics[width=80mm]{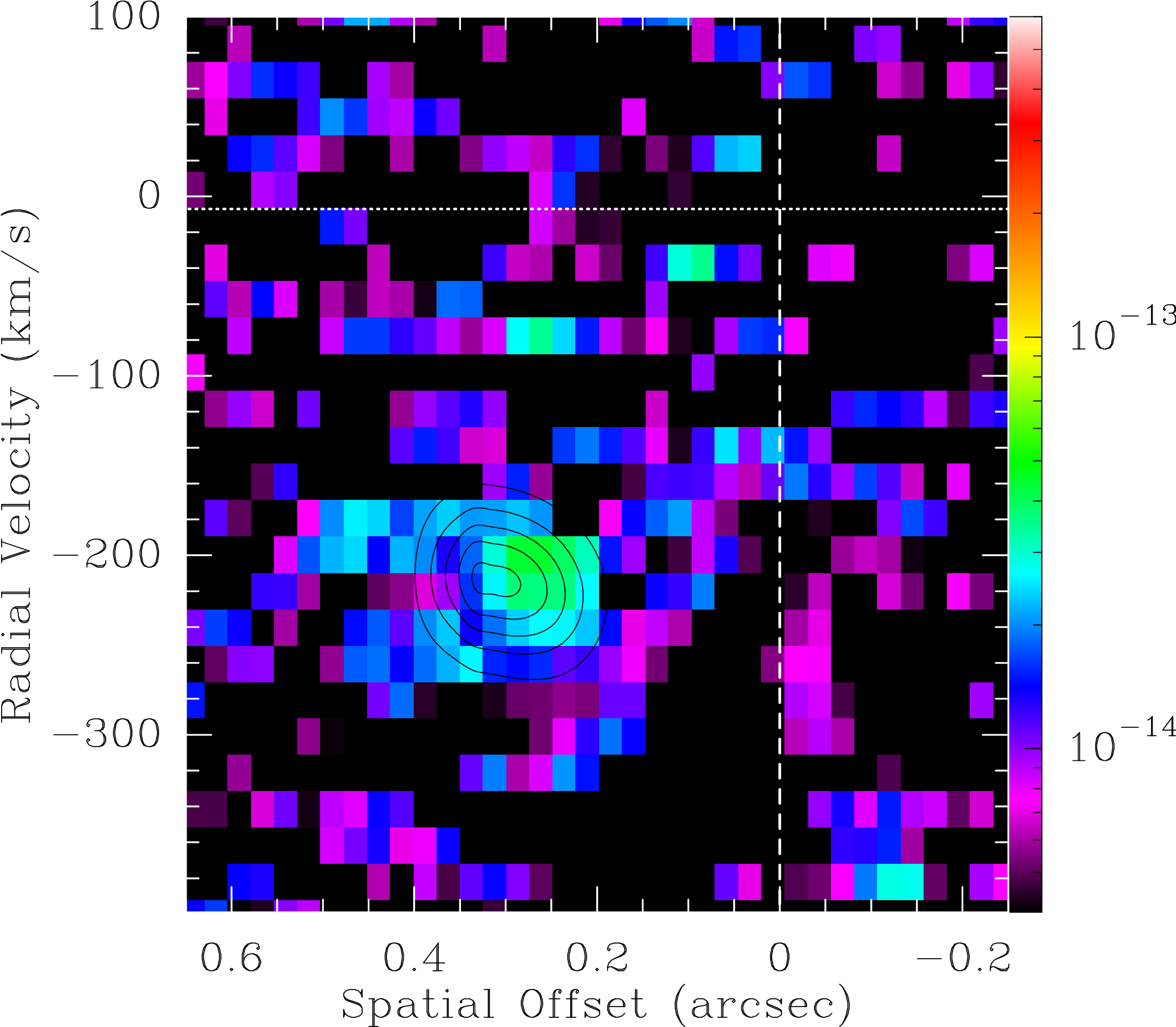}}
\subfigure[]{\label{9632offsrc0.1_lev1} \includegraphics[width=80mm]{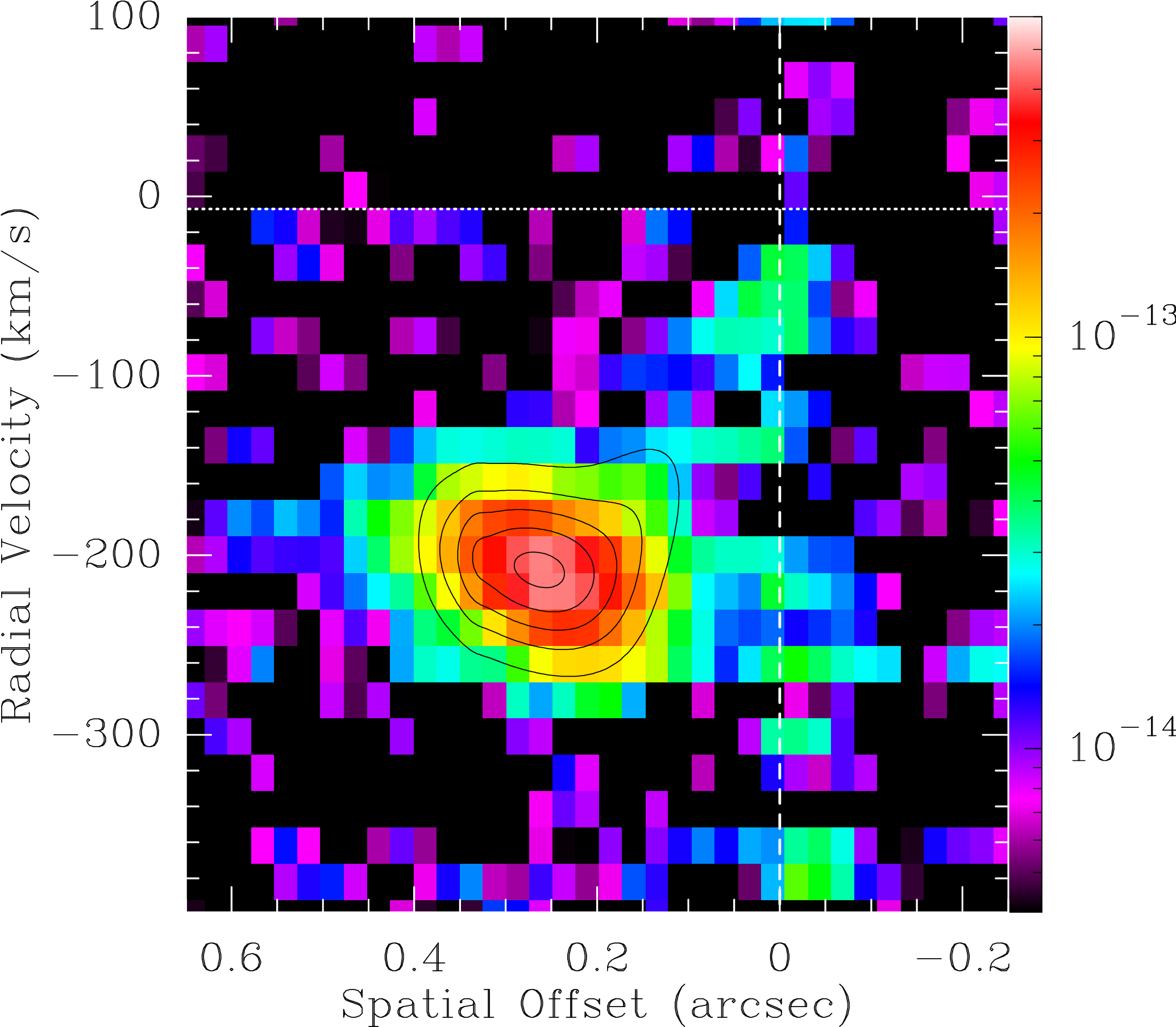}}
\subfigure[]{\label{9632onsrc_lev1}\includegraphics[width=80mm]{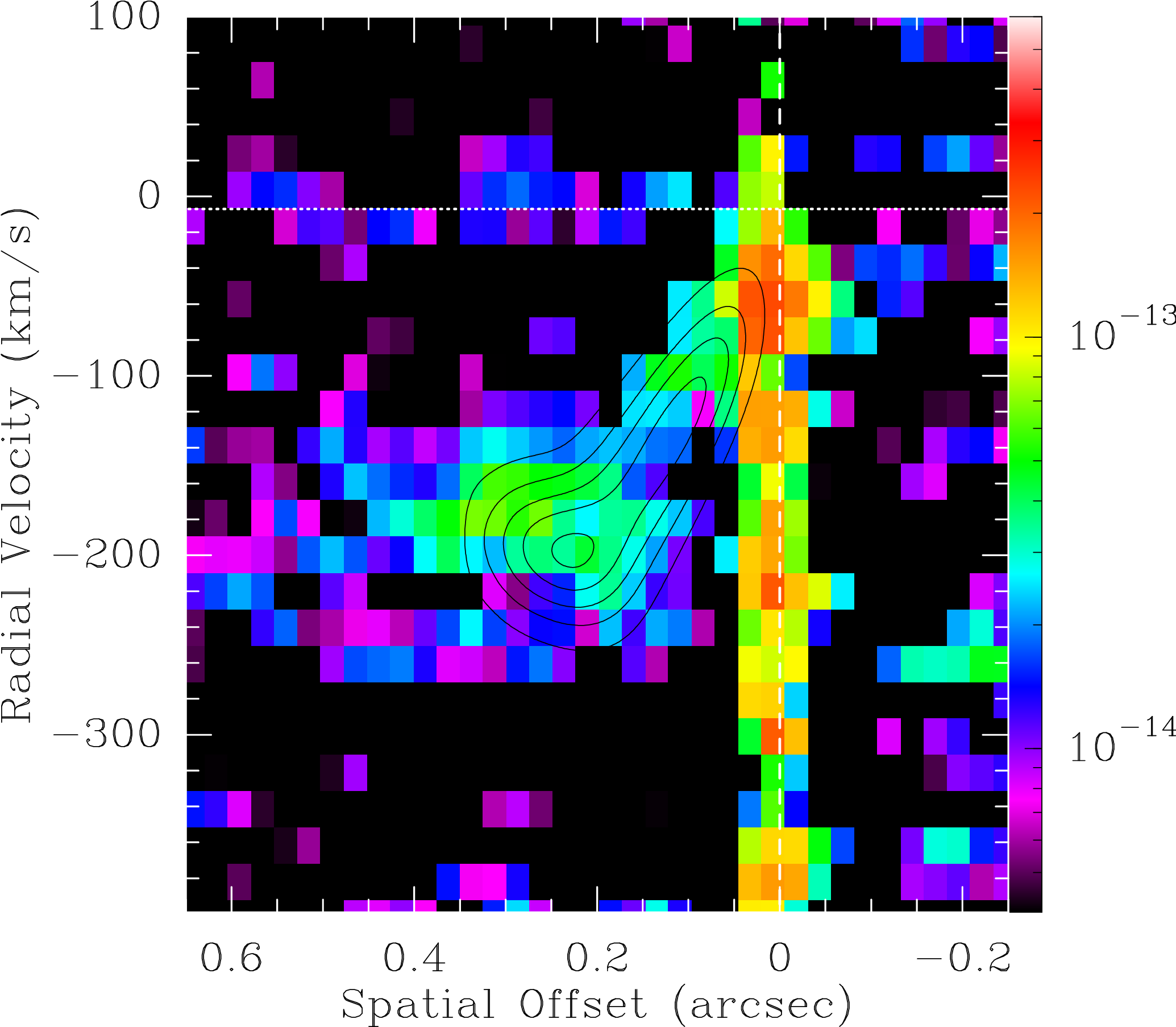}}
\caption{\label{9632model} PV plot of ``level 1" model for the detached emission blobs in the second epoch, 12-29-2002; colorscale shows the observations and the contours show the model.The 
observation and model results for slit $S_{-2t}$ are shown in (a), the results for slit $S_{-1t}$ are shown in (b), and the results for slit $S_{0t}$ are shown in (c).}
\end{figure}
\end{turnpage}

\begin{turnpage}
\begin{figure}
\subfigure[]{\label{9800offsrc_lev1} \includegraphics[width=80mm]{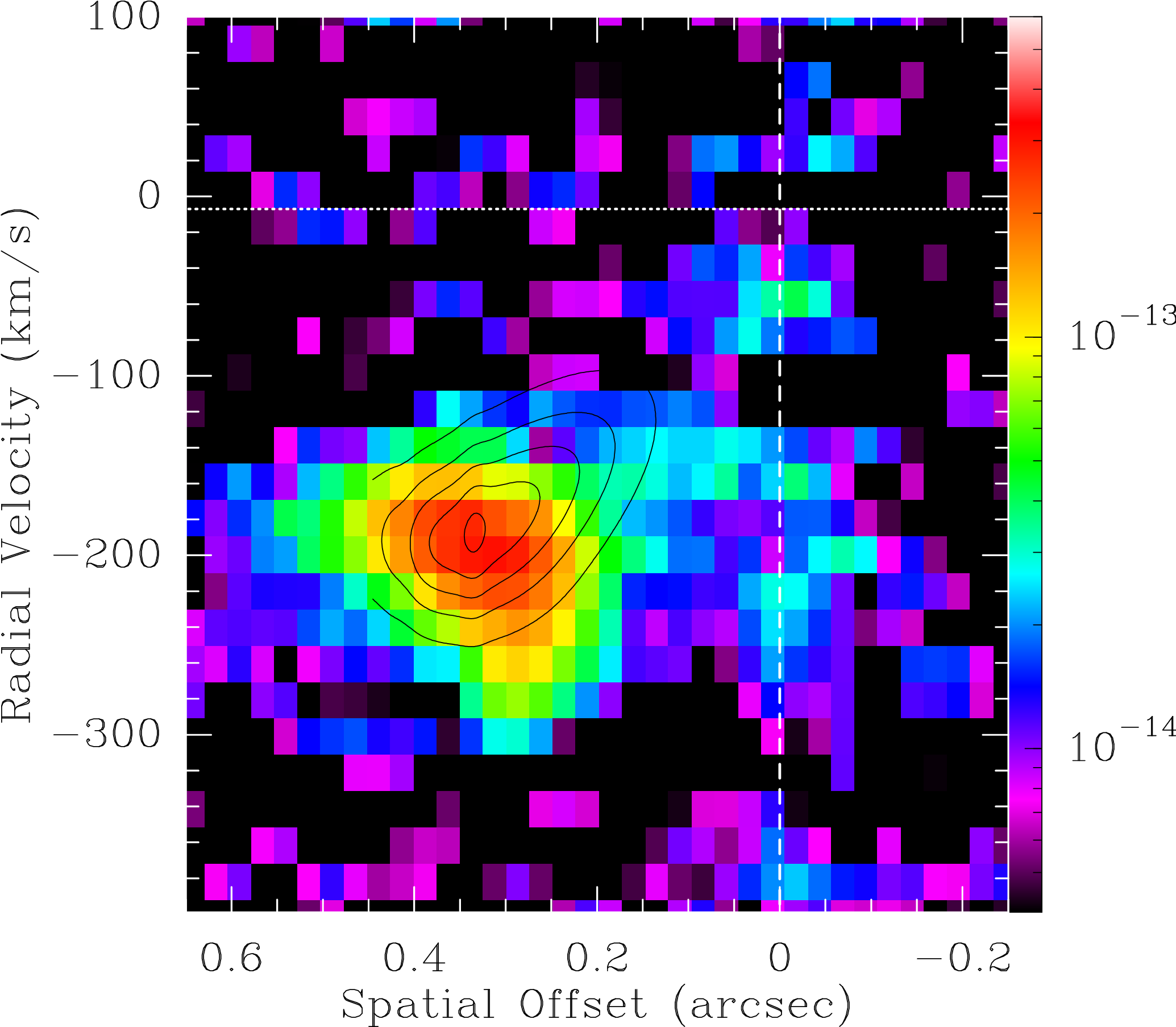}}
\subfigure[]{\label{9800onsrc_lev1} \includegraphics[width=80mm]{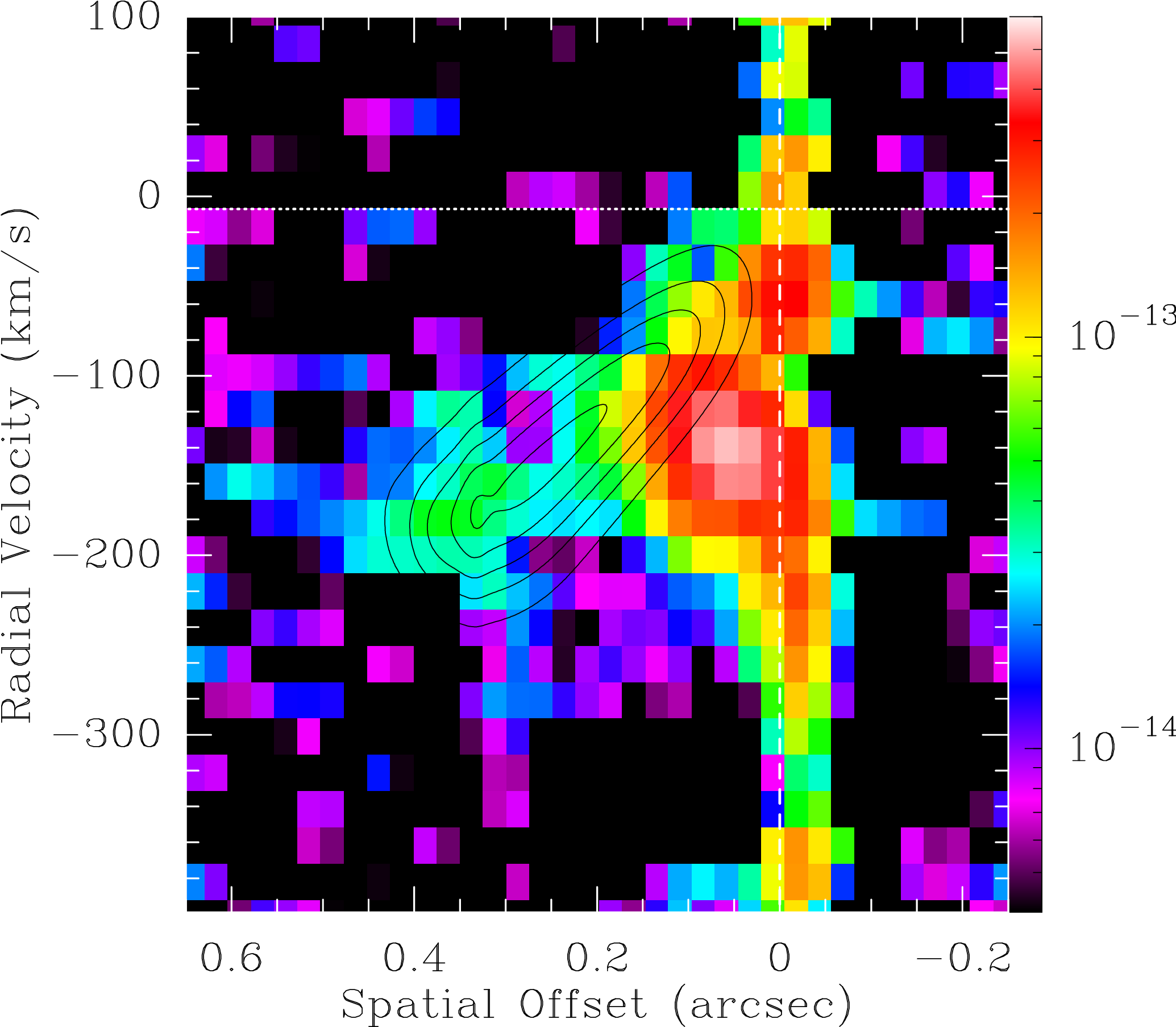}}
\caption{\label{9800model} PV plot of ``level 1" model for the detached emission blobs in the last epoch, 01-12-2004; colorscale shows the observations and the contours show the model. The observation and model results for slit $S_{-1t}$ are shown in (a), and results for slit $S_{0t}$ are shown in (b).
}
\end{figure}
\end{turnpage}

We now discuss how hydrodynamic interaction may help resolve some of the lingering deficiencies in our models. ``Level 2" intensity ratios are consistent within a factor of order unity compared to observations (Table\,\ref{intratios}), yet ``level 1" provides a better fit than ``level 2" for PV-morphology for epoch 3, slit S0$_t$ (compare Fig.\,\ref{9800model} to Fig.\,\ref{9800modellev2}), and epoch 2, slit S0$_t$ (compare Fig.\,\ref{9632model} and Fig.\,\ref{9632modellev2}). In all of our models bullets are cylindrically symmetric, therefore hydrodynamic interactions can break this symmetry making the side of the bullet closest to the cavity wall warmer and denser, therefore brighter in [SII] emission. 

If the side of bullet\#1 facing the cavity wall in V Hya were to be brighter, the emission from it would be brighter in the offset slits than in the on-source slit. Numerical studies support this idea, showing that emission from collimated outflows temporarily intensifies by an order of magnitude as these encounter the dense cavity wall of ambient AGB-wind \citep{2013ApJ...772...20B}. The amplitude of the brightness variations scales with the square of the varying density gradient across the shock as the bullet travels through the ambient medium. Considering these results, intensity ratios in our ``level 1" model would improve for epoch's 2 and 3, while worsening for epoch 1. We could, in principle, improve ``level 1" intensity ratios while also constructing an overall more robust model, with better PV-morphology than ``level 2". It is clear that a further study which provides an estimate of a density gradient in the surrounding medium is a warranted next step.

\section{Conclusion}\label{conclusion}

We present a quantitative spatio-kinematic model (using the SHAPE code) to fit the spectroscopic observational data of the bullet-like high-speed ejections from the carbon star, V Hya. Our main results are as follows:

\begin{enumerate}  
\item We set quantitative constraints on the physical parameters characterizing one of the bullets (\#1), such as it shape, mass, density, velocity and orientation. 
The general shape of the bullet is roughly a semi-ellipsoid that is tapered at both its ends and bulges in the middle.
  
The velocity of material within the bullet is not constant -- its magnitude increases from its origin (apex) towards its center, and then flattens out at about 245\,\kms. The model's velocity law is reasonably robust -- models in which the inclination angle and position angle were varied away from our best-fit model to accommodate different velocity laws were unsuccessful. The bullet density decreases away from the central axis towards its periphery. 
  
The bullet evolves via interaction with and entrainment of material from the ambient circumstellar environment, from epoch 1 to 3. First, its overall size increases: 
the bullet length increases from $0\farcs8$ (320\,au) to $1\farcs05$ (420\,au) and the cross-sectional diameter at its widest point increases from 
$0\farcs15$ (60\,au) to $0\farcs22$ (88\,au). Second, bullet mass increases from 10$^{27}$\,g to $2 \times 10^{27}$\,g, and then to $3 \times 10^{27}$\,g, and the volume  increases from $2 \times 10^{45}$\,cm$^{-3}$ to $3 \times 10^{45}$\,cm$^{-3}$, and then to $5 \times 10^{45}$\,cm$^{-3}$; however, the peak density remains roughly constant at $3.5 \times 10^{5}$ cm$^{-3}$. The (long) symmetry axis of the bullet tilts progressively towards the nebular axis, with the inclination angle of the axis increasing from 26$\arcdeg$ to 35$\arcdeg$ and its $PA'$ decreasing from 28$\arcdeg$ to 17$\arcdeg$.
  
\item We provide a detailed history of the movement of all four observed bullets, in terms of four spatio-kinematic parameters: LOS offset (x), sky-plane offset (y),  LOS velocity ($V_x$), and sky-plane velocity ($V_t$). This history shows that the bullets undergo both radial and tangential acceleration.
  
\item The tangential acceleration of newly ejected bullet\#3 may be caused by a magnetic field of 130\,mG, a value comparable to those derived from observations of H$_2$O masers found in other AGB stars. If so, the magnetic field must be due to the AGB star and not to the accretion disk around the companion, nor to the companion itself. Alternatively, bullet\#3 may have undergone a strong hydrodynamic interaction with the ambient circumstellar envelope.

\item Such hydrodynamic interactions, inferred from tangential acceleration, radial deceleration, and brightening of various bullets, are likely due to the latter moving within a dense-walled cavity that has been carved out by previous bullets in the ambient circumstellar envelope and deflecting off the cavity walls. 

\end{enumerate}

In the future we hope to obtain a more complete picture of V Hya by taking additional observations and running new simulations. An accepted proposal for cycle 25 Hubble observations will provide new data on the current status of V\,Hya's bullets. We aim to construct new SHAPE models for Period 2 observations. Recent 3D hydrodynamic simulations provide additional insight into the nature of the high-speed bullets; i.e., the bullets need to be confined by the hoop stresses of a toroidal magnetic field embedded within them, otherwise the bullets would expand laterally much faster than observed (Huang \& Sahai 2018, in prep). Such an embedded magnetic field has been inferred in the past for the high-speed bullets being ejected from the young PN, He\,2-90 (\citealt{2002ApJ...573L.123S}, \citealt{2004ApJ...606..483L}). In our continuing study of V Hya, we will investigate the disk that is the launch-site of the jet-like outflows that produce bipolar planetary nebulae, and test models for making such disks, including accretion onto a companion or common envelope ejection.

\acknowledgements We thank Wolfgang Steffen and Nico Koning for their continued support with our many questions in regards to the SHAPE code. We thank Neal Turner (JPL) for helpful discussions on magnetic fields. We also thank our anonymous referee for helpful comments. S.S.'s contribution was carried out during her tenure as a NASA Undergraduate Intern (UI) at the Jet Propulsion Laboratory (JPL). R.S.'s contribution to the research described here was carried out at JPL under a contract with NASA, with financial support provided by NASA, and in part from an STScI HST award (GO\,12227.01).

\clearpage

\renewcommand{\thesection}{\Alph{section}.\arabic{section}}
\setcounter{section}{0}

\begin{appendices}

\section{Appendix A} \label{appendixA}

\subsection{Instrumental Effects} \label{instrumeffects}
The bullet that we are modeling does not uniformly fill the slit, which produces an instrumental shift along the dispersion direction. If a point-source of monochromatic light with a certain wavelength, $\lambda_0$, is located at the center of the slit this will produce a finite-sized image centered at some specific location along the dispersion direction of the detector, which is calibrated to be at $\lambda_0$. 
If this same point of light is displaced away from the center of the slit towards the edge then its image will also move along the dispersion direction of the detector, even though its wavelength has not changed, thus introducing an artificial shift in the ``observed" wavelength.

\begin{figure}
\subfigure[]{\label{conemodeloffsrc} \includegraphics[width=80mm]{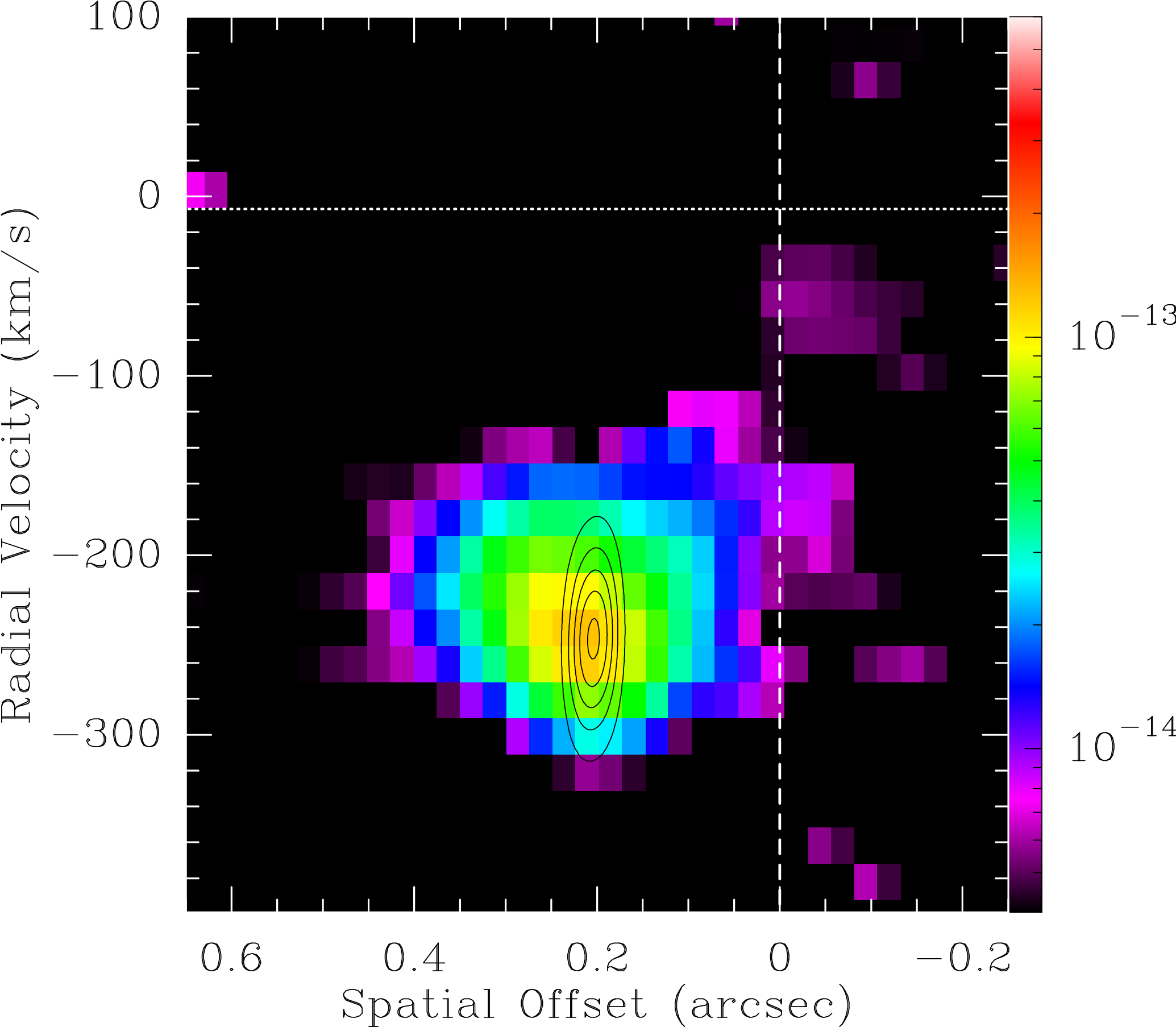}}
\subfigure[]{\label{conemodelonsrc} \includegraphics[width=80mm]{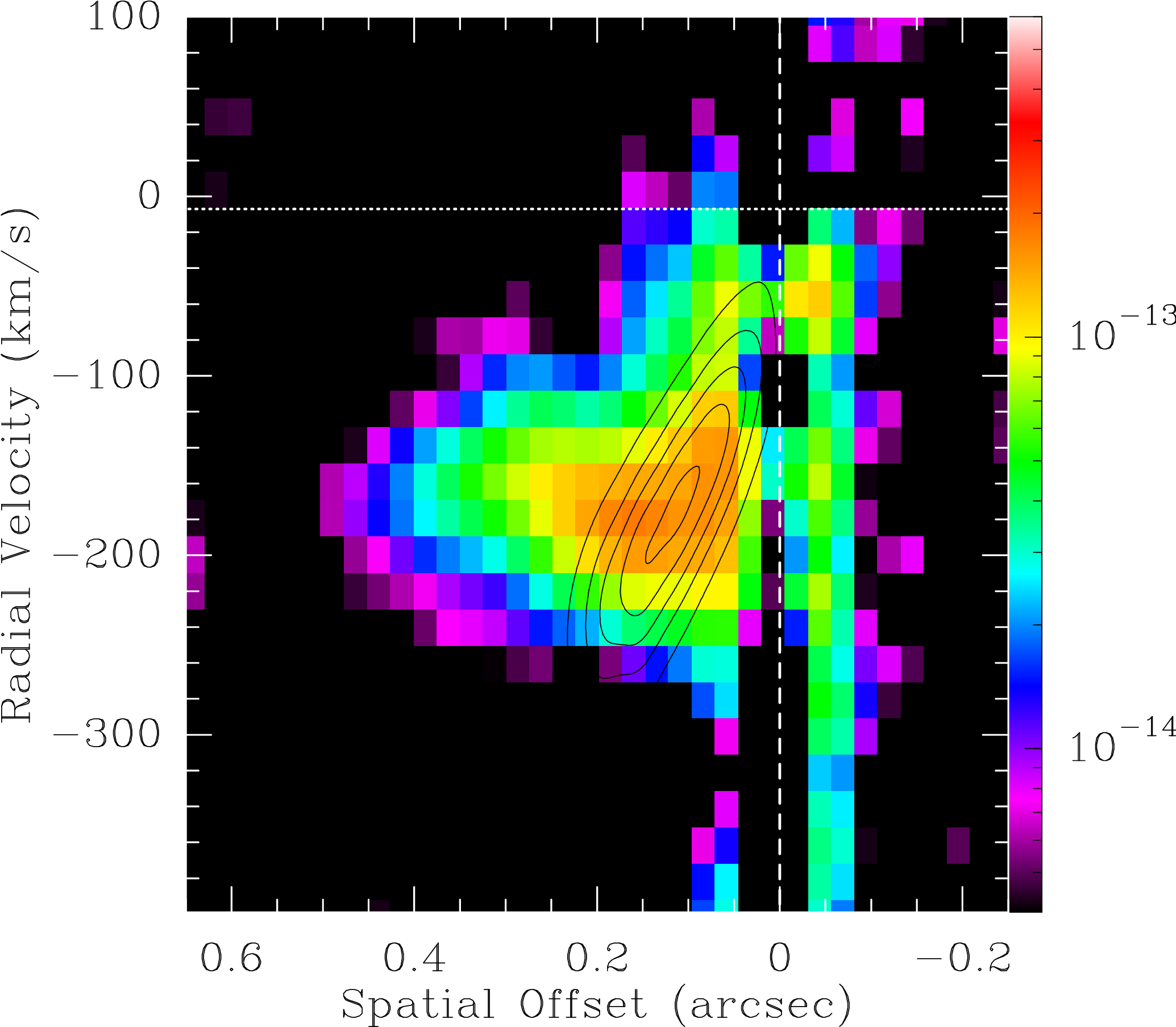}}\\
\subfigure[]{\label{conemodeloffsrcsh} \includegraphics[width=80mm]{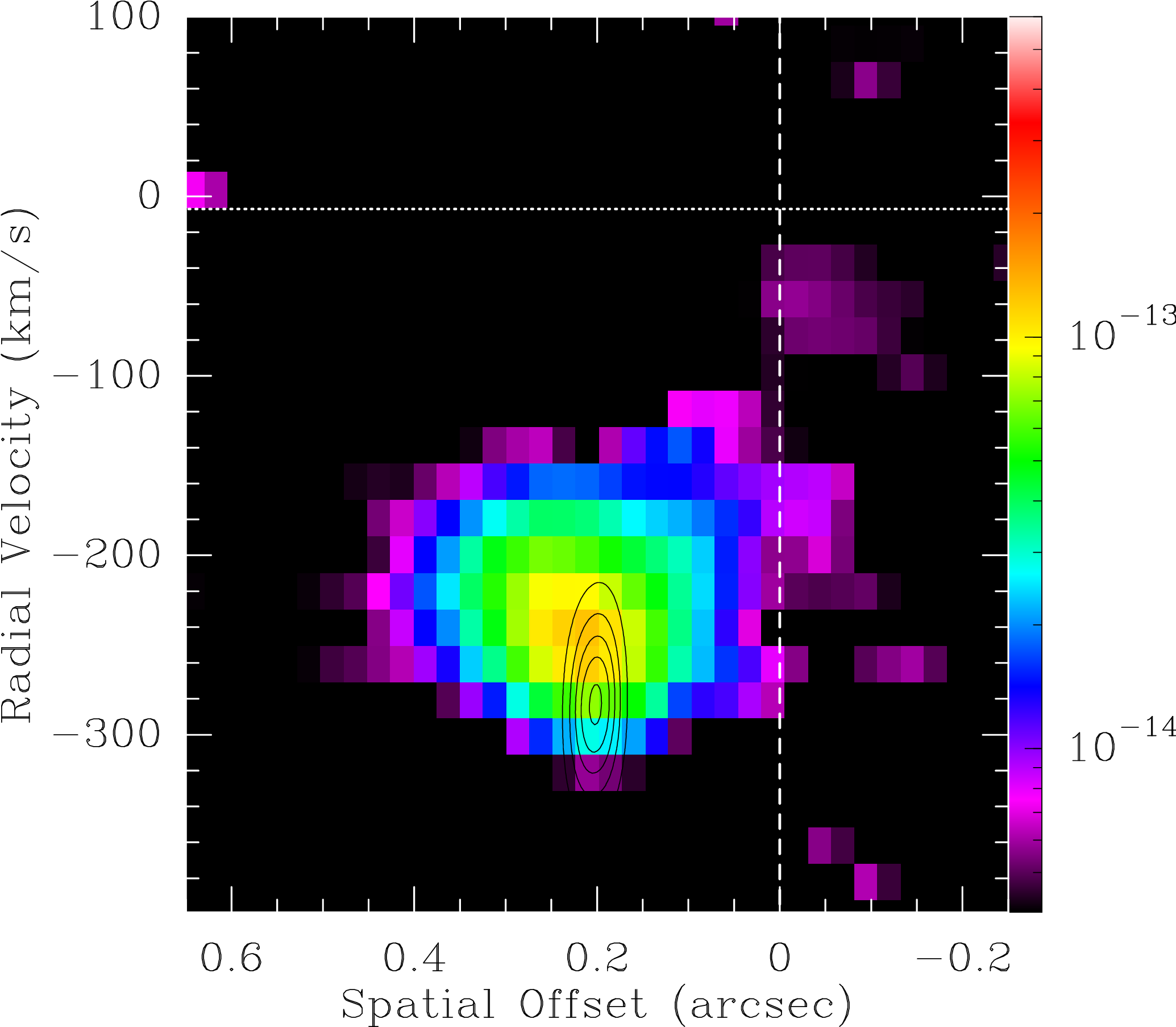}}
\subfigure[]{\label{conemodelonsrcsh} \includegraphics[width=80mm]{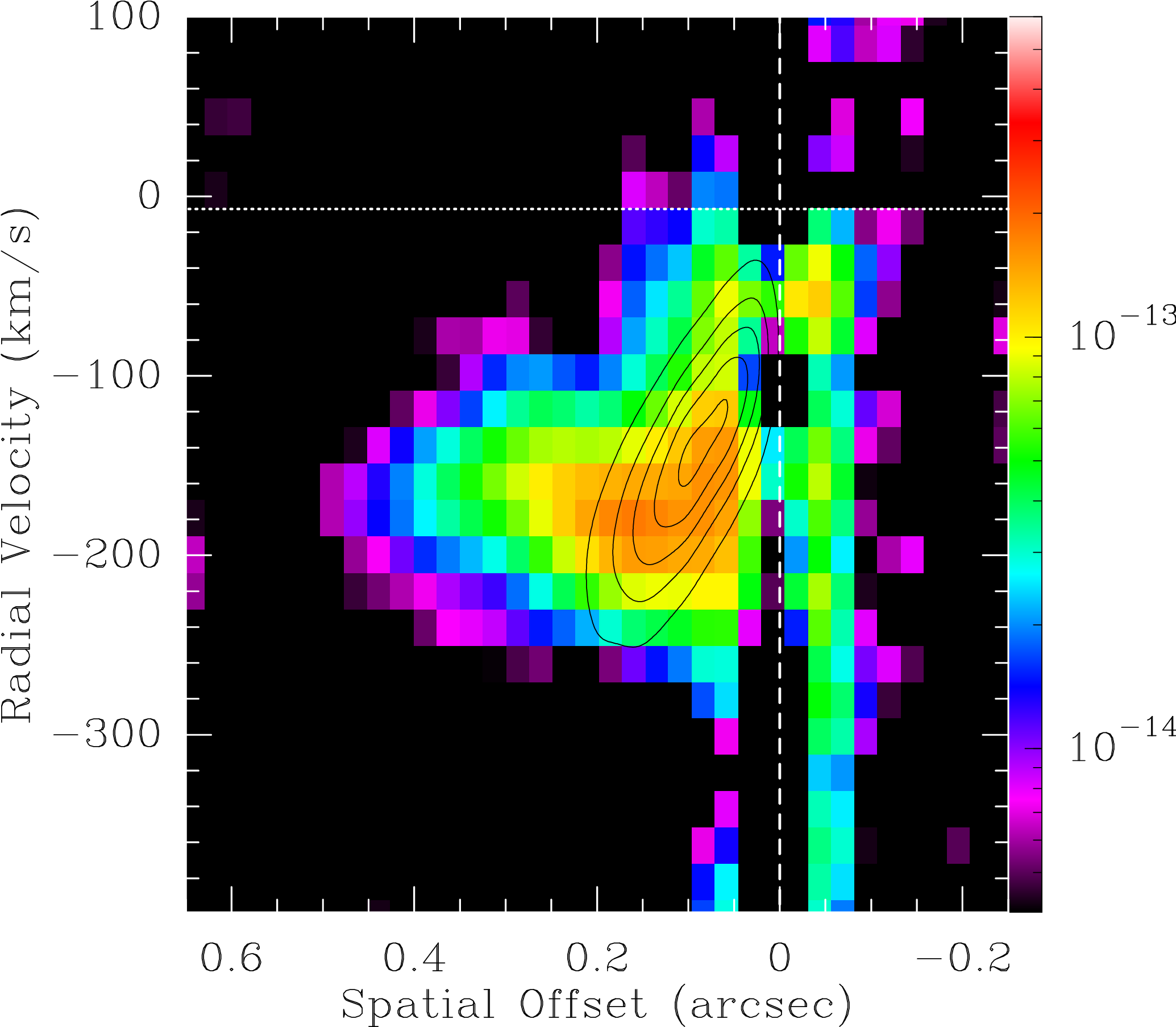}}
\caption{\label{conemodels}PV diagrams for a cone-shaped bullet geometry. The top row shows our original model, without the instrumental effect
correction (see \S\,\ref{instrumeffects}). The bottom row shows the same model but now including the instrumental shift.
Both are ``level 0'' models with $PA'$ = 20$\arcdeg$, i = 20$\arcdeg$, height of the cone, $h$ = $1\farcs0$, 
and radius r$_b$ = $0\farcs06$ (i.e. the radius of the base of the cone). The density and velocity are both linear functions, n($\rho$) = $1.0-1.2(\rho/h)$ for 0 $<$ $\rho$ $<$ $h$ and v($\rho$)(\,\kms) = $-75-310(\rho/h)$ respectively. The variable $\rho$ is the radial vector in spherical coordinates.}
\end{figure}

The plate scale of the STIS CCD is $0\farcs05078$ per pixel and has a dispersion of 0.276 Angstroms/pixel, which gives a shift of 5.4352 Angstroms per arc second of displacement from the slit-center. The SHAPE code was modified to incorporate this instrumental effect in order to allow us to specify the amount of wavelength shift caused as a function of offset from the center of the slit along the dispersion direction of the slit. Observed emission from epoch 1 has been overlaid with model contours in PV space from an initial cone-shaped geometry model (``level 0"), both before the instrumental effect was incorporated (Fig.\,\ref{conemodels} top row) and after (Fig.\,\ref{conemodels} bottom row).

\section{Appendix B} \label{appendixB}

\subsection{Level 0 Model} \label{lev0}
Our ``level 0" model was constructed from the same conical geometry as in the Setal03 model, but with altered parameters that improved our comparisons of morphology, peak intensity and relative intensities.
This model was created with a $PA'$ = 20$\arcdeg$, i = 20$\arcdeg$, a cone height $h$ = $1\farcs0$, and cone radius base r$_b$ = $0\farcs06$. The relative density\footnote{We do not have absolute units for density.} and 3D velocity are both linear functions,
n($\rho$) = $1.0-1.2(\rho/h)$ for 0 $<$ $\rho$ $<$ $0\farcs8$ and v($\rho$)(\,\kms) = $-75-310(\rho/h)$ respectively. The variable $\rho$ is the radial vector in spherical coordinates. Unless otherwise stated, the velocity vector is always directed radially away from the center of the system (see Fig.\,\ref{coordgrid} a).

The centroids of the ``level 0" emission model in both slits of epoch 1 align well with the data. However, the detailed morphology of the emission in the PV plot (PV-morphology) is clearly dissimilar to the model, as the model contours are much narrower (Fig.\,\ref{conemodels} top row). The assumed cone-shape of the bullet created diagonal contours in the on-center slit and horizontal contours in the off-center slit, inconsistent with the actual vertical structure (east/west smear) of the observed emission in both slits. Attempts to ``patch" the ``level 0" model by adopting variable velocity and density laws did not improve the fit of the PV-morphology.


\subsection{Level 1 Model} \label{lev1model}

\begin{figure}
\vspace{5in}
\centering
\includegraphics[width=0.8\textwidth]{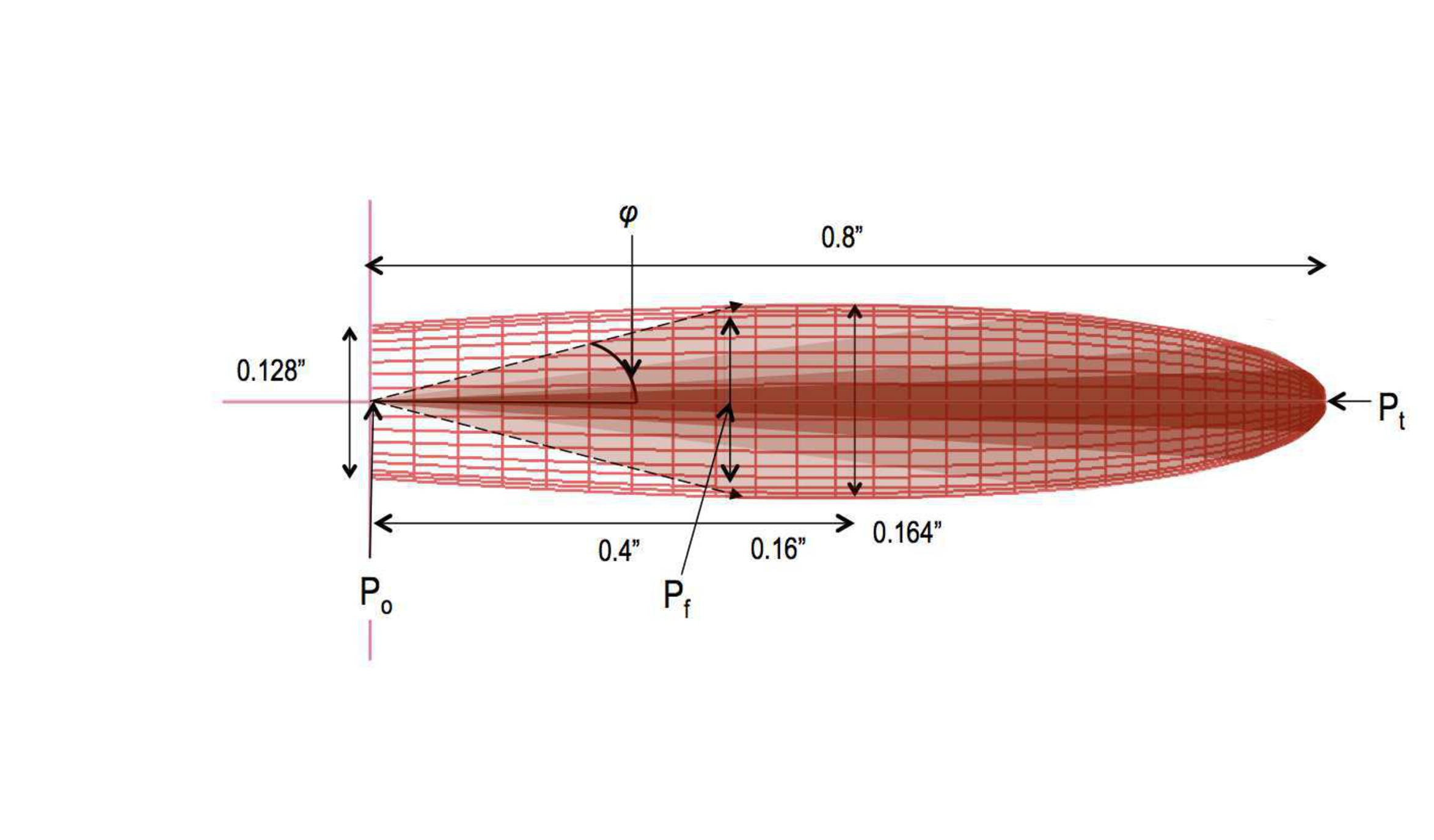}
\caption{\label{sketch}Schematic geometry of the axisymmetric bullet for the first epoch, January 2002, for our ``level 1" model. The length of the bullet is 
$0\farcs8$, and the cross-sectional diameter at its base is $0\farcs128$, increasing to $0\farcs164$ at its widest at 
$z=0\farcs4$. The tip of the bullet is labelled as the cardinal point $P_t$. The flattening point $P_f$ and the origin point $P_o$, which are each relevant to our flattened velocity law, are also labelled. The width of the bullet at $P_f$ is $0\farcs16$. The variable $\varphi$ is defined in the density law; the density decreases to zero at 30$^o$.}
\end{figure}

\begin{turnpage}
\begin{figure}
\subfigure[]{\label{9100offsrc_lev1}\includegraphics[width=80mm]{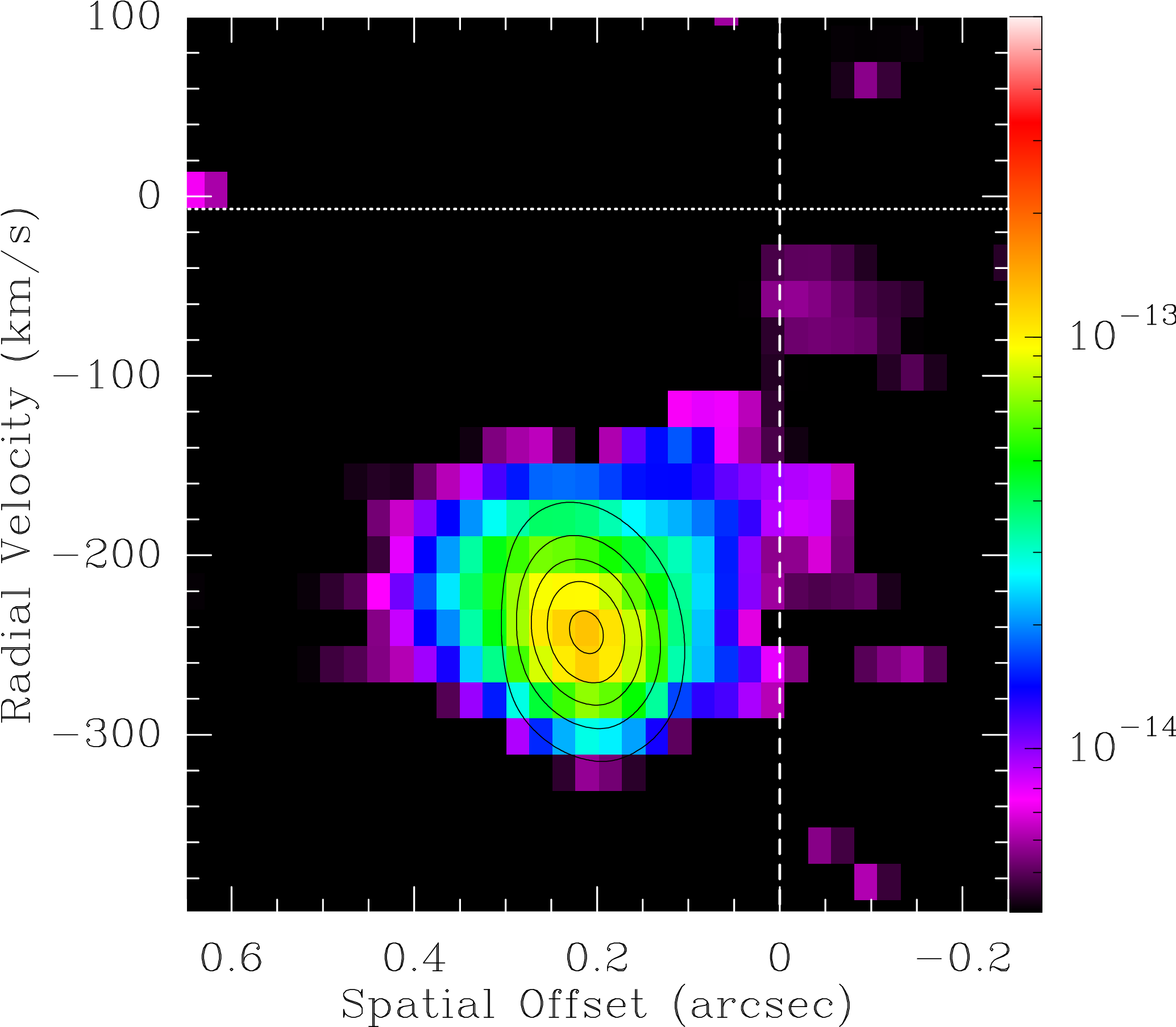}}
\subfigure[]{\label{9100onsrc_lev1}\includegraphics[width=80mm]{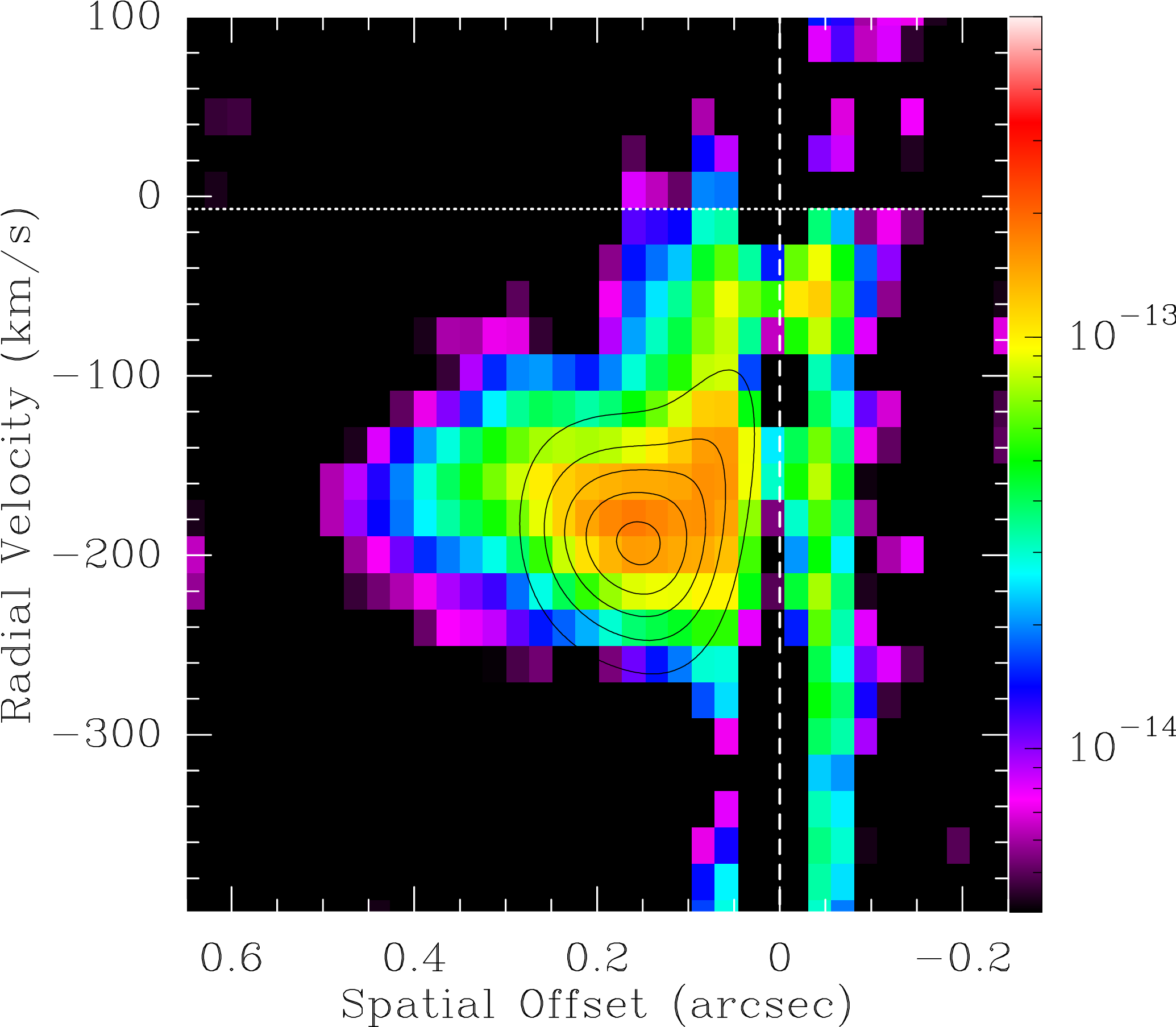}}
\caption{\label{9100model} A PV plot of ``level 1" model for the detached emission blobs in the first epoch, 01-28-2002; colorscale shows the observations and the contours show the model. The 
observation and model results for slit $S_{-2b}$ are shown in (a), and results for slit $S_{0b}$ are shown in (b).}
\end{figure}
\end{turnpage}

Because of the distorted PV-morphology, we modified the basic geometry of the bullet from a cone-shape (``level 0") to a semi-ellipsoidal shape (``level 1"). The cone-shaped geometry widens at the end, creating the extreme localized emission in the model.
Therefore, by switching to an ellipsoidal shape, which bulges in the middle and tapers off at the forward tip, we can improve the overall fit to the PV-morphology. 
The adopted semi-ellipsoid shape, based on first epoch observations, has a length of $\ell$ = $0\farcs8$. The cross-sectional diameter varies from the base to the tip as shown in Fig.\,\ref{sketch}, with a maximum diameter of  $0\farcs164$. Observed emission and ``level 1" model contours for the first three epochs are shown in Figures \,\ref{9100model}, \,\ref{9632model}, and \,\ref{9800model}. We found that the choice of parameters for this model produced a desirable fit, both in respect to PV-morphology and centroid position of the observed emission.

In this level we focused on how to quantitatively modify our model to account for change as a function of epoch. The geometrical shape of the semi-ellipsoidal bullet was kept the same for all three epochs, however the bullet was translated south-east in accordance to the velocity law. We use cardinal points, i.e., tip position $P_t$, flattening position $P_f$ and origin position $P_o$, of the bullet to aid in describing both the geometry and the velocity law. The width of the semi-ellipsoid at the flattening point $P_f$ is $0\farcs16$. The adopted ``level 1" model for each epoch shares a flattened velocity law that extends to $-245$\,\kms~at $P_t$ (i.e. Epoch 1 law in Fig.\,\ref{vellaw} ``Level 1 \& 2") and a flattened density law which reaches zero at $\varphi$ = 15$\arcdeg$ (Fig.\,\ref{densitylaw} ``Level 1"). 

\subsubsection{Bullet Evolution with Time} \label{evolution}

To roughly determine the evolution of the bullet's size and shape, we assume there is no radial acceleration when we calculate the location of the cardinal points labeled in Fig.\,\ref{sketch}. This assumption is observationally motivated (see \S\,\ref{radialacc}). The three cardinal points, labeled $P_t$, $P_f$ or $P_o$, are defined by coordinate pairs 
($\rho$, $v(\rho)$) where $\rho$ is the radial vector length and $v(\rho)$ is the velocity at that length in \,\kms~for each of the three epochs; i.e., in epoch 1 at the tip of the bullet the length is $0\farcs8$ and the velocity is -245\,\kms, recorded as (0.8, -245.0) in Table\,\ref{varymodelparams}. 
The width of the bullet scales with the length within our SHAPE model, since the geometry is a semi-ellipsoid.
We tabulate the adjusted lengths, $PA'$, inclination angle $i$ and translation offsets of the bullet toward the east and south of the emission blobs in each of the epochs in Table\,\ref{varymodelparams}. These variable parameters remained the same in the ``level 1" and ``level 2" models (see \S\,\ref{level2}). 

\begin{turnpage}
\begin{figure}
\subfigure[]{\label{9100blob_lev1} \includegraphics[width =80mm]{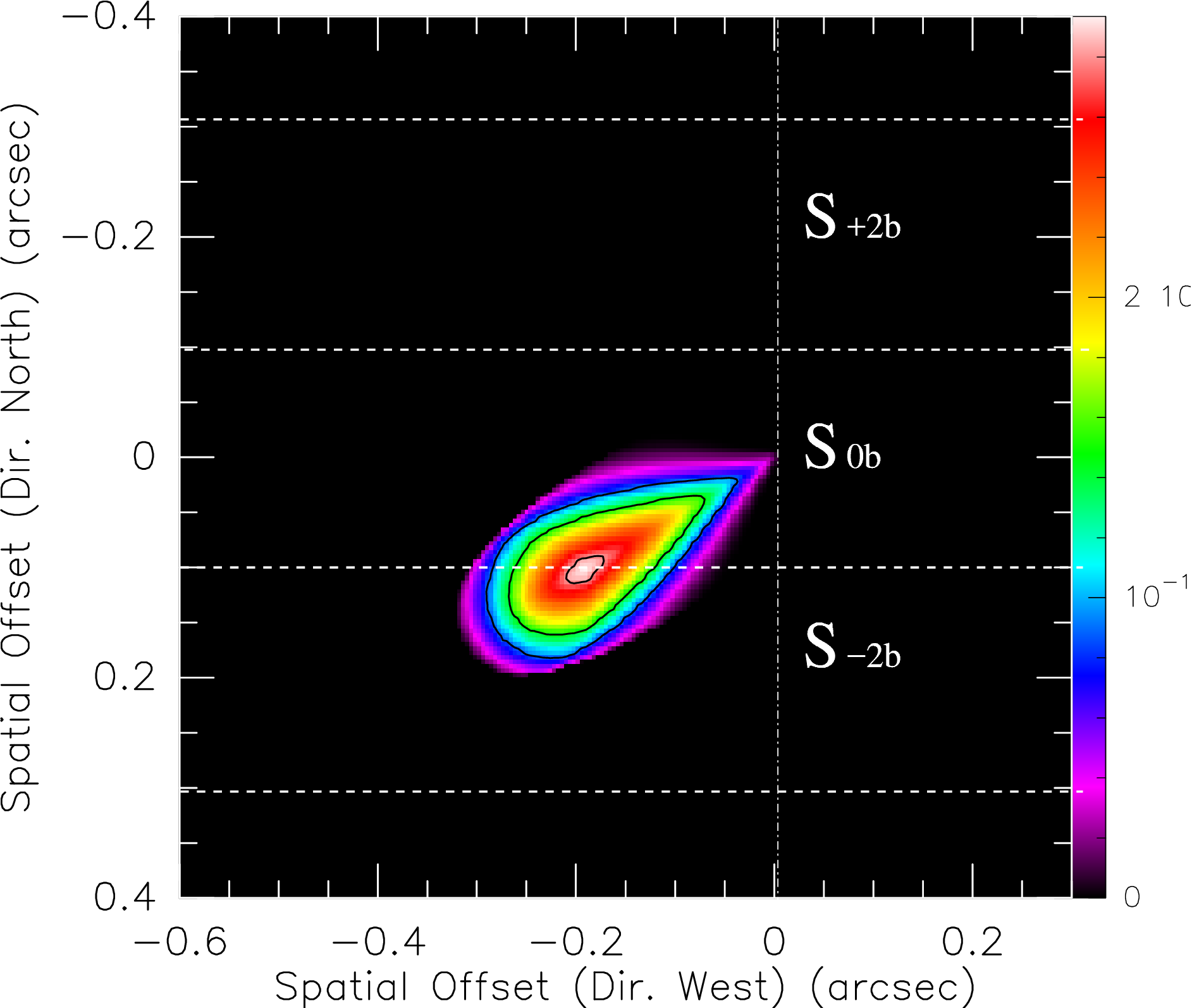}}
\subfigure[]{\label{9632blob_lev1} \includegraphics[width=80mm]{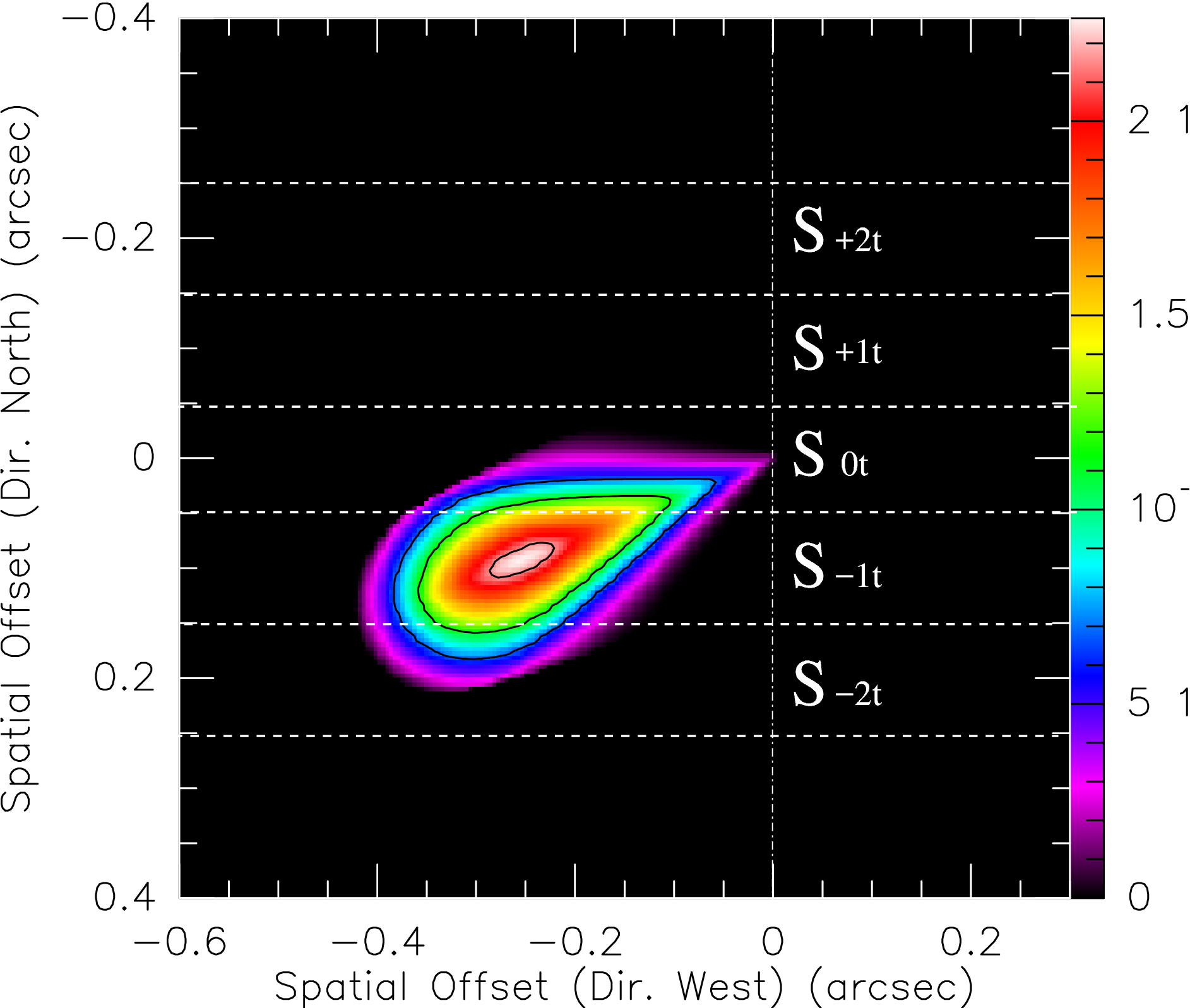}}
\subfigure[]{\label{9800blob_lev1} \includegraphics[width=80mm]{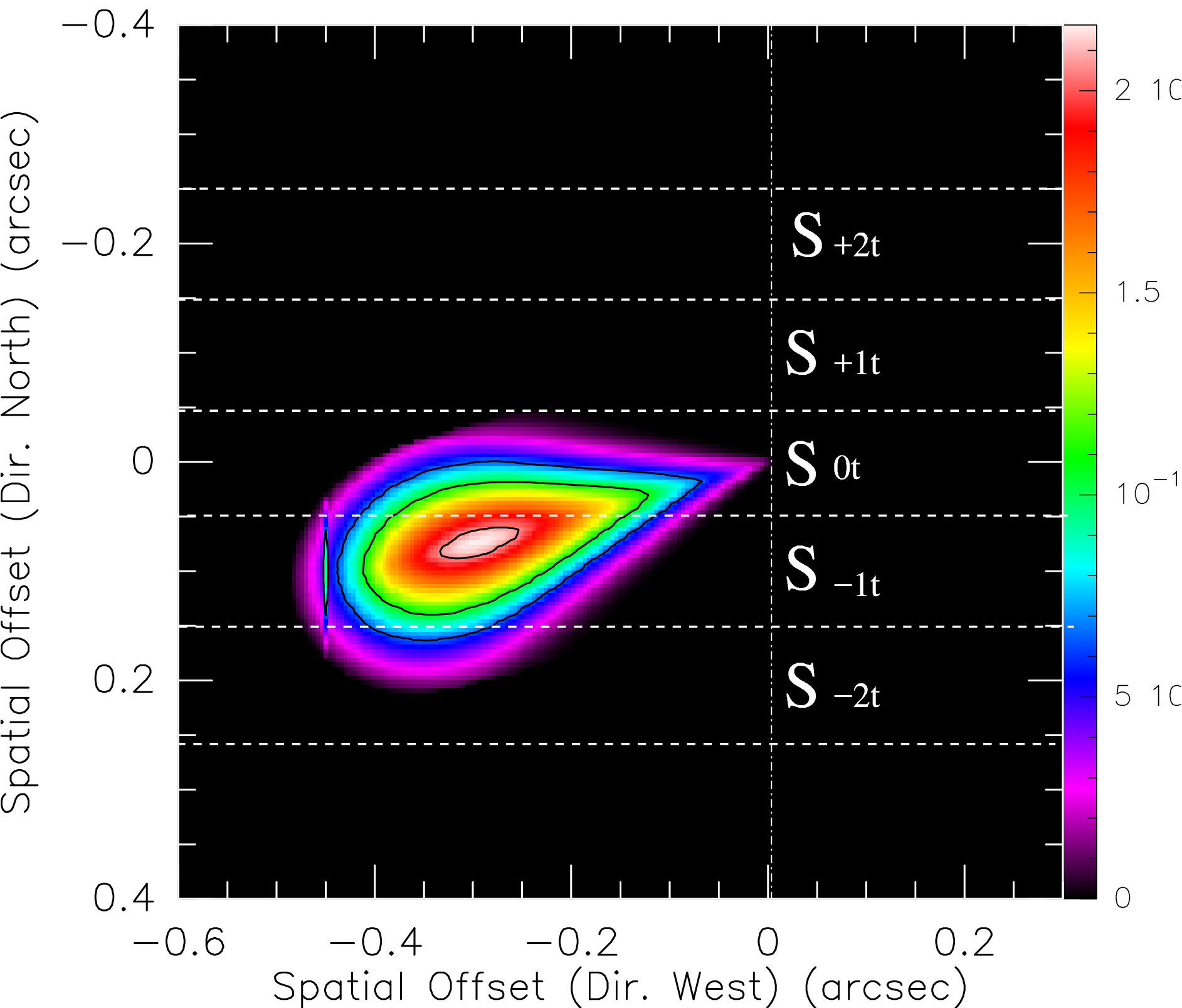}}
\caption{\label{blobs}``Level 1" projected surface brightness distribution of the high-velocity jet in the first epoch 01-28-2002 (a), the second epoch 12-29-2002 (b), and the third
epoch 01-12-2004 (c).}
\end{figure}
\end{turnpage}

The intensity distribution of the modeled bullet at each epoch, as well as the change in
$PA'$ as a function of epoch, is depicted in Fig.\,\ref{blobs}.
The $PA'$ and inclination $i$ for epochs 2 and 3 were adjusted because the peak-emission location of the emission blob gets progressively more distant from the center. As mentioned, decreasing the $PA'$, i.e. moving the lower tip toward negative north and east offsets, changes the spatial offset of the blob from the center in both the central and offset slits by roughly similar amounts. An increase in the inclination, which lengthens the projection of the blob on the plane of the sky, changes the spatial offset of the blob in the offset slit more strongly than in the central slit.

\subsubsection{Velocity Law} \label{vellawsec}

\begin{figure}
\subfigure[]{\label{constoff} \includegraphics[width=80mm]{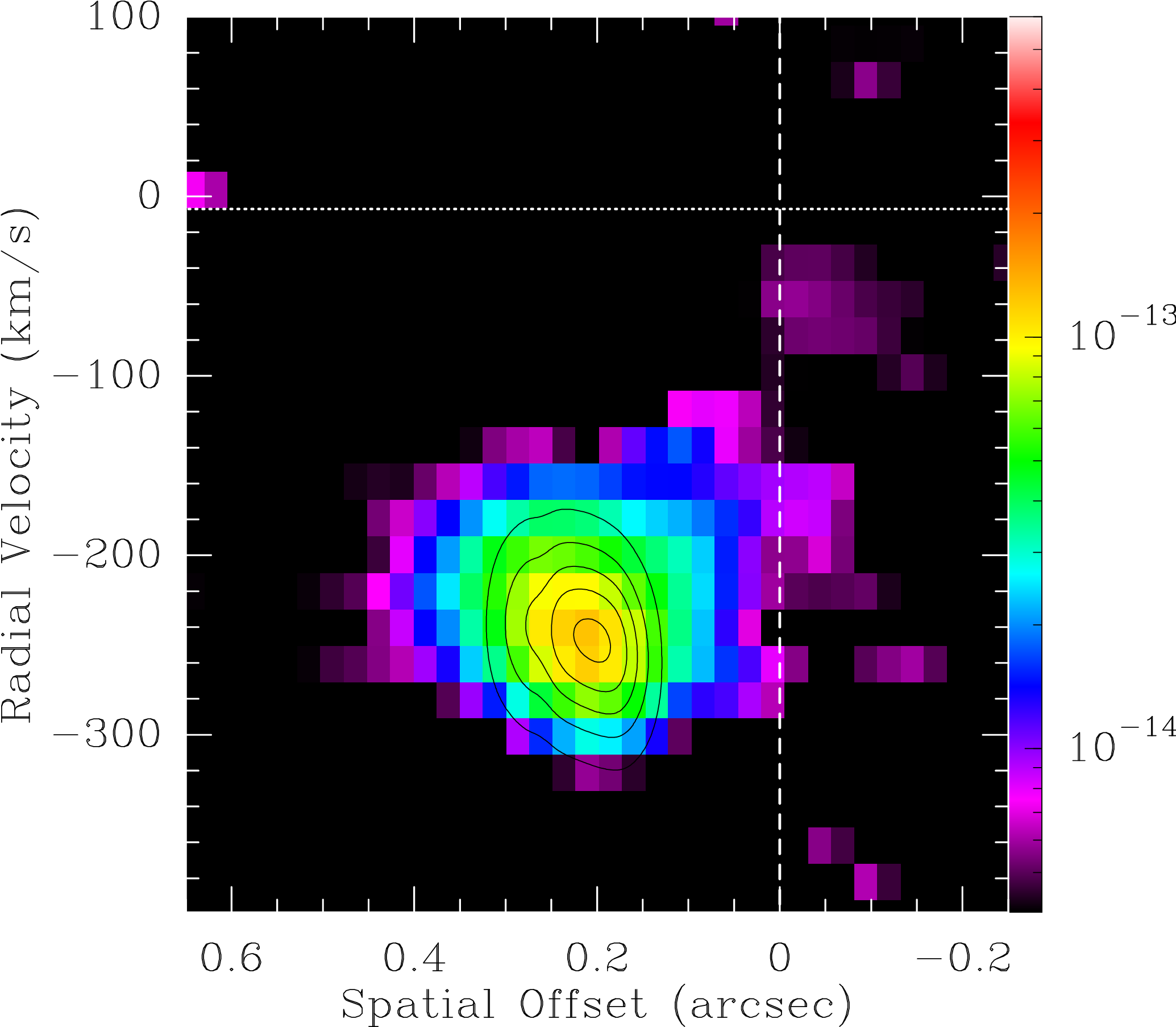}}
\subfigure[]{\label{conston}\includegraphics[width=80mm]{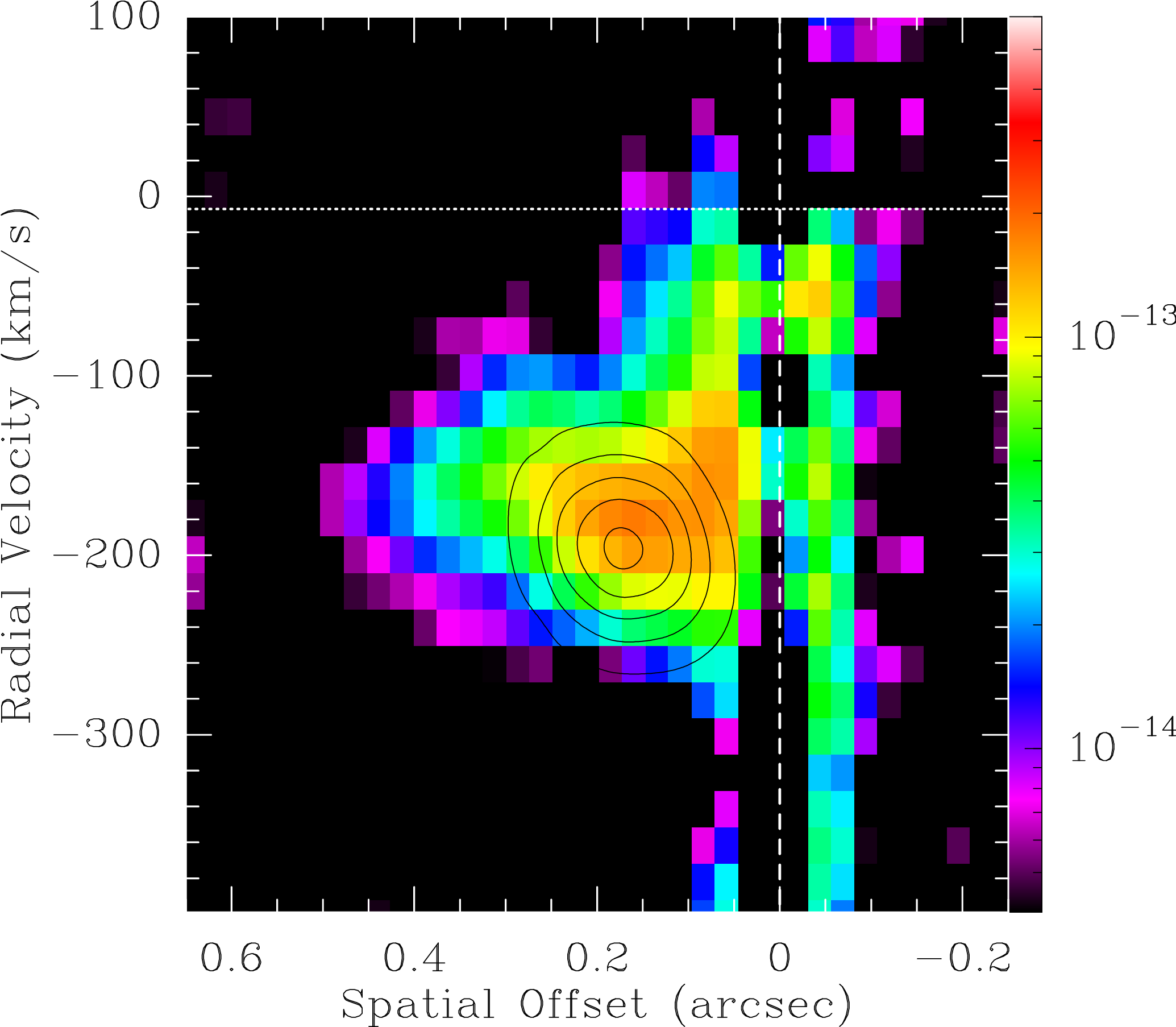}} \\
\subfigure[]{\label{linoff} \includegraphics[width=80mm]{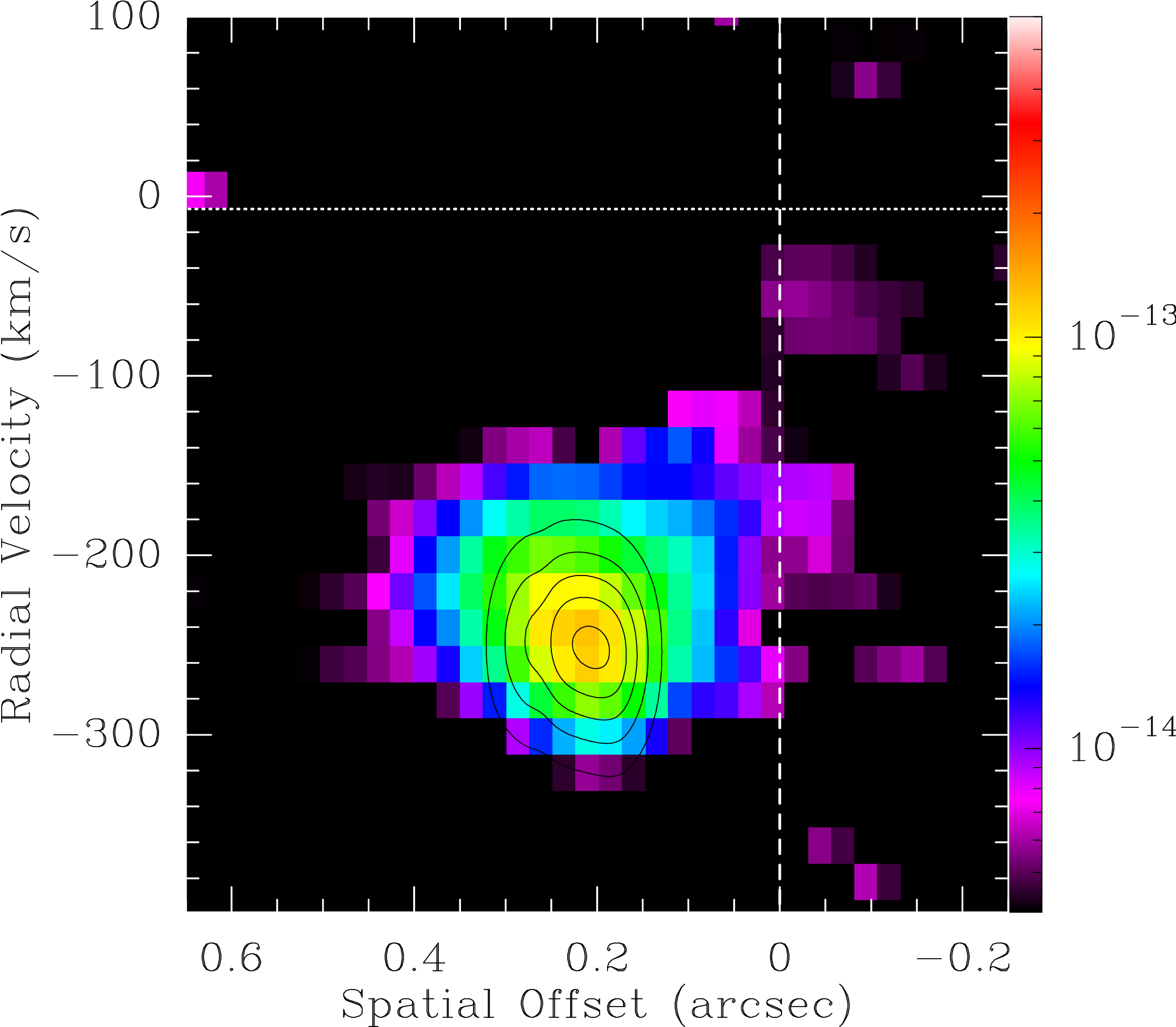}}
\subfigure[]{\label{linon}\includegraphics[width=80mm]{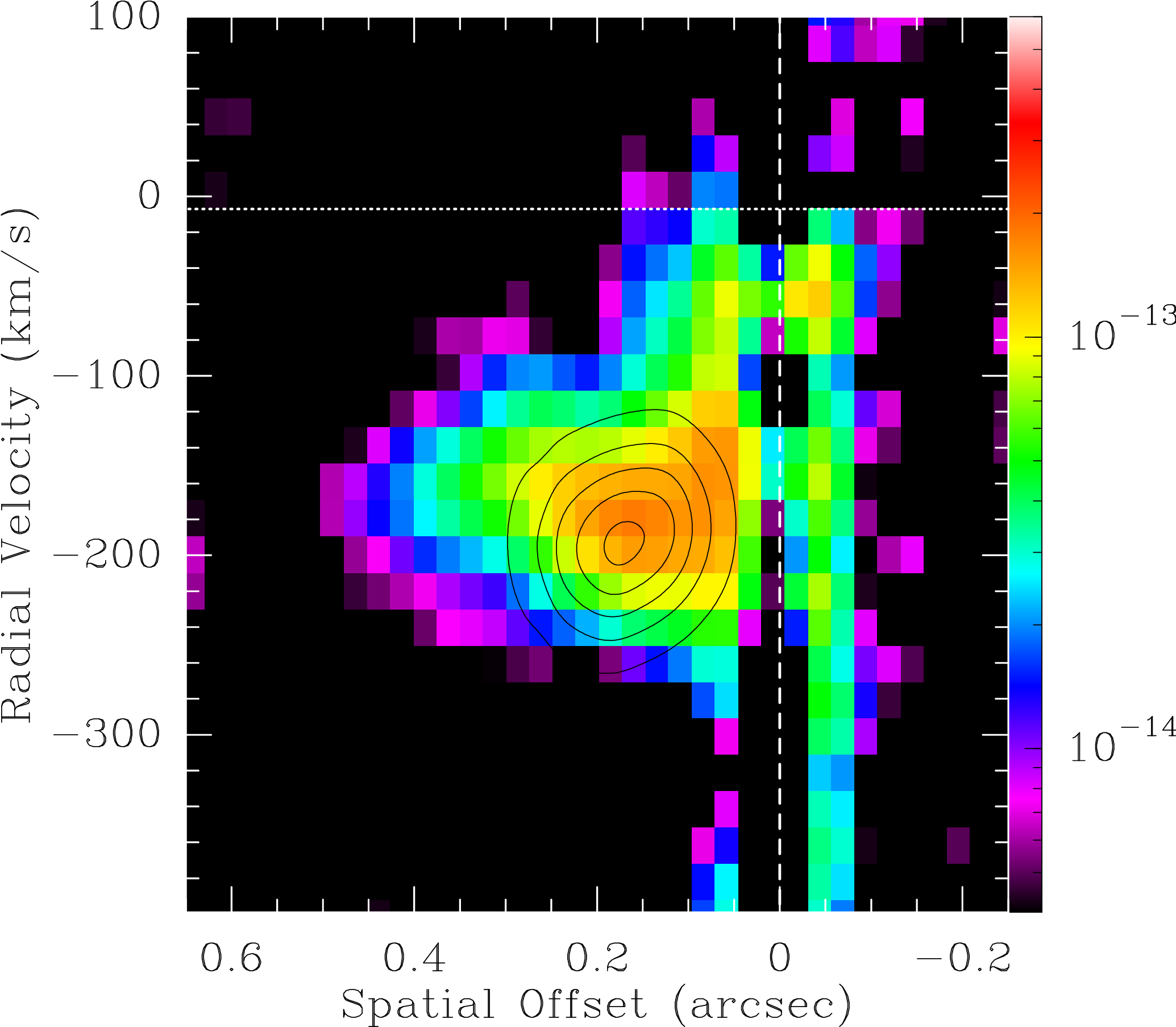}}\\
\subfigure[]{\label{flatoff}\includegraphics[width=80mm]{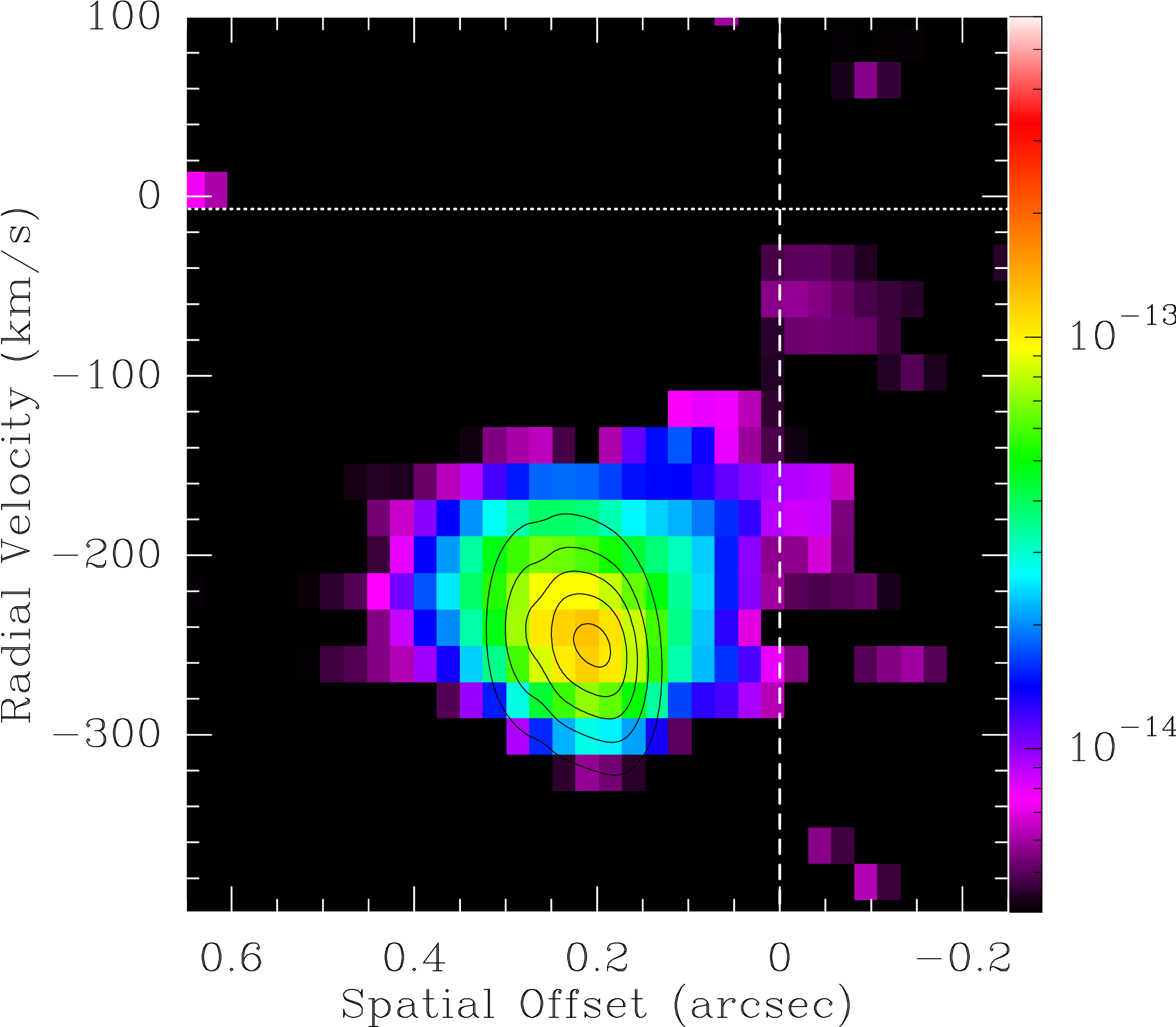}}
\subfigure[]{\label{flaton} \includegraphics[width=80mm]{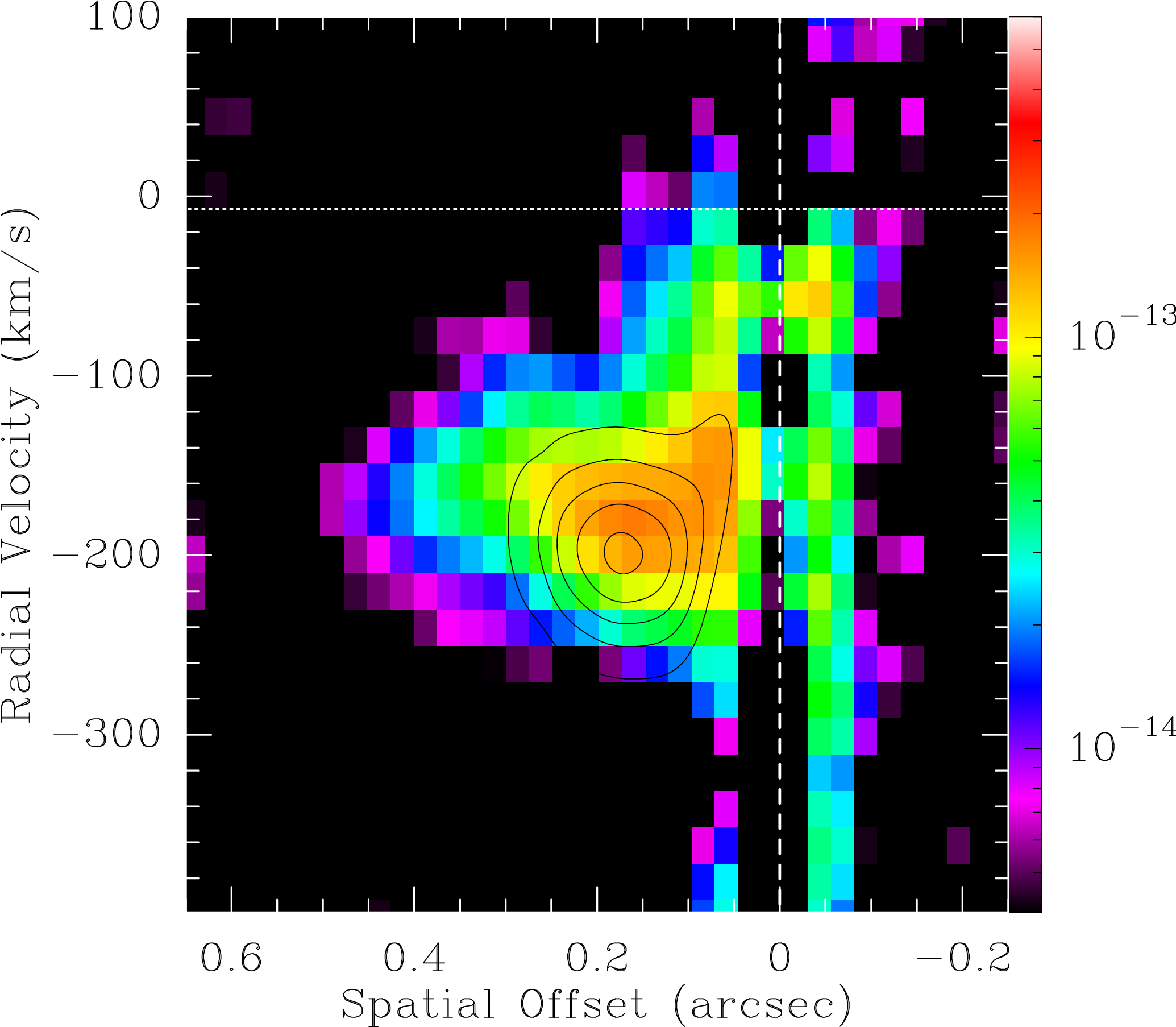}}
\caption{\label{velprogression} We see the progression, starting from top to bottom, from first a constant, linear then flattened velocity law.
Each of these models is a squeezed sphere of length $0\farcs8$ cut 
by an angle of 150 degrees (See Appendix C for more information). Also, in these models $PA'$ = 28$\arcdeg$, i = 26$\arcdeg$, and 
the density is a constant function, n($\varphi$) = 1. The constant velocity model is given a velocity of v($\rho$)(\,\kms) = -245\,\kms,
the linear velocity model is described by the function 
v($\rho$)(\,\kms) = $-200-70(\rho/\ell)$ and the flattened velocity model decreases linearly from -70\,\kms~to -248\,\kms~at $0\farcs3$ and then becomes constant.
}
\end{figure}

Using the same geometry, $PA'$, and inclination angle $i$ just described, we investigated three different velocity laws -- constant, linear and flattened -- and found that the flattened law created the most consistent results. A uniform density was used when different velocity laws were being tested. In Fig.\,\ref{velprogression} we show the progression, from top to bottom, of epoch 1 model contours created from a constant, linear, then flattened velocity law, respectively. 

\begin{figure}
\vspace{2in}
\centering
\includegraphics[width=0.60\textwidth]{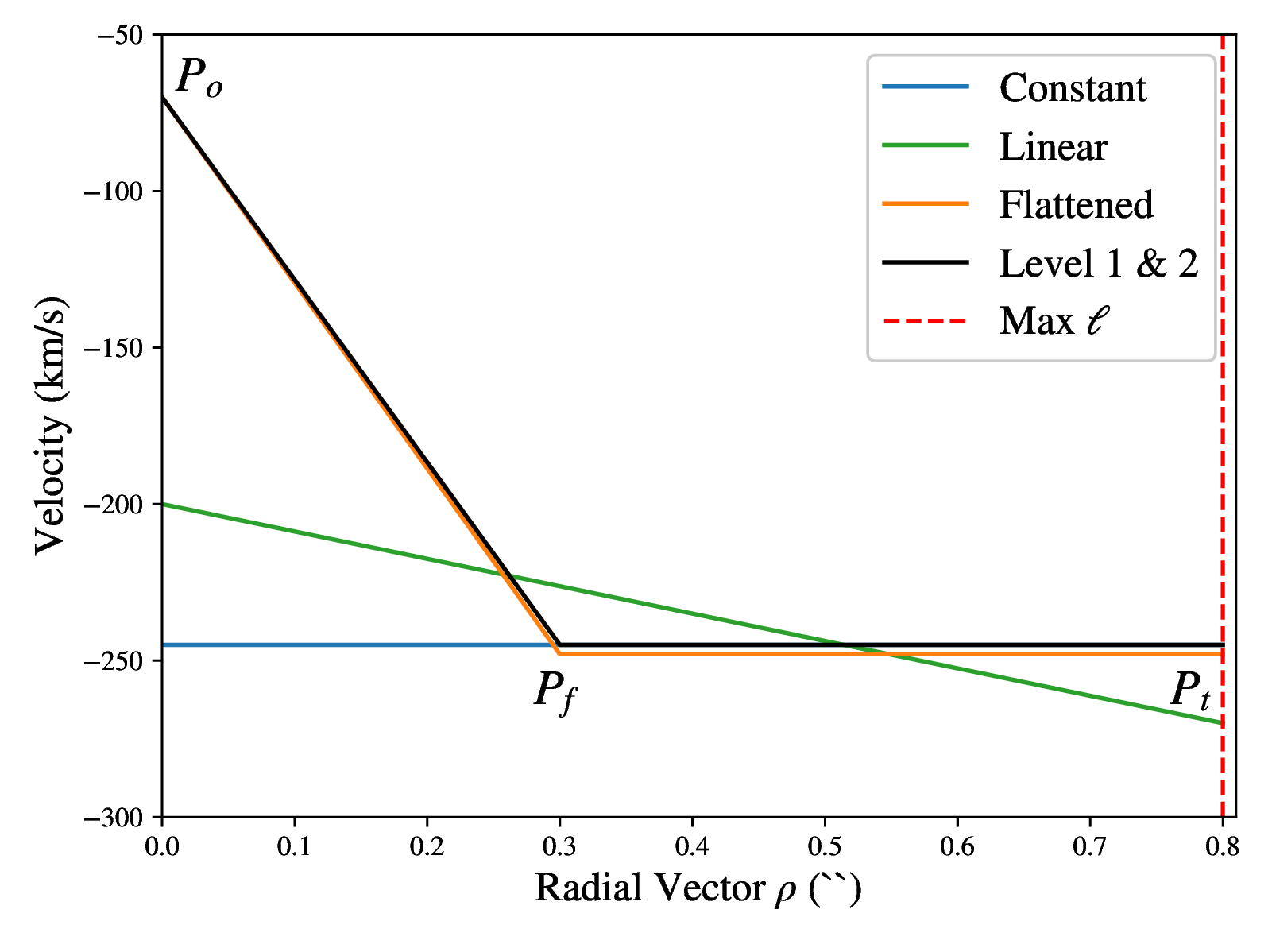}
\caption{\label{vellaw}Velocity laws in blue, green and orange correspond to those used for models shown in the top row, middle row, and bottom row of Fig.\,\ref{velprogression}, respectively. The black line describes the velocity law of both our ``level 1'' and ``level 2'' models (epoch 1) that we define as V$_{flat,1}$ (see Fig.\,\ref{vel_inc_rel}). }
\end{figure}

\begin{figure}
\centering
\includegraphics[width=1.0\textwidth]{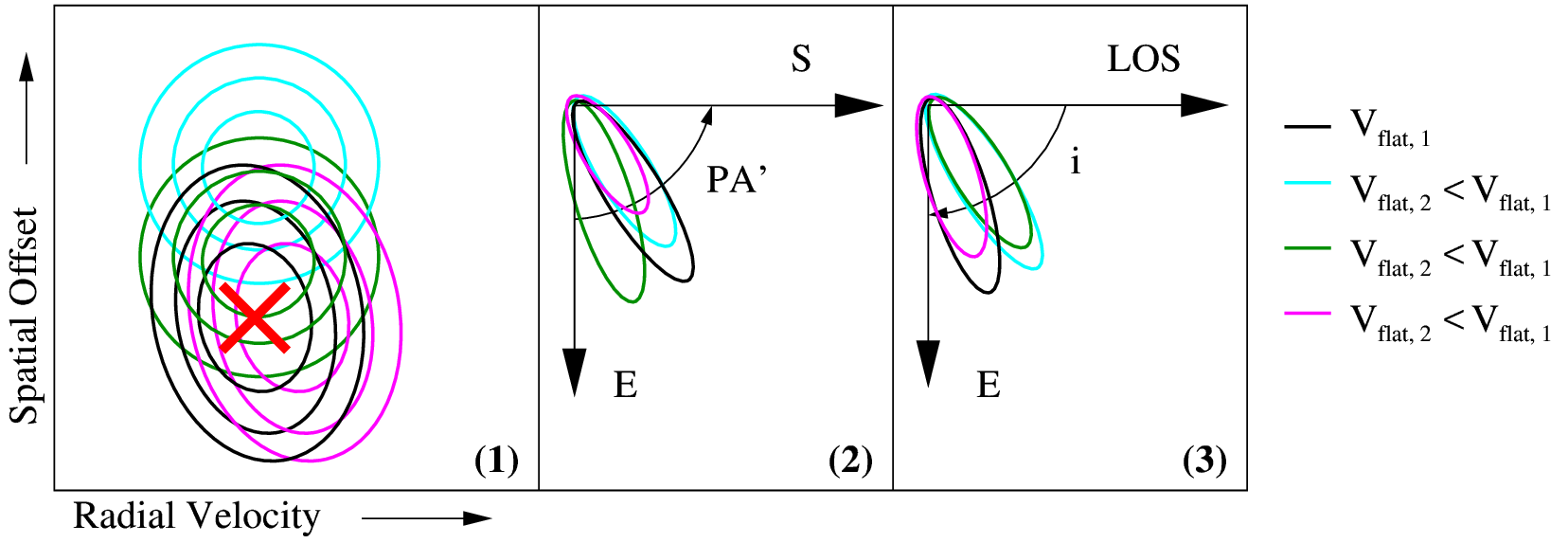} 
\caption{\label{vel_inc_rel} Model [SII] line emission contours as a function of changing $PA'$, inclination and the velocity at the flattening point $P_f$ that parametrises the velocity law in Fig.\,\ref{vellaw} (labelled V$_{flat,i}$, where i=1 for the best-fit model and i=2 for alternative models shown in cyan, green and pink). The red `X' represents the location of the observed emission peak. The ellipses in panels 2 and 3 represent the bullets at different $PA'$ and inclination (their sizes are arbitrary). Black contours are representative of our best-fit ``level 2" model, for epoch 2 and slit $S_{-2t}$. The cyan contours show the effect of choosing a smaller inclination angle towards the LOS (cyan ellipse in panel 3), with a compensating decrease in V$_{flat}$ in order to center the model peak at the correct velocity (cyan contours panel 1). V$_{flat,2}$ is now less than V$_{flat,1}$ and applies to the alternative models in order to show that our best-fit value for V$_{flat, 1}$ is robust. The $PA'$ can be lowered in order to better match the spatial location of the model to data (green ellipse panel 2), however this decreases the model intensity far below the observed one. Finally we show pink contours with the only difference being this new velocity law V$_{flat,2}$, finding the peak emission is still offset in velocity (pink emission contours panel 1). Note for models with the incorrect $PA'$ and inclination the observed morphology remains round (cyan, green ellipses in panel 1), whereas the observed one is elongated. }
\end{figure}

A model with a constant velocity law of v($\rho$) = $-245$\,\kms~(Fig.\,\ref{vellaw} `Constant') does not precisely reproduce the location of the peak emission in position-velocity space in the on-center slit of epoch 1. Specifically, the observations show that the peak emission in the central slit is 
red-shifted relative to that in the offset slit, $S_{-2b}$ by $\sim 65$\,\kms, whereas in the model, this is smaller,
$\sim 45$\,\kms~(Fig.\,\ref{velprogression} (b)).

A linearly-varying velocity of v($\rho$) $= -200 - 70(\rho/\ell)$\,\kms~(Fig.\,\ref{vellaw} `Linear') was also inconsistent with the observed emission. Despite a better fit to the central peak location of the emission in the center slit, the detailed spatio-kinematic PV-morphology is not well reproduced, i.e. the observation shows a near-vertical structure at the peak for the emission blob in the central-slit $S_{0b}$ moving towards more negative spatial offsets, whereas the linear velocity model shows a stronger velocity gradient (Fig.\,\ref{velprogression} (d)).

By altering our velocity law so that it remains constant after a flattening point, $P_f$, at a given length from the origin, $P_o$, the model's spatio-kinematic PV-morphology fits our observations much more precisely, yet it slightly re-shifts the centroid position of the on-center slit (Fig.\,\ref{velprogression} (f)). For the first epoch, the velocity law decreases from $-70$\,\kms~linearly out to a length of $0\farcs3$ and then flattens out to $-248$\,\kms~at $P_t$ (Fig.\,\ref{vellaw} `Flattened'). A flattened velocity law, in addition to fitting the data, is to be expected physically as the tip of the bullet is expected to slow as it encounters more mass (see \S\,\ref{masschange}). We acknowledge that there are only slight differences between these velocity laws, but we argue that it is still qualitatively evident that as we move from constant to linear to flattened, a better fit to the on-slit emission is found, whereas there is no significant change in the off-center slit case. After adopting the flattened velocity law, we then changed the density law to constrain intensity ratios, which were a factor of six off in the case of a uniform density law.


\subsubsection{Density Law} \label{density}

\begin{figure}
\centering
\includegraphics[width=0.60\textwidth]{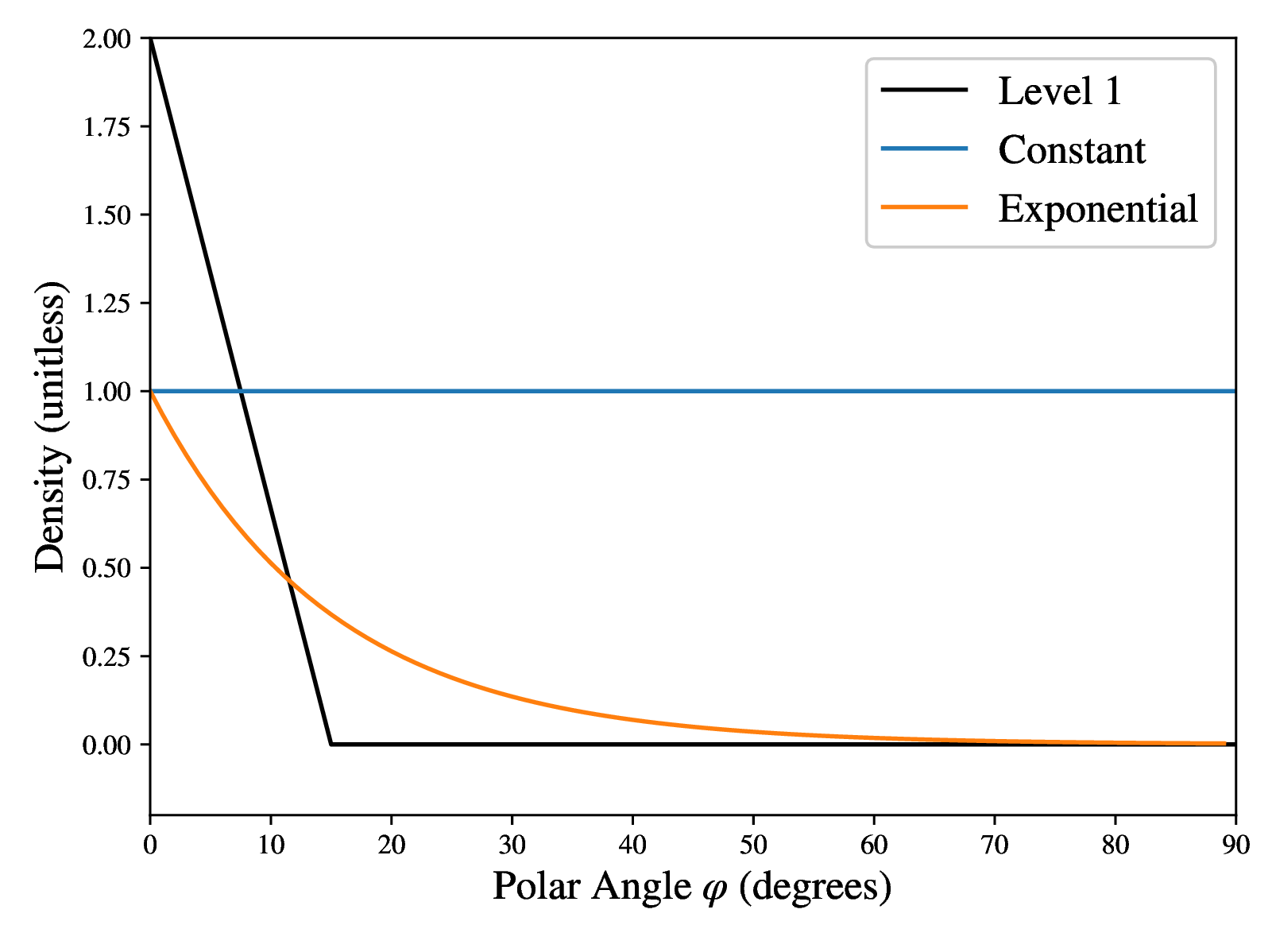}
\caption{\label{densitylaw}Density Laws in blue and orange correspond to the constant law, n($\varphi$) =1, and exponential law, n($\varphi$) = $\exp^{-/\varphi_0}$, respectively. The black line describes the density law of our ``level 1" model which varies away from the axis as a function of polar angle $\varphi$, where $n(\varphi) = 2 - \varphi/\varphi_o$ and
$\varphi_o$ = 15$\arcdeg$ .}
\end{figure}

The three different density laws we consider are only functions of the spherical coordinate $\varphi$ (not $\rho$), which is the ``zenith" angle, measured from the bullet's symmetry axis (see Fig.\,\ref{coordgrid} a). 
We define $R_I$ as the ratio of peak emission intensities in the central and offset slits, as it provides the main constraint on our density law. We find that $R_I$ had large discrepancies compared to the observed ones -- in the ``level 1" model, for example, when we investigated a constant density law and an exponential density law, we found $R_I$ was off by as much as a factor of 6 and 3, respectively. Due to the intensity ratio inconsistency, the constant and exponential laws were discarded. These two laws are depicted graphically in Fig.\,\ref{densitylaw} and labeled ``Constant" and ``Exponential," respectively. 

We find a better density law for our ``level 1" model, considering 
all three epochs, that varies away from the center of the bullet as a function of $\varphi$, where $n(\varphi) = 2 - \varphi/\varphi_o$ and
$\varphi_o$ = 15$\arcdeg$ (Fig.\,\ref{densitylaw} ``Level 1"). Using this density law we find better model $R_I$'s, e.g. model $R_I$ = 0.23 vs. observation $R_I$ = 0.14 for slit ratio $S_{0t}$/ $S_{-1t}$ and model $R_I$ = 0.32 vs. observation $R_I$ = 0.10 for slit ratio $S_{-2t}$/ $S_{-1t}$ . With this new density, the $R_I$'s of the models improve slightly from the exponential law (Table\,\ref{intratios}, ``level 1" varies w/ $\varphi$).

It is important to note that the $PA'$ also affects $R_I$; the $PA'$ can be adjusted to larger
values in order to match centroid emission observed in the next two epochs, i.e.,
progressively weaker emission in the center slit compared to that in the offset slit $S_{-2b}$.
However, we are not able to adjust the $PA'$ to yield correct $R_I$ 
while also keeping the correct center emission positions in all three epochs, hence only the density was adjusted. 

If the density varies more steeply with $\varphi_o$ then $R_I$ improves
for the last two epochs while worsening for the first epoch. This is because in the model a higher density gradient means less emission in the central slit -- at the origin of the emission blob. In epoch 1 our observations tells us there is actually more emission in the central slit while in epochs 2 and 3 there is more emission in the adjacent off-center slit, either $S_{-2b}$ or $S_{-1t}$ (Table \ref{intratios}). This trade-off was carefully balanced in our final density law. 

The ``level 1'' density law (Fig. \,\ref{densitylaw}, ``Level 1"),  i.e. $n(\varphi) = 2 - \varphi/\varphi_o$ where $\varphi_o$ = 15$\arcdeg$ for $0\le\varphi_o\le30$ and $n(\varphi) = 0$ for $\varphi>30$, is kept the same in all three epochs. This density law gives a suitable overall PV-morphology and center location for the emission blobs, as well as $R_I$'s that stay within a factor of three for all of the epochs (Table \ref{intratios}, ``level 1" varies w/ $\varphi$). Due to our density law for ``level 1'', the emission equals zero for $\varphi$ $>$ 30$^o$. Once we decided on this new density law, we implemented a slightly altered velocity law to re-match the centroid peaks (Fig.\,\ref{vellaw} `Level 1 $\&$ 2'). 

However, since the ``level 1'' density law is defined by the polar angle $\varphi$ (Fig.\,\ref{sketch}), additional modifications towards the tip of the bullet could not be achieved. This further justifies the need to create a new density law that has a more dramatic linear drop towards the tip of the bullet, which is beneficial since in later epoch, $R_I$ is strongly affected by weak emission in the off-center slit in addition to low emission in the on-center slit. See \S\,\ref{level2} for further detail on this new density law and our best-fit ``level 2" model. 



\section{Appendix C} \label{appendixC}

SHAPE\footnote{http://www.astrosen.unam.mx/shape/index.html} is comprised of various modules. Within those modules the user is allowed to adjust various parameters. We discuss in this section how to reproduce the ``level 1" model, focusing on parameters given for the first epoch.

\subsection{3D Geometry}

Within the 3D module, where objects are created and given various properties, we choose
a sphere as the ``level 1" starting shape and within this same module we set under the `Primitive' tab a radius, r = $0\farcs8$,
and an angle parameter, $\varphi_{min}$ = 90$\arcdeg$, which cuts the sphere in half. Squeeze, velocity, density, and displacement
modifiers are added by the `Modifier' tab in the 3D module to alter our object further. 

A squeeze modifier is chosen to flatten the semi-sphere into a semi-ellipsoid. Once the modifier was opened we added two points,(0.0,0.92),(10.0,0.0), which express the projection along the y-axis versus ``fraction of squeeze". As this SHAPE parameter ``fraction of squeeze" increases, say from our value of 0.92 to 1.0, the width of the bullet shrinks.
These two points were selected based purely on the qualitative appearance of the 
3D shape once the modifier was added, choosing a width roughly one-fifth of the length (Fig.\,\ref{sketch}).

Within the velocity modifier we chose `Custom' mode in spherical coordinates for a radial vector field. We manually adjusted the $v_r$ component (we label v($\rho$) in the text to avoid confusion with the cylindrical radial parameter $r$)
for the ``level 1" model, as shown in Fig.\,\ref{vellaw}, by adjusting an interactive graph. 

Next, within the density modifier we again choose the 
`Custom' mode in spherical coordinates and adjusted only the $n_{\varphi}$ polar angle component while keeping
the other components at a constant of 1, i.e. the constant multiplier
factor $n_0 = 1$, an azimuthal angle component $n_{\theta} = 1$, and a radial component $n_{r} = 1$.
Next, we entered the points used for the ``level 1" 
model: $(\varphi, n)$ = (0$\arcdeg$,2.0),(15$\arcdeg$,0.0),(90$\arcdeg$,0.0).

Note that for the next two epochs a displacement modifier was used; where negative `translation east' values were put in for $y$
and positive `translation south' values were put in for $x$; see Table\,\ref{varymodelparams} for specific values. 

\subsection{Rendering}

The render module in general controls the imaging output as well as the PV diagrams; it consists of sections for the image display, tool bars 
for the control of the image type and appearance, as well as a parameter panel from where to control rendering, seeing and velocity resolution, spectrograph slit properties, camera rotation, and other parameters. 
 
We chose to set the `World Units' to arcseconds in the render module under the tab `Units'. In the render module we also set under the `Camera' tab, for the case of epoch 1, $PA'$ = 17$\arcdeg$ and inclination, $i_{shape} = 145 \arcdeg$ (set to 180$\arcdeg$ minus the inclination given in Table\,\ref{varymodelparams}).
 
To account for the instrumental effect (\S\,\ref{instrumeffects}) we went to the modifiers parameter under the `Render' tab in the render module and added the Lambda-Shift parameter (in units of m/arcsec, i.e., $5.4352\times10^{-10}$, see \ref{instrumeffects}).
Additionally, we set the renderer parameter to `basic' in the render module under `Render' tab.
The convolution factor was applied by selecting both `PV diagrams' and `Images' (i.e. position diagrams) in SHAPE under the `Render' tab.

\end{appendices}





\clearpage
\newpage


\scriptsize
\begin{turnpage}
\begin{longtable}{p{0.7in}p{0.35in}p{0.35in}p{0.35in}p{0.6in}p{0.6in}}
\caption[]{Observed Intensity Values}\\
\hline \\[-2ex]
   \multicolumn{1}{l}{\textbf{Epoch}} &
   \multicolumn{1}{l}{\textbf{Slit}} &
   \multicolumn{1}{l}{\textbf{Offset}} &
   \multicolumn{1}{l}{\textbf{Width}} &
   \multicolumn{2}{c}{\textbf{Int.\footnotemark[1] (cgs)}}  \\[0.5ex]
   \multicolumn{1}{l}{\textbf{}} &
   \multicolumn{1}{l}{\textbf{Name}} &
   \multicolumn{1}{c}{\textbf{}} &
   \multicolumn{1}{l}{\textbf{}} &
   \multicolumn{1}{l}{\textbf{Detached}} &
   \multicolumn{1}{l}{\textbf{On-source}} \\[0.5ex] \hline
   \\[-1.8ex]
\endfirsthead

\multicolumn{3}{c}{{\tablename} \thetable{} -- Continued} \\[0.5ex]
\hline \\[-2ex]
   \multicolumn{1}{l}{\textbf{Epoch}} &
   \multicolumn{1}{l}{\textbf{Slit}} &
   \multicolumn{1}{l}{\textbf{Offset}} &
   \multicolumn{1}{l}{\textbf{Width}} &
   \multicolumn{2}{c}{\textbf{Int.\footnotemark[1] (cgs)}}  \\[0.5ex]
   \multicolumn{1}{l}{\textbf{}} &
   \multicolumn{1}{l}{\textbf{Name}} &
   \multicolumn{1}{c}{\textbf{}} &
   \multicolumn{1}{l}{\textbf{}} &
   \multicolumn{1}{l}{\textbf{Detached}} &
   \multicolumn{1}{l}{\textbf{On-source}} \\[0.5ex] \hline
   \\[-1.8ex]
\endhead
2002-01-28  & $S_{-2b}$ & $-0\farcs2$  &   $0\farcs2$    &  1.6e-13  &   
      \\
            & $S_{0b}$  & $0.0$        &   $0\farcs2$    &   2.2e-13 &   --      \\
            & $S_{+2b}$ & $0\farcs2$   &   $0\farcs2$    &    --      &    --    \\
2002-12-29  & $S_{-2t}$ & $-0\farcs2$  &   $0\farcs1$    &  4.6e-14 &   --    \\
            & $S_{-1t}$ & $-0\farcs1$  &   $0\farcs1$    &   4.3e-13 &   --     \\
            & $S_{0t}$  & $0.0$        &   $0\farcs1$    & 5.1e-14  &   --    \\
            & $S_{+1t}$ & $0\farcs1$   &   $0\farcs1$    &  --   &   --    \\
            & $S_{+2t}$ & $0\farcs2$   &   $0\farcs1$    &  --   &   --    \\
2004-01-12  & $S_{-1t}$ & $-0\farcs1$  &  $0\farcs1$     & 3.4e-13  &   --    \\
	    & $S_{0t}$  & $0.0$        &  $0\farcs1$     & 5.5e-14 & 7.9e-13 \\
	    & $S_{+1t}$ & $+0\farcs1$  &  $0\farcs1$     &  --    &  --  \\
	    & $S_{+2t}$ & $+0\farcs2$  &  $0\farcs1$     &   --   &  --  \\
2011-07-07  & $S_{+2b}$ & $-0\farcs2$  &  $0\farcs2$     & 2.5e-14 & -- \\
	    & $S_{0b}$  & $0.0$        &  $0\farcs2$     & --  &2.2e-13\\
	    & $S_{-2b}$ & $+0\farcs2$  &  $0\farcs2$     & --  &  -- 	\\
2012-07-14  & $S_{+2b}$ & $-0\farcs2$  &  $0\farcs2$     & 9.1e-15 & 2.4e-14\\
	    & $S_{0b}$  & $0.0$        &  $0\farcs2$     &  --   &   8.2e-13\\
	    & $S_{-2b}$ & $+0\farcs2$  &  $0\farcs2$     &    --  &  2.2e-14\\
2013-07-17  & $S_{+2b}$ & $-0\farcs2$  &  $0\farcs2$     &  1.2e-14& --    \\
	    & $S_{0b}$  & $0.0$        &  $0\farcs2$     & 1.3e-14     & 5.3e-13 \\
	    & $S_{-2b}$ & $+0\farcs2$  &  $0\farcs2$     & --   & --  \\
\label{modeltable}\\
\end{longtable}
\end{turnpage}
\footnotetext[1]{Line Intensity units are erg s$^{-1}$ cm$^{-2}$ arcsec$^{-2}$, obtained by integrating the line emission over velocity.}

\clearpage
\newpage 

\scriptsize
\begin{turnpage}
\begin{longtable}{p{1.0in}p{0.7in}p{0.7in}p{0.7in}p{0.9in}p{0.9in}}
\caption[]{Model Central Intensity Values }\\
\hline \\[-2ex]
   \multicolumn{1}{l}{\textbf{Epoch}} &
   \multicolumn{1}{l}{\textbf{Slit}} &
   \multicolumn{1}{l}{\textbf{Offset}} &
   \multicolumn{1}{l}{\textbf{Width}} &
   \multicolumn{1}{c}{\textbf{Int.}}  &
      \multicolumn{1}{c}{\textbf{Int.}} \\[0.7ex]
   \multicolumn{1}{l}{\textbf{}} &
   \multicolumn{1}{l}{\textbf{Name}} &
   \multicolumn{1}{c}{\textbf{}} &
   \multicolumn{1}{l}{\textbf{}} &
   \multicolumn{1}{l}{\textbf{Detached}} &
   \multicolumn{1}{l}{\textbf{On-Source}}  \\[0.7ex] \hline
   \\[-1.8ex]
\endfirsthead

\multicolumn{3}{c}{{\tablename} \thetable{} -- Continued} \\[0.7ex]
\hline \\[-2ex]
   \multicolumn{1}{l}{\textbf{Epoch}} &
   \multicolumn{1}{l}{\textbf{Slit}} &
   \multicolumn{1}{l}{\textbf{Offset}} &
   \multicolumn{1}{l}{\textbf{Width}} &
   \multicolumn{1}{c}{\textbf{Int.}}  &
      \multicolumn{1}{c}{\textbf{Int.}} \\[0.7ex]
   \multicolumn{1}{l}{\textbf{}} &
   \multicolumn{1}{l}{\textbf{Name}} &
   \multicolumn{1}{c}{\textbf{}} &
   \multicolumn{1}{l}{\textbf{}} &
   \multicolumn{1}{l}{\textbf{Detached}} &
   \multicolumn{1}{l}{\textbf{On-Source}}  \\[0.7ex] \hline
   \\[-1.8ex]
\endhead

2002-01-28  & $S_{-2b}$ & $-0\farcs2$  &   $0\farcs2$  &  2.9e-14                            & 1.6e-13\footnotemark[1]\\
            & $S_{0b}$  & $0.0$        &   $0\farcs2$    &   3.5e-14                         &  1.93e-13\\
            & $S_{+2b}$ & $0\farcs2$   &   $0\farcs2$    &    --        & --  \\
2002-12-29  & $S_{-2t}$ & $-0\farcs2$  &   $0\farcs1$    &  3.1e-15$|$6.5e-15\footnotemark[2]  & 1.71e-14$|$-- \\
            & $S_{-1t}$ & $-0\farcs1$  &   $0\farcs1$    &  3.8e-14                          &   2.10e-13 \\
            & $S_{0t}$  & $0.0$        &   $0\farcs1$    & 3.7e-15$|$5.4e-15\footnotemark[2]   &  2.04e-14$|$--\\
            & $S_{+1t}$ & $0\farcs1$   &   $0\farcs1$    &  --    & --  \\
            & $S_{+2t}$ & $0\farcs2$   &   $0\farcs1$    &  --    & --  \\
2004-01-12  & $S_{-1t}$ & $-0\farcs1$  &  $0\farcs1$     & 3.3e-14                           & 1.82e-13 \\
	    & $S_{0t}$  & $0.0$        &  $0\farcs1$     & 4.9e-15$|$1.7e-14\footnotemark[2]       & 2.71e-14$|$--\\
	    & $S_{+1t}$ & $+0\farcs1$  &  $0\farcs1$     &  --   & --    \\
	    & $S_{+2t}$ & $+0\farcs2$  &  $0\farcs1$     &   --  & -- \\
\label{modelparamtable}\\
\end{longtable}
\end{turnpage}
\footnotetext[1]{Scaled to Observed Intensity of Detached blob.}
\footnotetext[2]{Observed$|$Model: The first ratio was calculated from intensities at the location of the observed peak, the second ratio was calculated from the peak intensities in the model.}

\clearpage
\newpage
\scriptsize
\begin{turnpage}
\begin{longtable}{p{0.7in}p{1.2in}p{0.35in}p{0.35in}}
\caption[]{Intensity Ratios}\\
\hline \\[-2ex]
   \multicolumn{1}{l}{\textbf{Epoch}} &
   \multicolumn{1}{l}{\textbf{}} &
   \multicolumn{2}{c}{\textbf{$R_I$}} \\[0.5ex]
   \multicolumn{1}{l}{\textbf{}} &
   \multicolumn{1}{l}{\textbf{}} &
   \multicolumn{1}{l}{\textbf{$S_{0b}$/$S_{-2b}$}} &
   \multicolumn{1}{l}{\textbf{}} \\[0.5ex] \hline
   \\[-1.8ex]
\endhead
2002-01-28 & observation    & 1.4 & --\\
     & ``level 2"            & 1.2 & --\\
     & ``level 1" varies w/ $\varphi$ & 0.95 & --\\
     & ``level 1" $e^{-\varphi/\varphi_o}$ & 1.01 & -- \\
     & ``level 1" n(r) = 1    & 1.1  & -- \\  

\hline \\[-2ex]
   \multicolumn{1}{l}{\textbf{}} &
   \multicolumn{1}{l}{\textbf{}} &
   \multicolumn{1}{l}{\textbf{$S_{0t}$/$S_{-1t}$}} &
   \multicolumn{1}{l}{\textbf{$S_{-2t}$/$S_{-1t}$}} \\[0.5ex] \hline
   \\[-1.8ex]
 2002-12-29& observation & 0.14 & 0.10 \\
     & ``level 2"            & 0.10$|$0.14\footnotemark[1] & 0.08$|$0.17\footnotemark[1] \\
     & ``level 1" varies w/ $\varphi$  & 0.23 & 0.32 \\
     & ``level 1" $e^{-\varphi/\varphi_o}$ & 0.34 & 0.37 \\
     & ``level 1" n(r) = 1 & 0.61 & 0.52 \\
\hline \\[-2ex]
   \multicolumn{1}{l}{\textbf{}} &
   \multicolumn{1}{l}{\textbf{}} &
   \multicolumn{1}{l}{\textbf{$S_{0t}$/$S_{-1t}$}} &
   \multicolumn{1}{l}{\textbf{}} \\[0.5ex] \hline
   \\[-1.8ex]
2004-01-12 & observation & 0.16 & --\\
     & ``level 2"     & 0.15$|$0.52\footnotemark[1] & -- \\
     & ``level 1" varies w/ $\varphi$       & 0.31 & -- \\
     & ``level 1" $e^{-\varphi/\varphi_o}$ & 0.39 & -- \\
     & ``level 1" n(r) = 1 & 0.94 & -- \\
\label{intratios}\\
\end{longtable}
\end{turnpage}
\footnotetext[1]{Observed$|$Model: The first ratio was calculated from intensities at the location of the observed peak, the second ratio was calculated from the peak intensities in the model.}

\clearpage
\newpage 
\scriptsize
\begin{turnpage}
\begin{longtable}{p{0.55in}p{0.35in}p{0.35in}p{0.75in}p{0.85in}p{0.85in}p{0.30in}p{0.30in}p{1.0in}}
\caption[]{Model Parameters for Level 1 and 2}\\
\hline \\[-2ex]
   \multicolumn{1}{l}{\textbf{Epoch}} &
   \multicolumn{1}{l}{\textbf{$\ell$} ($\arcsec$)} &
   \multicolumn{1}{l}{\textbf{r$_{max}$} ($\arcsec$)} &
   \multicolumn{1}{l}{\textbf{Velocity Law (\,\kms)\footnotemark[1]}} &
   \multicolumn{1}{l}{\textbf{}} &
   \multicolumn{1}{l}{\textbf{}} &
   \multicolumn{1}{l}{\textbf{$PA'$ ($\arcdeg$)}} &
   \multicolumn{1}{l}{\textbf{i ($\arcdeg$)}} &
   \multicolumn{1}{l}{\textbf{Transl.}} \\[0.5ex] 
   \multicolumn{1}{l}{\textbf{}} &
   \multicolumn{1}{l}{\textbf{}} &   
   \multicolumn{1}{l}{\textbf{}} &
   \multicolumn{1}{l}{\textbf{($\rho, v$) at $P_o$}} &
   \multicolumn{1}{l}{\textbf{($\rho, v$) at $P_f$}} &
   \multicolumn{1}{l}{\textbf{($\rho, v$) at $P_t$}} &
   \multicolumn{1}{c}{\textbf{}} &
   \multicolumn{1}{l}{\textbf{}} &
    \multicolumn{1}{l}{\textbf{(east, south) ($\arcsec$)}} \\[0.5ex] \hline
   \\[-1.8ex]
\endhead
2002-01-28 & 0.8 & 0.075  & (0.0, -70.0) & (0.3, -245.0) & (0.8, -245.0) &28 & 26 & (0.0,  0.0)     \\
2002-12-29& 0.9185 & 0.097 & (0.0338, -70.0) & (0.4185, -245.0) & (0.9185, -245.0) & 23 & 32 & (0.01683, 0.00613) \\
2004-01-12& 1.0497 & 0.110 & (0.0721, -70.0) & (0.6682, -245.0) & (1.0497, -245.0) & 17 & 35  & (0.04013, 0.01000)\\
\label{varymodelparams}\\
\end{longtable}
\end{turnpage}
\footnotetext[1]{Coordinates of cardinal points $P_t$, $P_f$ and $P_o$ for each of the three epochs correspond to the radial position $\rho$ and the velocity at that point (i.e. our flattened velocity law).}

\clearpage
\newpage 
\scriptsize
\begin{turnpage}
\begin{longtable}{p{1.0in}p{0.5in}p{1.9in}}
\caption[]{Model Parameters for Level 2\footnotemark[1]}\\
\hline \\[-2ex]
   \multicolumn{1}{l}{\textbf{`Squeeze'\footnotemark[2]}} &
   \multicolumn{1}{l}{\textbf{$\varphi_{min}$ ($\arcdeg$)}} &
   \multicolumn{1}{l}{\textbf{Density}} \\[0.5ex] \hline
   \\[-1.8ex]
\endhead
(0,0.98),(5.0,0.0) & 90 & $n(r) = 1-9.0r$; 0 $<$ $r$ $<$ r$_{max}$ \\
\label{constantmodelparams}\\
\end{longtable}
\end{turnpage}
\footnotetext[1]{These model parameters remain constant from epoch to epoch in ``level 2" (see \S\,\ref{level2}), except $r_{max}$ varies as the geometry changes in each epoch (i.e. in epoch 1 $r_{max}$=$0\farcs075$).}
\footnotetext[2]{The coordinates are defined as the projection along the y-axis and the `fraction of squeeze' defined as `1' being compressed to a line and `0' being completely unsqueezed.}

\clearpage
\newpage
\scriptsize
\begin{turnpage}
\begin{longtable}{p{0.5in}p{0.5in}p{0.5in}p{0.5in}p{0.5in}p{0.5in}p{0.5in}p{0.5in}}
\caption[]{Radial and Perpendicular Offsets of Bullets from V\,Hya \footnotemark[1]}\\
\hline \\[-2ex]
   \multicolumn{1}{l}{\textbf{Bullet \#}} &
   \multicolumn{1}{l}{\textbf{Epoch \#}} &
   \multicolumn{1}{l}{\textbf{x\footnotemark[2] (AU)}} &
   \multicolumn{1}{l}{\textbf{y (AU)}} &
   \multicolumn{1}{l}{\textbf{x$_0$\footnotemark[3] (AU)}} &
   \multicolumn{1}{l}{\textbf{$t_0$ (yr)}} &
   \multicolumn{1}{l}{\textbf{$t_1$-$t_0$\footnotemark[3] (yr)}} &
   \multicolumn{1}{l}{\textbf{V$_r$ (\kms)}} \\[0.5ex] \hline
   \\[-1.8ex]
\endhead
0 &  1 &  664 & 400 & 0 &  0.0 & 16.0 & 157 \\
\hline
1 &  1 & 234 & 64 &  0 & 0.0 & 7.6 & 146 \\
-- & 2 & 275 & 94 &  234 & 7.6 & 1.0 & 193 \\
-- & 3 & 313 & 122 & 275 & 8.6 & 1.0 & 183 \\
-- & 4 &  589 &  300 & 313 & 9.6 & 7.6 & 172 \\
-- & 5 &  624 & 320  &  589 & 17.2 & 1.0 & 165 \\ 
\hline
2 & 3 &  28  & 22 & 0 & 0.0 & 1.0 & 132  \\
-- & 4 &  244 & 94 &  28 & 1.0 & 7.6 & 153 \\
-- & 5 &   281 & 125 & 244 & 8.6 & 1.0 & 173 \\
-- & 6 &  315  & 144 & 281 & 15.2 & 1.0 & 162 \\ 
 \hline
 3  & 4 &  0 & 13 &   0 & 0.0 & 0.0 & 153\\
 -- & 5 & 36  & 13  & 0  & 1.0 & 1.0 & 173 \\
 -- & 6 &  77 & 28 &  36  & 2.0 & 1.0 & 183 \\
\label{centroid}\\
\end{longtable}
\end{turnpage}
\footnotetext[1]{We report offsets calculated from observational constraints, which are used to construct Fig.\,\ref{distantblobsch}}
\footnotetext[2]{There are uncertainties in the absolute radial (x) locations of these bullets. Detached bullet\#2 x position was estimated using only one on-source measurement, whose radial position is not precisely known. Detached bullet\#1's x position is estimated from our model (see \S\,\ref{radialacc}. }
\footnotetext[3]{x$_0$ is the initial offset at initial time $t_0$ when bullet is ejected, $t_1-t_0$ is the time 
interval between ejection and the epoch in Col.\,2. We estimate that bullet\#2's epoch 3 is $\sim$1 year from initial ejection with same initial velocity (153\,\kms). }

\clearpage
\newpage

\clearpage
\newpage

\clearpage
\newpage

\clearpage
\newpage

\clearpage
\newpage

\clearpage
\newpage

\clearpage
\newpage

\clearpage
\newpage

\clearpage
\newpage

\clearpage
\newpage

\clearpage
\newpage
\begin{figure}
\centering
\includegraphics[width=1.0\textwidth]{bullet_trajectory.ps}
\caption{\label{distantblobsch} Schematic representation of the calculated bullet trajectory from V Hya observations. The paths of bullets \#0 (brown), \#1 (blue), \#2 (red) and \#3 (cyan) are also shown. The elliptical symbols show the bullet locations. Non-italicized numbers in bold are the associated LOS velocities in \,\kms, relative to V Hya's systemic radial velocity, $V_{hel}$ = -7\,\kms.  Tangential velocities are also displayed in italics. Equatorial East is to the left and bullet inclination increases away from the LOS (as shown). We provided the estimated radial position of of bullet \#2 when it was newly ejected and of distant bullet\#0 (single LOS velocity measurement for these points). The placement of the first location for bullet\#1 (in blue) was calculated using our model.}
\end{figure}

\clearpage
\newpage

\clearpage
\newpage

\clearpage
\newpage

\clearpage
\newpage

\clearpage
\newpage

\clearpage
\newpage

\clearpage
\newpage

\clearpage
\newpage

\clearpage
\newpage

\clearpage
\newpage

\clearpage

\end{document}